\documentclass[aps,prd,reprint,nofootinbib,longbibliography]{revtex4-1}

\usepackage{amsmath}
\usepackage{mathtools}
\usepackage{graphicx}
\usepackage[normalem]{ulem}
\usepackage[com pat=1.1.0]{tikz-feynman}
\usepackage{tikz}
\usetikzlibrary{external}
\usepackage{comment}
\usetikzlibrary{arrows}
\usepackage[colorlinks=true, allcolors=blue]{hyperref}

\AtBeginDocument{%
    \newwrite\bibnotes
    \def\bibnotesext{Notes.bib}
    \immediate\openout\bibnotes=\jobname\bibnotesext
    \immediate\write\bibnotes{@CONTROL{REVTEX41Control}}
    \immediate\write\bibnotes{@CONTROL{%
    apsrev41Control,author="08",editor="1",pages="1",title="0",year="1"}}
     \if@filesw
     \immediate\write\@auxout{\string\citation{apsrev41Control}}%
    \fi
}%

\begin{document}

\title{Neutrino Quantum Kinetics in Compact Objects}

\author{Sherwood A. Richers}
\email{sricher@ncsu.edu}
\author{Gail C. McLaughlin}
\author{James P. Kneller}
\author{Alexey Vlasenko}
\affiliation{North Carolina State University}

\begin{abstract}
Neutrinos play a critical role of transporting energy and changing the lepton density within core-collapse supernovae and neutron star mergers. The quantum kinetic equations (QKEs) combine the effects of neutrino-matter interactions treated in classical Boltzmann transport with the neutrino flavor-changing effects treated in neutrino oscillation calculations. We present a method for extending existing neutrino interaction rates to full QKE source terms for use in numerical calculations. We demonstrate the effects of absorption and emission by nucleons and nuclei, electron scattering, electron-positron pair annihilation, nucleon-nucleon bremsstrahlung, neutrino-neutrino scattering. For the first time, we include all these collision terms self-consistently in a simulation of the full isotropic QKEs in conditions relevant to core-collapse supernovae and neutron star mergers. For our choice of parameters, the long-term evolution of the neutrino distribution function proceeds similarly with and without the oscillation term, though with measurable differences. We demonstrate that electron scattering, nucleon-nucleon bremsstrahlung processes, and four-neutrino processes dominate flavor decoherence in the protoneutron star (PNS), absorption dominates near the shock, and all of the considered processes except elastic nucleon scattering and neutrino-neutrino processes are relevant in the decoupling region. Finally, we propose an effective decoherence opacity that at most energies predicts decoherence rates to within a factor of 10 in our model PNS and within $20\%$ outside of the PNS.
\end{abstract}
\maketitle

\section{Introduction}
Neutrinos interacting with matter are central to the mechanism by which massive ($\gtrsim 10 M_\odot$) stars explode as core-collapse supernovae (CCSNe, see, e.g., \cite{Colgate1966,Bethe1990,Janka2012}). When the iron core of such a star exceeds its effective Chadrasekhar mass, it becomes unstable to collapse and the center reaches nuclear densities within a couple hundred milliseconds. At this point, nuclear forces cause the equation of state to quickly stiffen, halting collapse and launching a shock wave into the still supersonically-infalling iron core. This initial shock is not strong enough to lead to an explosion. Neutrinos emitted as matter accretes onto the protoneutron star (PNS) have the potential to transport enough energy to the fluid under the shock front to lead to an explosion. However, a detailed accounting of precisely how this happens remains elusive.

Numerical simulations have become the primary tool for studying the nonlinear and multi-physics dynamics in CCSNe. Current state of the art simulations account for neutrino transport by treating neutrinos as classical radiation and employing a variety of implementations of the relativistic Boltzmann equation or approximations thereof (e.g., \cite{OConnor2018,OConnor2018a,Summa2017,Richers2017a,Vartanyan2018,Kotake2018,Cabezon2018}). Over the years since the first numerical model of a neutrino-driven supernova by \citet{Colgate1966}, many different neutrino-matter interaction processes have been found to be important, including absorption and emission by nucleons and nuclei, scattering by nucleons and electrons, pair production and annihilation, nucleon-nucleon bremsstrahlung radiation, and neutrino-neutrino pair annihilation and scattering (see \cite{Bruenn1985,Burrows2006} for overviews).

However, it has long been known that neutrinos are able to change flavor in-flight\cite{Pontecorvo1968,Wolfenstein1978,Mikheyev1989,Pantaleone1992}, though many of the relevant parameters are not yet well constrained \cite{Capozzi2018}. Understanding how the neutrino flavor evolves from the neutrino's emission in the CCSN to its detection on earth is necessary in order to use neutrino signals to gain valuable insight into the depths of the CCSN explosion mechanism (e.g., \cite{Horiuchi2018}). Historically, calculations of flavor conversion of CCSN neutrinos have been performed by evolving neutrinos moving outward from the neutrinosphere (see \cite{Duan2010,Bellini2013} for recent reviews). In contrast to neutrino transport calculations, simulations of neutrino flavor conversion neglect collisions but include some aspects of neutrino-matter interactions in the form of a potential through which neutrinos propagate. Neutrino-neutrino interactions also contribute to this potential, making the evolution equations very nonlinear and difficult to simulate over timescales relevant to the CCSN.

If neutrinos are experiencing flavor oscillations and collisions in widely separated regions of the supernovae then the no-oscillation approximation in neutrino transport calculations and the no-collision approximation in neutrino flavor conversion calculations at first glance appear very reasonable. Neutrinos are emitted in flavor states inside of the supernova shock wave and the large matter-induced contribution to the neutrino potential suppresses flavor oscillations in regions where collisions are important \cite{Wolfenstein1979}. However, including the contribution of neutrinos themselves to this potential (namely, the "self-interaction" part of the potential) makes the problem much more rich. The first calculations including the full potential but only one neutrino energy and propagation direction suggested that flavor oscillations could occur within the shock \cite{Duan2006}. Subsequent calculations that include multiple energies and/or propagation directions indicate that, though still occurring inside the shock, oscillations begin at too large of a radius to significantly affect the CCSN explosion mechanism \cite{Chakraborty2011,Duan2011,Dasgupta2012}. Once the neutrinos are beyond the strong matter potential, the matter density is so low that scattering events are exceedingly rare, making collision terms unimportant where flavor oscillations take place.

One could take this as evidence that a full quantum kinetic treatment including both collisions and oscillations is not needed, but recent developments have suggested that there may be scenarios where neutrino oscillations and scattering may not be so separable. So-called fast flavor conversions may occur deep within the supernova shock wave if the electron neutrino and electron antineutrino angular distributions differ enough that they intersect in momentum space. In this way, in some directions the contribution to the potential from antineutrinos overwhelms that from neutrinos, allowing neutrino flavor states to mix \cite{Sawyer2005,Sawyer2016,Chakraborty2016,Izaguirre2017,Capozzi2017,Dasgupta2018,Abbar2018}. Though the lepton-emission self-sustained asymmetry \cite{Tamborra2017,OConnor2018a} may be conducive to fast flavor conversions\cite{Dasgupta2017,Chakraborty2016}, searches for conditions conducive to fast flavor conversions in simulations of CCSNe have had mixed results \cite{Tamborra2017,Abbar2018a,Azari2019}.

The neutrino halo effect is another such scenario where collision and oscillation physics are simultaneously relevant. Even though neutrinos are decoupled outside of the shock, the small changes to the neutrino distribution from rare scattering events can enhance the neutrino self-interaction potential and significantly change the location and strength of the flavor conversions \cite{Cherry2012,Sarikas2012,Mirizzi2012,Cherry2013,Cirigliano2018}. Though the halo effect is not expected to have an impact on the CCSN explosion mechanism, it could significantly change the neutrino signal we might expect to see from the next nearby CCSN. 

Neutron star mergers also create an environment with complex neutrino radiation fields, and the details of this radiation field have profound effects on the amount of mass ejected from the merger, the eventual composition of this ejecta, the fate of the central remnant, and  potential launching of a relativistic jet (e.g., \cite{Rosswog2003,Dessart2008,Metzger2014,Richers2015,Fujibayashi2017,Perego2017,Wu2017a,Radice2018,Foucart2018}). It has been shown that the neutrinos are also likely subject to a number of interesting flavor transformation effects that are expected to occur near the decoupling region (e.g., \cite{Zhu2016,Malkus2016,Vaananen2016,Chatelain2017,Chatelain2018,Vlasenko2018,Deaton2018,Cirigliano2017,Tian2017a,Wu2017}).  However, the effects of quantum kinetics in these systems is yet to be investigated in detail.

To account for both neutrino flavor conversions and collisions consistently, one must evolve the neutrino quantum kinetic equations (QKEs) \cite{Sigl1993,Vlasenko2014,Volpe2015,Cirigliano2015,Blaschke2016}. Though the collision physics for neutrinos in flavor eigenstates has been extensively explored (see, e.g., \cite{Bruenn1985,Burrows2006} for overviews) and the general form of the collision integral was worked out in \cite{Sigl1993}, how neutrinos in general flavor states interact with matter through the specific collisional processes relevant to CCSNe and neutron star mergers was only recently fleshed out by \cite{Blaschke2016}. However, these detailed interaction physics have yet to be explored thoroughly in numerical simulations.

In this paper, we begin in Sec.~\ref{sec:QKE} with a brief introduction of the QKEs and the concept of distribution flavor vectors that are critical in visualizing the results. We discuss our numerical method for solving the QKEs and our choice of initial conditions in Sec.~\ref{sec:isotropicsqa}. Then in Sec.~\ref{sec:collisionterms} we describe a method by which the existing wealth of understanding of interaction rates for neutrinos in flavor eigenstates can be extended in a straightforward way to the full collision kernel for the QKEs.  We also show results from nonoscillating calculations in this section to build intuition about the impact of each process on the flavor dynamics. We present our main results in Sec.~\ref{sec:results}. Sec.~\ref{sec:QKE_noosc} shows the evolution of flavor decoherence resulting from a combination of all of the interactions discussed in this paper without the oscillation term. In Sec.~\ref{sec:QKE_osc} we describe the same with the oscillation term turned on and compare it to the nonoscillating calculation. Finally, we perform nonoscillating calculations using input parameters from many points along a 1D core-collapse supernova simulation snapshot and suggest and approximate decoherence length scale formula in Sec.~\ref{sec:supernova}. To facilitate future simulations of the QKEs in core-collapse supernovae and neutron star mergers, we write the general relativistic moment form of the QKEs and the corresponding oscillation and collision source terms in Appendix~\ref{app:moment}.

\section{Quantum kinetic equations}
\label{sec:QKE}
In this section, we present the QKEs in general relativistic form to allow future implementations in codes that account for spacetime curvature and velocity dependence. The QKEs describe the evolution of the (dimensionless) Wigner transform of the neutrino two-point correlation function \cite{Vlasenko2014}
\begin{equation}
\mathcal{F}_{ABab}(\nu,\mathbf{\Omega},x^\mu) \coloneqq \begin{bmatrix}
f_{LLab} & f_{LRab} \\
f_{LRab}^* & f_{RRab}
\end{bmatrix}\,,
\end{equation}
where $f_{LL}$ and $f_{RR}$ are $N_f\times N_f$ matrices representing left- and right-handed neutrinos, respectively and $N_f$ is the number of neutrino flavors. Throughout this paper, uppercase latin indices represent handedness indices, lowercase latin indices are flavor indices, greek indices are spacetime indices, and we use the $(-,+,+,+)$ sign convention for the metric. The flavor-diagonal ($a=b$) components of $f_{LL}$ and $f_{RR}$ are the occupation probabilities at particular spacetime coordinates $x^\mu$ of neutrinos with frequency $\nu$ and direction 3-vector $\mathbf{\Omega}$ in a comoving orthonormal tetrad. That is, $f_{AAaa}$ are the ordinary neutrino distribution functions and the number density $n_{Aa}$ of neutrino flavor $a$ and handedness $A$ in such a frame is
\begin{equation}
    n_{Aa}(x^\mu) = \int \frac{d^3\nu}{c^3} f_{AAaa}\,,
\end{equation}
where $d^3\nu=\nu^2 d\nu d\mathbf{\Omega}$. The off-diagonal elements of $f_{LL}$ and $f_{RR}$ describe quantum flavor coherence. $f_{LR}$ is also a $N_f\times N_f$ matrix that represents quantum coherence between left- and right-handed neutrinos. In the ultrarelativistic limit, the QKEs can be written to order $\epsilon^2$, where $\epsilon\ll 1$ is the ratio of the neutrino mass, mass splitting, forward-scattering potentials, or gradients to the neutrino frequency, as
\begin{equation}
\label{eq:QKE}
    \frac{d\mathcal{F}}{d\lambda} +\mathrm{force}+\mathrm{drift} = -p^\mu u_\mu\left(\mathcal{C} - \frac{i}{\hbar c}[\mathcal{H},\mathcal{F}]\right),
\end{equation}
where $u^\alpha$ is the dimensionless fluid four-velocity. $\lambda$ is an affine parameter such that the neutrino four-momentum is $p^\mu = d x^\mu / d\lambda$. In the comoving orthonormal tetrad, the momentum is also $p^\mu=(1,\mathbf{\Omega})h\nu/c$, where the first component is the time component. The derivative expands in a general curved spacetime to $d/d\lambda = p^\mu \partial/\partial x^\mu - \Gamma^\mu_{\alpha\beta}p^\alpha p^\beta \partial/\partial p^\mu$, where $\Gamma^\mu_{\alpha\beta}(x^\mu)$ are connection coefficients (units of $\mathrm{cm}^{-1}$). The $2N_f \times 2N_f$ collision integral  $\mathcal{C}_{ABab}(\nu,\mathbf{\Omega},x^\mu)$ (units of $\mathrm{cm}^{-1}$) can be decomposed as \cite{Vlasenko2014}
\begin{equation}
    \mathcal{C} = \mathcal{C}^+-\mathcal{C}^- = \left\{1-\mathcal{F},\widetilde{\Pi}^+\right\} - \left\{\mathcal{F},\widetilde{\Pi}^-\right\}\,,
    \label{eq:collisionintegral}
\end{equation}
where the calculation and use of the self-energies $\widetilde{\Pi}^\pm_{ABab}(\nu,\mathbf{\Omega},x^\mu)$ will be described in more detail in Sec.~\ref{sec:collisionterms}. The oscillation potential $\mathcal{H}_{ABab}(\nu,\mathbf{\Omega},x^\mu)$ (units of ergs) is described below. If neutrinos are Majorana particles, only $\mathcal{F}$ needs to be evolved, but if neutrinos are Dirac particles, an additional antineutrino field $\bar{\mathcal{F}}$ must be evolved with an analogous equation. Throughout the rest of this paper, we will refer to the $N_f\times N_f$ matrices $f_{ab}$ and $\bar{f}_{ab}$ as simply the neutrino and antineutrino distribution functions, respectively.

The terms labeled \lq\lq force" and \lq\lq drift" in Eq.~\ref{eq:QKE} are $\mathcal{O}(\epsilon^2)$ corrections to the $d/d\lambda$ term that account for refractive effects due to the finite masses of neutrinos (see, e.g., \cite{Vlasenko2014,Cirigliano2015}). We neglect them in this work for simplicity, and more work needs to be done to assess the importance of these terms in the context of core-collapse supernovae. We will also assume for simplicity that there is no spin coherence ($f_{LR}=\bar{f}_{LR}=0$) and, in the case of Dirac neutrinos, there are no right-handed neutrinos and no left-handed antineutrinos ($f_{RR}=\bar{f}_{LL}=0$). Spin coherence effects are not expected to be important in CCSNe, since the neutrino-antineutrino-mixing contributions to the potential are suppressed by an additional factor of $\epsilon$ and neutrinos in typical CCSN profiles likely pass through the associated resonance too rapidly \cite{Serreau2014,Cirigliano2015,Tian2017}. However, spin coherence effects may not be negligible in other environments such as neutron star mergers \cite{Chatelain2017}. Under the assumption of no spin coherence, we can separate left-handed neutrinos from right-handed (anti)neutrinos, and Dirac and Majorana neutrinos evolve identically. The QKEs for $f=f_{LL}$ and $\bar{f}=f_{RR}$ (Majorana) or $\bar{f} = \bar{f}_{RR}$ (Dirac) then become
\begin{equation}
\begin{aligned}
p^\mu \frac{\partial f}{\partial x^\mu} - \Gamma^\mu_{\alpha\beta}p^\alpha p^\beta \frac{\partial f}{\partial p^\mu} &=  -p^\mu u_\mu\left(C-\frac{i}{\hbar c}[H,f] \right)\\
p^\mu \frac{\partial \bar{f}}{\partial x^\mu} - \Gamma^\mu_{\alpha\beta}p^\alpha p^\beta \frac{\partial \bar{f}}{\partial p^\mu} &=  -p^\mu u_\mu\left(\bar{C}-\frac{i}{\hbar c}[\bar{H},\bar{f}] \right),\\
\end{aligned}
\label{eq:DecomposedQKEs}
\end{equation}
where $C_{ab}$ and $H_{ab}$ are the $N_f \times N_f$ collision integral and Hamiltonian, respectively, from the $LL$ quadrants of $\mathcal{C}$ and $\mathcal{H}$. Similarly, $\bar{C}_{ab}$ and $\bar{H}_{ab}$ come from the $RR$ quadrants of $\mathcal{C}$ and $\mathcal{H}$ (Majorana) or $\bar{\mathcal{C}}$ and $\bar{\mathcal{H}}$ (Dirac).

The Hamiltonian operator is often decomposed as
\begin{equation}
\label{eq:hamiltonian}
    H = H_\mathrm{vacuum} + H_\mathrm{matter} + H_\mathrm{neutrino}.
\end{equation}
The Hamiltonian for antineutrinos is related to that for neutrinos by $\bar{H}_\mathrm{vacuum} = H_\mathrm{vacuum}^*$, $\bar{H}_\mathrm{matter} = -H_\mathrm{matter}^*$, and $\bar{H}_\mathrm{neutrino} = -H^*_\mathrm{neutrino}$. The vacuum Hamiltonian is 
\begin{equation}
    H_\mathrm{vacuum} = U H_\mathrm{vacuum}^{(m)} U^\dagger\,,
\end{equation}
where $H_\mathrm{vacuum}^{(m)}=\mathrm{diag}(\sqrt{h^2 \nu^2+m_l^2c^4})$ is the vacuum Hamiltonian in the neutrino mass basis, $m_l$ is the mass of the neutrino corresponding to lepton flavor $l$. The unitary matrix $U$ describes the mixing between the flavor and mass bases \cite{Maki1962,PDG2018}. The matter potential in the local comoving frame is
\begin{equation}
    H_\mathrm{matter} = \sqrt{2}G_F\hbar^3 c^3\mathrm{diag}(n_l-n_{\bar{l}})\,,
\end{equation} 
where $n_l$ and $n_{\bar{l}}$ are the number density of charged lepton and antilepton of flavor $l$, though in the astrophysical systems of interest electrons are the only lepton with a significant abundance. Neutral current interactions with nucleons also technically contribute to the potential, but since they affect all flavors equally, the potential offset does nothing to modify oscillations and can be ignored. Finally, the neutrino self-interaction potential is 
\begin{equation}
    H_\mathrm{neutrino}=\sqrt{2}G_F\hbar^3 \int d^3 \nu' (1-\cos\theta) (f'-\bar{f}^{\prime *})\,,
\end{equation}
where $\cos\theta=\mathbf{\Omega\cdot\Omega'}$. This is analogous to the matter potential, except that neutrinos are not in general isotropic and the $\cos\theta$ term is needed to account for the angular dependence of the anisotropic neutrino distributions.

\subsection{Distribution flavor vector}
\label{sec:isospin}
It is useful to visualize a neutrino quantum state as a vector in flavor isospin space. We can also visualize the distribution functions $f_{ab}$ and $\bar{f}_{ab}$ with a similar flavor isospin vector, but unlike that for an individual neutrino spin 1/2 quantum state, the distribution flavor vector can represent states with a trace different from unity. When working in a two-flavor system, the vector components are written as
\begin{equation}
    \begin{aligned}
    f_{ab} &= f_{(t)}\delta_{ab} + f_{(x)}\sigma^{(x)}_{ab} + f_{(y)}\sigma^{(y)}_{ab} + f_{(z)}\sigma^{(z)}_{ab} \\
    f_{(t)} &= \frac{1}{2}(f_{ee} + f_{\mu\mu}) \\
    f_{(x)} &= \frac{1}{2}(f_{e\mu} + f_{\mu e}) \\
    f_{(y)} &= \frac{-i}{2}(f_{e\mu} - f_{\mu e}) \\
    f_{(z)} &= \frac{1}{2}(f_{ee} - f_{\mu \mu}) \\
    \end{aligned}\,,
\end{equation}
where $\sigma^{(\alpha)}$ are Pauli matrices. We can define the length of the spatial part of the distribution flavor vector as
\begin{equation}
    \label{eq:isospin_length}
    L \coloneqq \sqrt{\vec{f}\cdot\vec{f}}\,,
\end{equation}
where in components $\vec{f}=(f_{(x)},f_{(y)},f_{(z)})$. We recall that for two arbitrary matrices $A$ and $B$, one can derive the following identities:
\begin{equation}
\label{eq:vector_identities}
\begin{aligned}
    -\frac{i}{2}\left[A,B\right] &= \left(0,\vec{A}\times \vec{B}\right) \\
    \frac{1}{2}\left\{A,B\right\} &= A_{(t)}B_{(t)}\left(1 + \frac{\vec{A}\cdot\vec{B}}{A_{(t)}B_{(t)}}, \frac{\vec{A}}{A_{(t)}} + \frac{\vec{B}}{B_{(t)}}\right)\,,
\end{aligned}
\end{equation}
where the first term in the parentheses is the \lq\lq time" component of the distribution flavor vector. These identities are useful in gaining intuition for how the oscillation and collision terms affect the neutrino distribution. Applying the first identity to Eq. \ref{eq:DecomposedQKEs}, we see that the Hamiltonian term is incapable of changing the number of neutrinos (the $t$ component of the anticommutator is zero) and that it is incapable of changing the length of the flavor vector ($d\vec{f}/d\lambda$ is perpendicular to $\vec{f}$). The second identity applied to the same equation together with Eq.~\ref{eq:collisionintegral} shows that the collision terms can change both.

Finally, given values for $f_{ee}$ and $f_{\mu\mu}$, one can determine the maximum magnitude of the off-diagonal component of the distribution function (i.e., $\sqrt{f_{(x)}^2+f_{(y)}^2}$) by requiring that no diagonal component of the distribution function be smaller than 0 (i.e., $L_f \leq f_{(t)}$) or larger than 1 (i.e., $L_f \leq 1-f_{(t)}$) in any basis. Since $f_{(z)}$ is determined by $f_{ee}$ and $f_{\mu\mu}$, the off-diagonal components must satisfy
\begin{equation}
\label{eq:maxmix}
    f_{(x)}^2 + f_{(y)}^2 \leq \min(f_{(t)}, 1-f_{(t)})^2 - f_{(z)}^2\,.
\end{equation}
This will be useful for setting the initial conditions for our test calculations.

\section{IsotropicSQA}
\label{sec:isotropicsqa}
We evolve the QKEs assuming isotropy, homogeneity, and flat spacetime in a stochastic, operator-split manner with the new, open-source code IsotropicSQA\footnote{\url{https://github.com/srichers/IsotropicSQA}}\footnote{\url{http://doi.org/10.5281/zenodo.3236833}}. Under these assumptions, the neutrino QKEs [Eqs.~\ref{eq:QKE}] take the form of
\begin{equation}
  \frac{1}{c}\frac{\partial f}{\partial t} = C - \frac{i}{\hbar c} \left[H,f\right]\,.
\end{equation}
In the absence of collisions, this equation can be rewritten as 
\begin{equation}
    \frac{1}{c}\frac{\partial S}{\partial t} = -\frac{i}{\hbar c} H S\,,
    \label{eq:S_evolution}
\end{equation}
where the (unitary) evolution operator $S$ determines the distribution function at a later time as 
\begin{equation}
\label{eq:Sf_mapping}
f(t)=S f(0) S^\dagger\,.    
\end{equation}
The (Hermitian) Hamiltonian operator $H$ was described in Sec.~\ref{sec:QKE}. The corresponding equations for antineutrinos are exactly analogous. We assume a two-flavor system under the normal hierarchy with a mass splitting of $\Delta m_{21}=2.43\times10^{-3}\,\mathrm{eV}^2$ and mass eigenstates rotated from flavor eigenstates by $9^\circ$. The assumptions of isotropy and homogeneity preclude important spatial transport and multiangle effects (e.g., \cite{Vlasenko2018,Cirigliano2018}), but allow us to get a handle on the effects of the various contributions to the collision term on the flavor evolution of a neutrino distribution.

\subsection{Maximally mixed initial conditions}
To demonstrate the effects of each of the contributions to the collision term, we will evolve a flavor-mixed isotropic neutrino distribution function in time without oscillations. We configure the initial distribution function to be
\begin{equation}
\label{eq:FDmaxmix}
    \begin{aligned}
        f(\nu) &= \begin{bmatrix}
        \mathrm{FD}(T,\mu_{\nu_e},\nu) & f_{(x),\mathrm{max}}(\nu) \\
        f_{(x),\mathrm{max}}(\nu) & \mathrm{FD}(T,0,\nu)
        \end{bmatrix} \\
        \bar{f}(\nu) &= \begin{bmatrix}
        \mathrm{FD}(T,-\mu_{\nu_e},\nu) & \bar{f}_{(x),\mathrm{max}}(\nu) \\
        \bar{f}_{(x),\mathrm{max}}(\nu) & \mathrm{FD}(T,0,\nu)
        \end{bmatrix} \,,\\
    \end{aligned}
\end{equation}
where the Fermi-Dirac distribution is given by
\begin{equation}
    \mathrm{FD}(T,\mu,\nu) = \frac{1}{e^{(h\nu-\mu)/k_BT}+1}\,.
\end{equation}
The electron neutrino chemical potential is $\mu_{\nu_e}=\mu_e+\mu_p-\mu_n$, where $\mu_n$, $\mu_p$, and $\mu_e$ are the neutron, proton, and electron chemical potentials determined by an input equation of state (EOS). We use the HShen equation of state \cite{Shen2011a} to determine the chemical potentials used in the initial conditions and collision reactions because it is consistent with the Furusawa EOS \cite{Furusawa2011,Furusawa2013} used to produce the core-collapse supernova simulation snapshots used in Sec.~\ref{sec:results}. We set the imaginary part of the off-diagonal components to zero and give the real parts positive values that maximally mix the distribution function at each energy according to Eq.~\ref{eq:maxmix}. Our fiducial fluid parameters for the test calculations are $\rho=10^{12}\,\mathrm{g\,cm}^{-3}$, $T=10\,\mathrm{MeV}$, and $Y_e=0.3$, which yields $\mu_n=1.39\,\mathrm{MeV}$, $\mu_p=-8.56\,\mathrm{MeV}$, $\mu_e=10.1\,\mathrm{MeV}$, and $\mu_{\nu_e}=0.0977\,\mathrm{MeV}$ (including contributions from masses).

\subsection{Stochastic integration}
\begin{figure*}
    \centering
    \includegraphics[width=\linewidth]{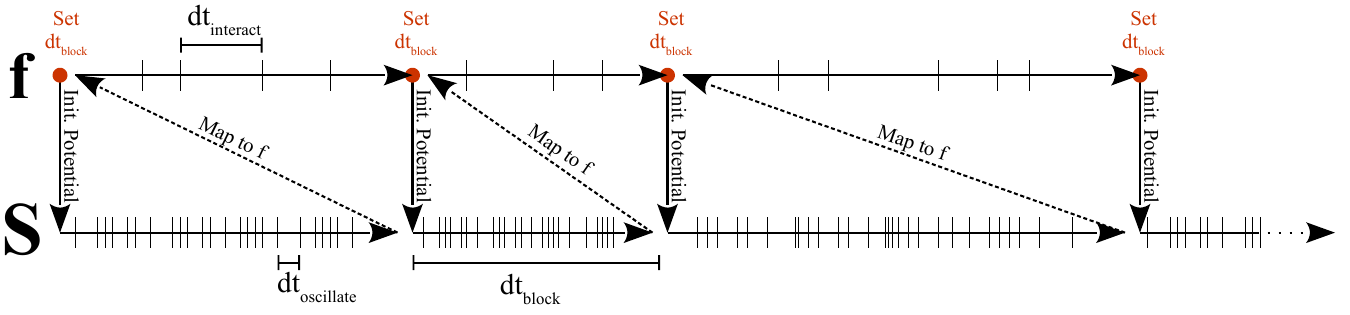}
    \caption{IsotropicSQA Method Summary. Neutrino oscillations are evolved explicitly via the evolution matrix $S$ (bottom horizontal lines) separately from collisions (top horizontal lines) and the effects are combined at time intervals of $dt_\mathrm{block}$. This allows us to take very large time steps for the expensive and slowly-acting interactions, and very short steps for the inexpensive and rapidly varying oscillations.}
    \label{fig:method}
\end{figure*}

In our implementation of the QKEs depicted in Fig.~\ref{fig:method}, we evolve neutrino oscillations without collisions using Eq.~\ref{eq:S_evolution} punctuated by interaction steps at random intervals, as described below. We discretize the distribution functions and evolution matrices into 50 energy bins centered on integer multiples of $2\,\mathrm{MeV}$, each with a width of $2\,\mathrm{MeV}$. This choice allows a direct implementation of the neutrino-neutrino collision terms presented in \cite{Blaschke2016}. We initialize the evolution matrices $S$ and $\bar{S}$ at each energy to the identity and calculate the Hamiltonian based on $f$ and $\bar{f}$ [Eq.~\ref{eq:hamiltonian}, depicted as vertical arrows in Fig.~\ref{fig:method}]. We then evolve $S$ and $\bar{S}$ for a block in time $dt_\mathrm{block}$ without collisions using an adaptive sixth-order Cash-Karp Runge Kutta integrator (depicted as the bottom horizontal arrows in Fig.~\ref{fig:method}). At the end of this block, we map the evolution matrix back onto the distribution function using Eq.~\ref{eq:Sf_mapping} (depicted as diagonal dashed arrows in Fig.~\ref{fig:method}). We then evolve $f$ starting from this updated value for an identical time $dt_\mathrm{block}$ using the same sixth-order integrator (depicted as the upper line in Fig.~\ref{fig:method}). Then the process repeats for subsequent blocks until the end of the simulation.

The oscillation term time step $dt_\mathrm{oscillate}$, the collision term time step $dt_\mathrm{interact}$, and the block time step $dt_\mathrm{block}$ are all independent. $dt_\mathrm{oscillate}$ and $dt_\mathrm{interact}$ are dynamically adjusted to keep the maximum relative difference between fifth- and sixth-order solutions for each step within a critical value of $10^{-12}$. Such high accuracy is required to maintain an accurate solution over the large number of time steps in our simulations. The time step $dt_\mathrm{oscillate}$ is typically much shorter than $dt_\mathrm{block}$, but $dt_\mathrm{interact}$ is typically very large and is restricted by $dt_\mathrm{block}$. This separation of oscillation and collision integration allows us to take much larger time steps for the collision term than for the oscillation term, both for computational efficiency and for preventing the conversion between $f$ and $S$, which suffers from truncation errors, from ruining the accuracy of the solution (see Appendix~\ref{app:tests}).

If $dt_\mathrm{block}$ is left constant or is changed deterministically, there will be artificial correlations between successive interaction steps. For example, if $dt_\mathrm{block}$ is a multiple of the oscillation timescale, the collision step will always act on the same point in the oscillation cycle, and this would not represent the physical solution where collisions affect the distribution continuously throughout the oscillations. To prevent such aliasing errors, we randomize $dt_\mathrm{block}$ based on a target time step $dt_\mathrm{block,target}$ as
\begin{equation}
    dt_\mathrm{block} = \min(-\log(U), 5)dt_\mathrm{block,target}\,,
\end{equation}
where $U$ is a uniform random number between 0 and 1. The resulting exponential random time step makes the interaction sampling a Poisson process, which causes the times of the collision events to be uncorrelated and uniformly sampled in time. The limiter of $5$ prevents an excessively large time step. The first  $dt_\mathrm{block,target}$ is an input parameter, and subsequent values of $dt_\mathrm{block}$ are determined in order to keep the impact of the collision term within a block relatively small. We define the impact over the previous block as
\begin{equation}
    I = \max\left\lvert\frac{f(t) - f(t-dt_\mathrm{block})}{L(t)}\right\rvert\,,
\end{equation}
where $L(t)$ is the length of the distribution flavor vector [Eq.~\ref{eq:isospin_length}] at the same energy and helicity as $f$, and the maximum is taken over all matrix components, helicities, and energies. Comparing to $L$ rather than the trace allows us to adapt $dt_\mathrm{block}$ to quantities relevant to oscillations, since the oscillation term rotates this distribution flavor vector. After the first block, subsequent values of $dt_\mathrm{block,target}$ are set to 
\begin{equation}
    dt_\mathrm{block,target} \leftarrow dt_\mathrm{block,target}\times \min\left(1.1, \frac{I_\mathrm{target} dt_\mathrm{block}}{I\,dt_\mathrm{block,target}}\right)\,
\end{equation}
in order to drive the impact of the collisions over $dt_\mathrm{block}$ toward the target impact. The factor of $1.1$ prevents $dt_\mathrm{block}$ from growing quickly following a step with serendipitously low impact. We choose a target impact of $I_\mathrm{target}=10^{-4}$. See Appendix~\ref{app:tests} for convergence tests for the integrator accuracy, the target impact, and the number of neutrino energy bins.

We use detailed flavor-incoherent neutrino interaction rates from the open-source neutrino rate library {\tt NuLib} \cite{OConnor2015}. However, we must generalize these interaction rates to flavor-coherent neutrinos. This is the subject of Sec.~\ref{sec:collisionterms}.

\section{Collision terms}
\label{sec:collisionterms}
Calculating the neutrino self-energies in Eq.~\ref{eq:collisionintegral} requires evaluating two-point irreducible Feynman diagrams starting at two loops as in \cite{Vlasenko2014,Blaschke2016}. As in Sec.~\ref{sec:QKE}, we neglect spin coherence and calculate $\Pi^\pm_{ab}=\widetilde{\Pi}^\pm_{LLab}$ and $\bar{\Pi}^\pm_{ab}=\widetilde{\Pi}^\pm_{RRab}$ (Majorana) or $\bar{\Pi}^\pm_{ab}=\widetilde{\bar{\Pi}}^\pm_{RRab}$ (Dirac), though without spin coherence the Dirac and Majorana terms are identical. In the following sections, we break down the various contributions to the collision term and evaluate their effects individually. In many cases, these diagrams have already been evaluated for applications in core-collapse supernovae assuming neutrinos remain in flavor eigenstates (see, e.g., \cite{Bruenn1985,Burrows2006}). We will demonstrate how to extend them to full QKE source terms without reevaluating the diagrams.
\subsection{Absorption and emission}
\label{sec:abs}
\begin{figure}
\begin{center}
\includegraphics[width=0.65\linewidth]{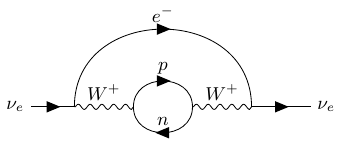}
\end{center}
\caption{Two-point diagram for electron neutrino absorption and emission processes. The equivalent diagram for electron antineutrinos is the same diagram in reverse. The lack of internal neutrino lines makes this a particularly straightforward process to transform into QKE collision terms. }
\label{diagram:abs}
\end{figure}

Let us first consider charged-current neutrino absorption by free neutrons ($\nu + n \rightarrow e^- + p$) and the reverse emission process (Fig.~\ref{diagram:abs}). Typical CCSN temperatures of around $10\,\mathrm{MeV}$ are not high enough to produce significant amounts of $\mu$ or $\tau$ leptons. Thus, we are able to ignore all charged-current interactions involving leptons other than electrons. This is accounted for by only considering the $ee$ component of $\Pi^\pm$, such that
\begin{equation}
    \Pi^+ \sim \Pi^- \sim I_{(e)ab}\,,
\end{equation}
where $I_{(c)ab} = \delta_{ca}\delta_{cb}$. Expanding out Eq.~\ref{eq:collisionintegral}, we find that
\begin{equation}
\begin{aligned}
    C^+
    &\sim \begin{bmatrix}
    (1-f_{ee}) & -f_{e\mu}/2 & -f_{e\tau}/2\\
    -f_{\mu e}/2 & 0 & 0 \\
    -f_{\tau e}/2 & 0 & 0 \\
    \end{bmatrix}\\
    C^- 
    &\sim \begin{bmatrix}
        f_{ee} & f_{e\mu}/2 & f_{e\tau}/2\\
        f_{\mu e}/2 & 0 & 0 \\
        f_{\tau e}/2 & 0 & 0
    \end{bmatrix}\,.\\
\end{aligned}
\end{equation}
Recall that the contributions to the standard collision integral for the $ee$ component in the absence of phase coherence are
\begin{equation}
\begin{aligned}
    C^+_{ee} &= j_{(\nu_e)}(1-f_{ee})\\
    C^-_{ee} &= \kappa_{(\nu_e)}f_{ee}\,,
\end{aligned}
\end{equation}
where $j_{(\nu_e)}(\nu,x^\mu)$ is the electron neutrino emissivity and $\kappa_{(\nu_e)}(\nu,x^\mu)$ is the electron neutrino absorption opacity, both in units of cm$^{-1}$. After matching terms and applying the same process to other lepton species, we arrive at
\begin{equation}
\label{eq:C_abs}
    C_{ab} = j_{(\nu_a)}\delta_{ab} - (\langle j\rangle_{ab}+\langle \kappa\rangle_{ab})f_{ab}\,,\\
\end{equation}
where $\langle j\rangle_{ab}=(j_{(\nu_a)}+j_{(\nu_b)})/2$, and similarly for $\langle\kappa\rangle_{ab}$. In line with the above assertion that there are very few heavy leptons present in conditions relevant to core-collapse supernovae and neutron star mergers, it is often assumed that $j_{(\nu_\mu)}=j_{(\nu_\tau)}=\kappa_{(\nu_\mu)}=\kappa_{(\nu_\tau)}=0$. The collision integral for antineutrinos is exactly analogous. While for this process it is particularly straightforward to write down the collision terms, it serves to illustrate the matching procedure that we will apply to the more complex processes in the next sections.

\begin{figure}
    \centering
    \includegraphics[width=\linewidth]{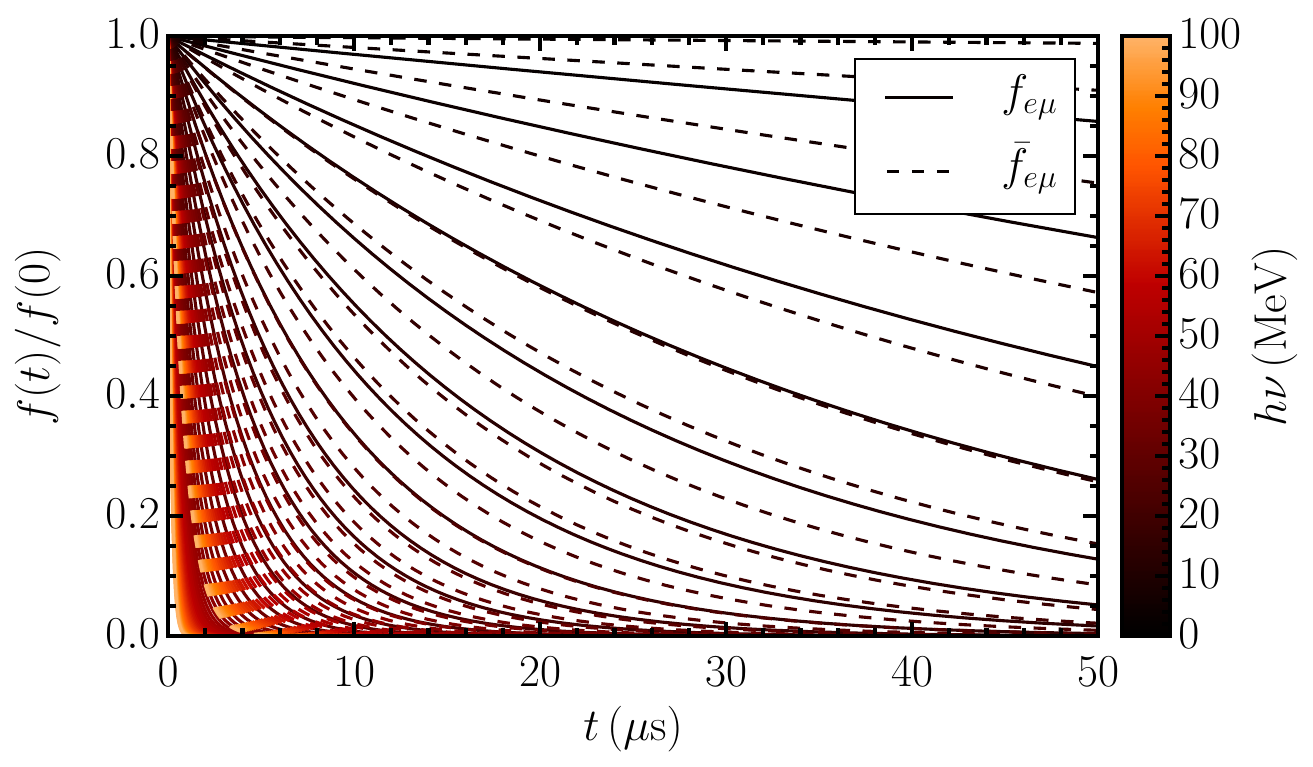}
    \caption{\textit{Absorption} - Evolution of the flavor-off-diagonal components of a maximally mixed Fermi-Dirac neutrino distribution [Eq.~\ref{eq:FDmaxmix}] due to absorption and emission by nucleons and nuclei. On-diagonal components are not shown since they simply remain at their initial values. Interaction rates are based on a background described by $\rho=10^{12}\,\mathrm{g\,cm}^{-3}$, $T=10\,\mathrm{MeV}$, and $Y_e=0.3$. We will show in Sec.~\ref{sec:supernova} that absorption dominates decoherence of higher-energy neutrinos outside of the PNS.}
    \label{fig:noosc_abs}
\end{figure}
The on-diagonal collision terms are exactly identical to the standard transport collision terms, while the off-diagonal ones contain only the absorption component with interaction rates averaged between two flavors. Thus, absorption and emission will always cause flavor coherence to decay. Fig.~\ref{fig:noosc_abs} shows the evolution of the off-diagonal components of our fiducial maximally mixed Fermi-Dirac distribution [Eq.~\ref{eq:FDmaxmix}] relative to their initial values due only to absorption onto and emission from free nucleons and nuclei. All curves demonstrate that the off-diagonal components decay exponentially with a timescale determined by the absorption opacities according to Eq.~\ref{eq:C_abs}. As one would expect, neutrinos at higher energies (yellow curves) interact more strongly than those at low energies (black cuves), resulting in more rapid flavor decoherence. Similarly, neutrinos (solid lines) interact more strongly than antineutrinos (dashed lines), resulting in more rapid flavor decoherence. The on-diagonal components (not plotted) remain constant at their initial Fermi-Dirac values, since the collision terms for these components do not depend on the off-diagonal components of $f$.

\subsection{Electron scattering}
\label{sec:escat}
\begin{figure}
\begin{center}
\includegraphics[width=0.65\linewidth]{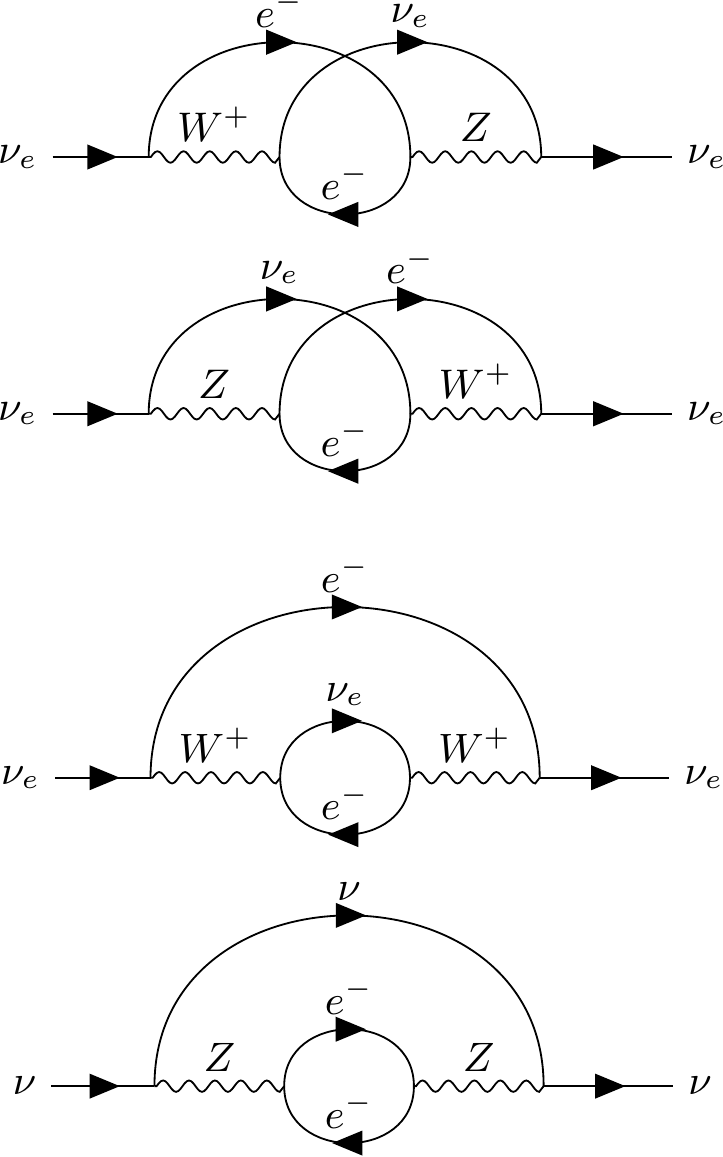}
\end{center}
\caption{Two-point diagrams for neutrino-electron scattering processes. The charged-current processes lead to flavor decoherence and all processes allow redistribution in energy.}
\label{fig:feynman_escat}
\end{figure}

Charged-current reactions are once again kinematically suppressed for heavy-lepton neutrinos, so we only consider scattering by electrons. The two-point diagrams contributing to electron processes are shown in Fig.~\ref{fig:feynman_escat}. These give rise to four terms, each with a different flavor structure. They are \cite{Blaschke2016}
\begin{equation}
\begin{aligned}
    &\begin{aligned}
    \Pi^- = \int \frac{d^3 \nu'}{c^4} \{ &A^- Y_L (1-f')Y_L + B^-Y_R(1-f')Y_R \\ & + D^-[Y_L(1-f')Y_R+Y_R(1-f')Y_L]\}\\
    \end{aligned}\\
    &\begin{aligned}
    \Pi^+ = \int \frac{d^3 \nu'}{c^4} \{&A^+ Y_L f'Y_L + B^+Y_Rf'Y_R \\ & + D^+[Y_Lf'Y_R+Y_Rf'Y_L]\}\\
    \end{aligned}
\end{aligned}
\label{eq:escatPi}
\end{equation}
where $Y_R=\sin^2\theta_W$ and $Y_L=\sin^2\theta_W-1/2 + I_{(e)}$, and we use $\sin^2\theta_W\approx 0.22343$ \cite{PDG2018}. We will now use our matching procedure to find the values of the constants $A^\pm$, $B^\pm$, and $D^\pm$, These constants can depend on both the ingoing and outgoing neutrino momenta. In the absence of flavor coherence, the collision rates have long been used for neutrino transport simulations (\cite{Yueh1977} and corrections by \cite{Bruenn1985}). They are usually written in the form of
\begin{equation}
\begin{aligned}
    C_{(\nu_a)} = \int \frac{d^3\nu'}{c^4} &\left[ f'_{(\nu_a)} R^+_{(\nu_a)} (1-f_{(\nu_a)}) \right. \\
     - &\left.\phantom{[}f_{(\nu_a)} R^-_{(\nu_a)}(1-f'_{(\nu_a)}) \right] \\
    \end{aligned}
\end{equation}
where primed quantities refer to the final state of the neutrino after scattering. $R^\pm_{(\nu_a)}(\nu,\nu',\cos\theta,x^\mu)$ (units of cm$^3$s$^{-1}$) are the inscattering (+) and outscattering (-) kernels for neutrino species $a$ that depend on both the ingoing and outgoing neutrino frequencies and the cosine of the angle between the ingoing and outgoing neutrinos in a comoving orthonormal tetrad. We indicate dependence on the background fluid quantities generally with a dependence on the spacetime coordinates $x^\mu$. Detailed balance requires that $R^+_{(\nu_a)}(\nu,\nu',\cos\theta,x^\mu)=R^-_{(\nu_a)}(\nu',\nu,\cos\theta,x^\mu)=\exp[h(\nu'-\nu)/kT]R^-_{(\nu_a)}(\nu,\nu',\cos\theta,x^\mu)$. When arguments are suppressed, we assume that $\nu'$ is the second argument

Following \cite{Bruenn1985}, we note that the coefficients $D^\pm$ are smaller than $A^\pm$ and $B^\pm$ by approximately a factor of the ratio of the electron mass to the electron energy, so we neglect them here. When $f$ and $\bar{f}$ are flavor diagonal, the standard neutrino rates and the QKE collision terms should be identical, so we can simply match terms to arrive at
\begin{equation}
\begin{aligned}
    R^\pm_{(\nu_e)} &= (A^\pm Y_L Y_L + B^\pm Y_R Y_R)_{ee} \\
    R^\pm_{(\nu_\mu)} &= (A^\pm Y_L Y_L + B^\pm Y_R Y_R)_{\mu\mu}.
    \end{aligned}
\end{equation}
Note that the scattering kernels for $\nu_\tau$ are the same as for $\nu_\mu$  in the absence of $\mu$ and $\tau$ leptons. We can solve for $A^\pm$ and $B^\pm$.  The result is
\begin{equation}
\label{eq:escatAB}
\begin{aligned}
    A^\pm &= \frac{R^\pm_{(\nu_e)}-R^\pm_{(\nu_\mu)}}{2\sin^2\theta_W} \\
    B^\pm &= \frac{(2\sin^2\theta_W+1)^2R^\pm_{(\nu_\mu)}-(2\sin^2\theta_W-1)^2 R^\pm_{(\nu_e)}}{8 \sin^6\theta_W}. \\
\end{aligned}
\end{equation}
We can then write the self-energies very compactly as
\begin{equation}
\label{eq:escatPiR}
\begin{aligned}
    \Pi^+_{ab} &= \int \frac{d^3 \nu'}{c^4} R_{ab}^+ f_{ab}'  \\
    \Pi^-_{ab} &= \int \frac{d^3 \nu'}{c^4} R^-_{ab}(\delta_{ab} - f'_{ab})\,,\\
\end{aligned}
\end{equation}
where
\begin{equation}
\label{eq:escat_Rdeconstruct}
\begin{aligned}
R^\pm_{ab}(\nu,\nu',\cos\theta) &\coloneqq \langle R\rangle^\pm_{ab} - \widetilde{R}^\pm_{ab}\,, \\
\widetilde{R}^\pm_{ab}(\nu,\nu',\cos\theta) &\coloneqq \epsilon_{ab}\frac{R^\pm_{(\nu_a)} - R^\pm_{(\nu_b)}}{4\sin^2\theta_W}\,,\\
\langle R\rangle^\pm_{ab}(\nu,\nu',\cos\theta) &\coloneqq \frac{R^\pm_{(\nu_a)}+R^\pm_{(\nu_b)}}{2}\,, \\
\end{aligned}
\end{equation}
and $\epsilon_{ab}$ is the rank-two Levi-Civita symbol that makes the kernel symmetric on the flavor indices. 

Plugging this into the collision term [Eq.~\ref{eq:collisionintegral}], we get
\begin{equation}
\label{eq:C_escat}
\begin{aligned}
    C_{ab}^{+} &= \int \frac{d^3 \nu'}{c^4} \left[
    R^+_{ab}f_{ab}' - \varsigma^+_{ab}\right] \\
    C_{ab}^{-} &= \int \frac{d^3 \nu'}{c^4} \left[
    \langle R\rangle^-_{ab}f_{ab} - \varsigma^-_{ab} \right],\\
\end{aligned}
\end{equation}
where
\begin{equation}
\label{eq:varsigma}
    \varsigma^\pm_{ab}(\nu,\nu',\cos\theta,x^\mu) \coloneqq  \frac{1}{2}\sum_c\left(R^\pm_{cb}f_{ac}f'_{cb}+R^\pm_{ac}f'_{ac}f_{cb}\right)
\end{equation}
(units of $\mathrm{cm}^{3}\,\mathrm{s}^{-1}$). The first term in each line in Eq.~\ref{eq:C_escat} accounts for unblocked in- and outscattering, respectively, and the second term accounts for Pauli blocking. As with absorption processes, the outscattering rate in these off-diagonal elements is based on the average of the outscattering rates of two neutrino flavors. For the flavor-diagonal elements ($a=b$), the $c=a$ terms in Eq.~\ref{eq:varsigma} sum to the ordinary flavor-incoherent blocking terms, but we see that there is an additional blocking or enhancement [depending on the sign of $\mathrm{Re}(f'_{ac}f_{ca})$] coming from flavor coherence. 

\begin{figure}
    \centering
    \includegraphics[width=\linewidth]{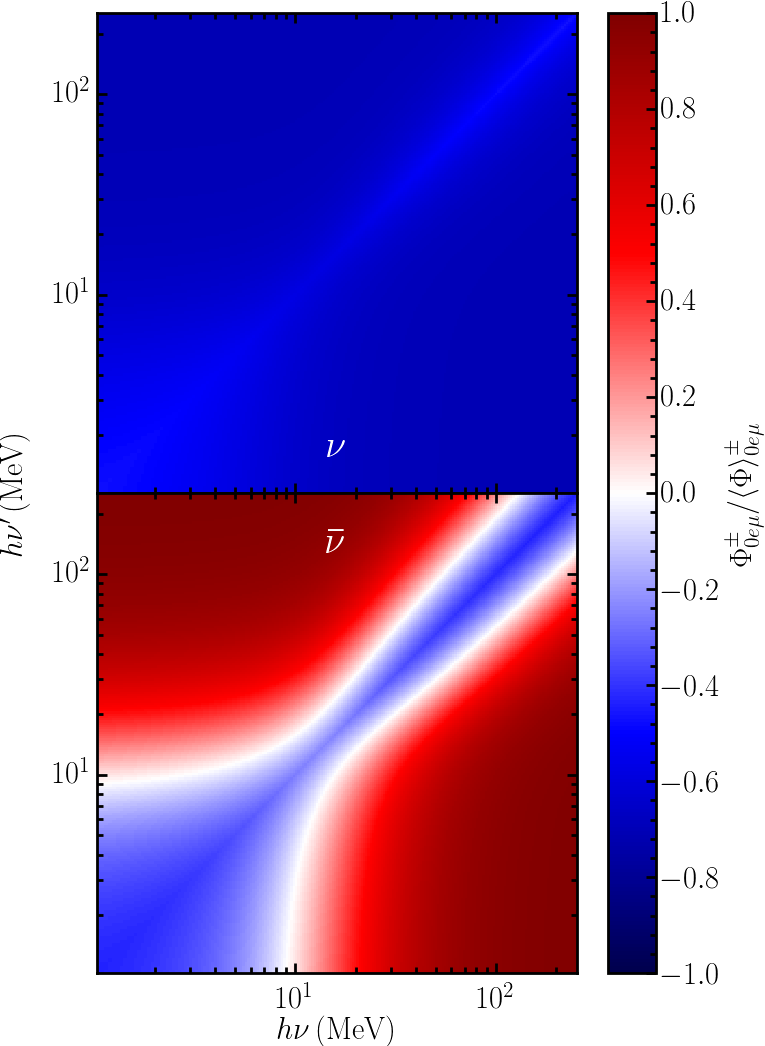}
    \caption{Flavor structure of the isotropic part $\Phi_0^\pm$ of electron scattering collision kernel $R^\pm$ [Eq.~\ref{eq:legendre}] as used in Eq.~\ref{eq:C_escat}. The reaction rates are calculated at an electron chemical potential of $\mu_e=3.16\,\mathrm{MeV}$ and a temperature of $kT=2.74\,\mathrm{MeV}$ as well as using $\theta_W=0.2223$. The top plot is for neutrinos and the bottom for antineutrinos. Red regions indicate that scattering events mostly preserve the sign and magnitude of the scattered neutrino's off-diagonal elements, white regions collapse them to zero, and blue regions flip their sign.}
    \label{fig:iscat_Phitilde0}
\end{figure}

For numerical applications, we expand the interaction kernels in Legendre polynomials as in \cite{Bruenn1985} and consider only the first two terms. That is,
\begin{equation}
\label{eq:legendre}
    R^\pm_{ab} \approx \frac{1}{2}\Phi_{0ab}^\pm(\nu,\nu',x^\mu) + \frac{3}{2}\Phi_{1ab}^\pm(\nu,\nu',x^\mu) \cos\theta\,.
\end{equation}
Inelastic scattering rates in this form are available for flavor-incoherent neutrinos as part of the open-source neutrino rate library {\tt NuLib} \cite{OConnor2015}. Fig.~\ref{fig:iscat_Phitilde0} shows $\Phi_{0e\mu}/\langle\Phi\rangle_{0e\mu}$ over a range of incoming and outgoing neutrino frequencies at a particular temperature of $2.74\,\mathrm{MeV}$ and electron degeneracy $\mu_e=3.16\,\mathrm{MeV}$ chosen to clearly show the features. A value of 1 (red) means that a scattered neutrino carries over its quantum state to its new direction and energy. A value of 0 (white) means the off-diagonal flavor component is erased and that a neutrino collapses to a flavor eigenstate when scattering. Negative values (blue) mean the scattered quantum state receives a phase inversion.

In the case of neutrinos scattering on electrons, the scattering kernel for a neutrino of ingoing momentum $p_\nu$ and outgoing momentum $p_\nu'$ involves an integral over the ingoing electron momentum $p_e$ and outgoing electron momentum $p_e'$. To interpret Fig.~\ref{fig:iscat_Phitilde0}, we first note that in computing $R_{(\nu)}$, there are three terms in this integral that come from combinations of the momentum structures of the different diagrams in Fig.~\ref{fig:feynman_escat}. One term is proportional to $(p_e \cdot p_\nu)(p_e'\cdot p_\nu')$, another is proportional to $(p_e'\cdot p_\nu)(p_e\cdot p_\nu')$, and a third that we neglect (again, because it is suppressed by a factor of $m_e/E_e$) is proportional to $p_\nu\cdot p_\nu'$ (see, e.g., the integrand in Eq.~C49 in \cite{Bruenn1985}). If the reaction rate is dominated by the first integral, one can use the coefficients of this term for electron and muon neutrinos to show that $\Phi^\pm_{0e\mu}/\langle\Phi\rangle^\pm_{0e\mu}\approx(4\sin^4\theta_W-1)/(4\sin^4\theta_W)\approx-0.67$. Similarly, if the reaction is dominated by the second term, the appropriate combination of coefficients yields $\Phi^\pm_{0e\mu}/\langle\Phi\rangle^\pm_{0e\mu}=1$. As can clearly be seen in the top panel of Fig.~\ref{fig:iscat_Phitilde0}, the first integral dominates for large energy transfers, causing the top left and bottom right parts of the plot to be blue. The reaction rates for antineutrinos can be determined by replacing $p_\nu\leftrightarrow -p_\nu'$, effectively swapping the coefficients of the two energy integrals. Thus, the top left and bottom right parts of the bottom panel are red. When there are equal contributions from both, $\Phi_{0e\mu}/\langle\Phi\rangle_{0e\mu}=(8\sin^4\theta_W-1)/(8\sin^4\theta_W+1)=-0.43$. This is visible along the diagonal in both plots. 

\begin{figure}
    \centering
    \includegraphics[width=\linewidth]{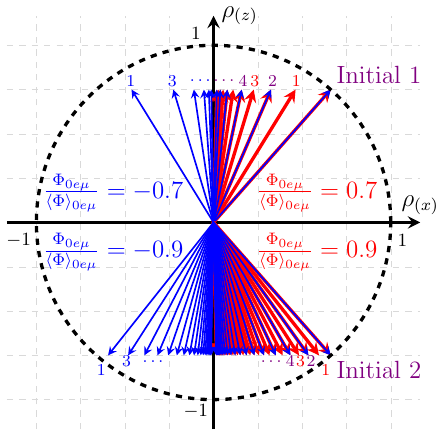}
    \caption{Diagram demonstrating how the first term in Eq.~\ref{eq:C_escat} (electron scattering collision integral) affects the phase of 2-flavor mixed-state neutrinos after scattering when we follow the neutrino from its initial to its final energy with each scatter. We show a maximal-length single particle flavor isospin vector $\vec{\rho}$ for an initial neutrino flavor state described by the density matrix $\rho_{ab}$ (unrelated to the fluid density $\rho$) that is mostly electron neutrino (Initial 1) and one that is mostly muon neutrino (Initial 2) for illustration purposes ($\rho_{(x)}=\sqrt{7}/4$, $\rho_{(y)}=0$ and $\rho_{(z)}=\pm3/4$). We set the oscillation Hamiltonian to zero and only consider electron scattering. Red regions in Fig.~\ref{fig:iscat_Phitilde0} drive the distribution function closer to the flavor axis with each scatter, as shown by the series of red arrows. Blue regions in Fig.~\ref{fig:iscat_Phitilde0} follow the same pattern, except that the neutrino phase is negated on each scatter, as shown by the series of red arrows. In both cases, the density matrix approaches the flavor states more quickly as $\Phi_{0e\mu}^\pm/\langle\Phi\rangle^\pm_{0e\mu}$ approaches 0. To see this, compare the Initial 1 case to the Initial 2 case, where the magnitude of the off-diagonal component of the scattering kernel is larger for the latter. The dashed circle represents the sphere along which oscillations can rotate the vector, described by $\rho_{(x)}^2+\rho_{(y)}^2+\rho_{(z)}^2 = 1$.}
    \label{fig:scatter_diagram}
\end{figure}
The phase inversion process is depicted in Fig.~\ref{fig:scatter_diagram} using flavor vectors for an initial single particle flavor state described by a density matrix $\rho_{ab}$ (unrelated to the fluid density $\rho$). We define vector components of $\vec{\rho}$ based on $\rho_{ab}$ in the same way that we define the distribution flavor vector components in Sec.~\ref{sec:isospin}. Focus first on the series of lines in the top half starting at initial state \lq\lq Initial 1". One can interpret this initial flavor vector as representing a neutrino at a single frequency that is more likely to be measured as a neutrino than an antineutrino ($\rho_{(z)}>0$) and has some flavor coherence ($|\rho_{(x)}|>0$). The series of red lines depict the neutrino flavor structure after a series of scattering events where $0<\Phi_{0e\mu}^\pm/\langle\Phi\rangle_{0e\mu}^\pm<1$. This could be realized, for instance, by antineutrinos scattering back and forth between high and low energies (red regions in the bottom panel of Fig.~\ref{fig:iscat_Phitilde0}) without blocking. During each scatter, the neutrino is not destroyed and the relative probability of finding the neutrino in an electron or muon flavor remains unchanged ($\rho_{(t)}$ and $\rho_{(z)}$ are constant). However, since the ratio is positive but not 1, some of the flavor coherence decays away with each scattering event ($\rho_{(x)}$ decreases). Similarly, the series of blue lines depict a scattering event where $-1<\Phi_{0e\mu}^\pm/\langle\Phi\rangle_{0e\mu}^\pm<0$. Once again, the neutrino is not destroyed and the relative probability of being measured as each flavor remains unchanged, but the flavor coherence ($\rho_{(x)}$ in this diagram) is negated and loses magnitude with each scattering event. In both cases, the distribution decays down to a flavor-diagonal state at the same rate. The series of lines in the bottom half start at a similar initial condition \lq\lq Initial 2" that is mostly muon neutrino for illustration purposes. The evolution of the flavor vectors proceed very similarly to those in the top half, but decay to flavor diagonal states more slowly since the magnitude of $\Phi_{0e\mu}^\pm/\langle\Phi\rangle_{0ab}^\pm$ is closer to unity. In realistic QKE calculations, the impact of phase inversions on the distribution function will depend on the relative directions of the distribution flavor vectors at the initial and final momenta.

The feature in the bottom left corner of Fig.~\ref{fig:iscat_Phitilde0} (below $h\nu\sim h\nu'\sim 10\,\mathrm{MeV}$) is due to the degeneracy of electrons. Along the diagonal ($\nu\approx\nu'$), where scattering events exchange no energy, electron blocking for inscattering exactly cancels that for outscattering. However, in scattering from high energy to low energy (bottom right corner), electron blocking strongly reduces the reaction rate, whereas in scattering from low to high energies the terms that are traditionally associated with blocking effectively amplify the rate. This is to say that the feature is located precisely where one expects degeneracy effects to be important. Increasing the temperature or electron degeneracy increases the size of the feature.

\begin{figure*}
    \centering
    \includegraphics[width=\linewidth]{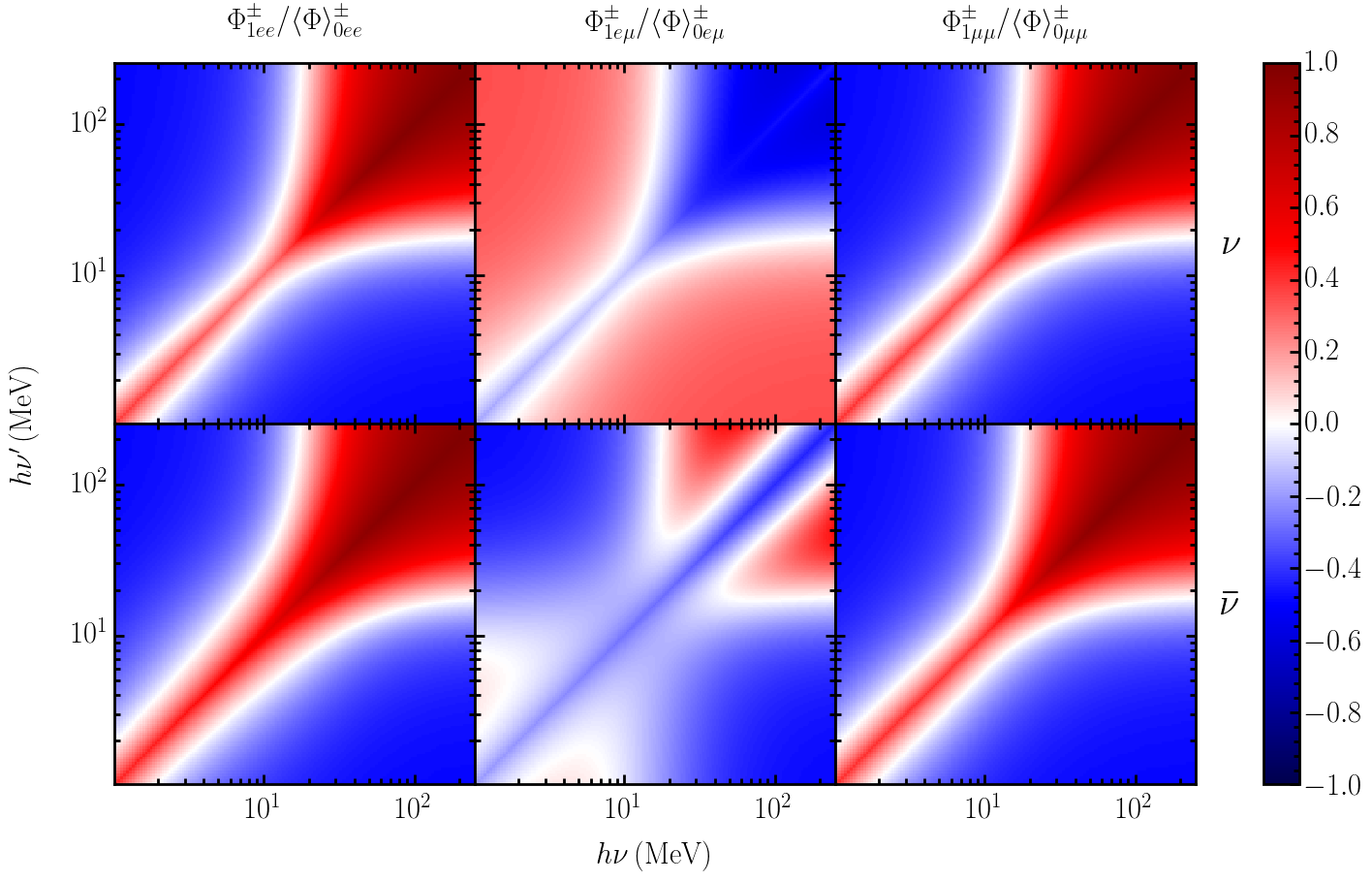}
    \caption{Flavor structure of the linearly anisotropic part $\Phi_1^\pm$ of electron scattering collision kernel $R^\pm$ as used in Eq.~\ref{eq:C_escat}. The reaction rates are calculated at an electron chemical potential of $\mu_e=3.16\,\mathrm{MeV}$ and a temperature of $kT=2.74\,\mathrm{MeV}$ using $\theta_W=0.2223$. The top plot is for neutrinos and the bottom for antineutrinos. The anisotropic collision term for off-diagonal elements (middle panels) reflect both the anisotropy of the flavor-diagonal parts (left and right panels) and the flavor structure of Fig.~\ref{fig:iscat_Phitilde0}. }
    \label{fig:iscat_Phitilde1}
\end{figure*}
The leftmost and rightmost plots in Fig.~\ref{fig:iscat_Phitilde1} show the standard anisotropic component of the scattering kernel for the $ee$ (left panels) and $\mu\mu$ (right panels) components of the neutrino (top panels) and antineutrino (bottom panels) distributions. As we would expect, the large positive values along the diagonal indicate that for small energy transfer, electron scattering is largely forward peaked, and the negative values away from the diagonal mean that scattering events with large energy transfer are backward peaked. However, the center plot describing the anisotropy of the scattering kernel for off-diagonal components is strikingly different because it contains both angular information similar to the plots on the left and right and flavor information similar to Fig.~\ref{fig:iscat_Phitilde0}. Thus, in the collision term for $f_{e\mu}$ (top center panel), large energy transfers are both backward peaked and induce a phase flip, resulting in a net positive (red) value. Similarly for antineutrinos (bottom center panel), the collision term is backward peaked and does not induce a phase flip, resulting in a net negative (blue) value. For small energy transfers (along the diagonal), the scattering is forward peaked and induces a phase flip, resulting in a net negative value for both neutrinos and antineutrinos. The anisotropic ($\Phi_1$) terms do not enter into the isotropic evolution equations in this paper, but will be important for quantum kinetics calculations with spatial transport.

\begin{figure}
    \centering
    \includegraphics[width=\linewidth]{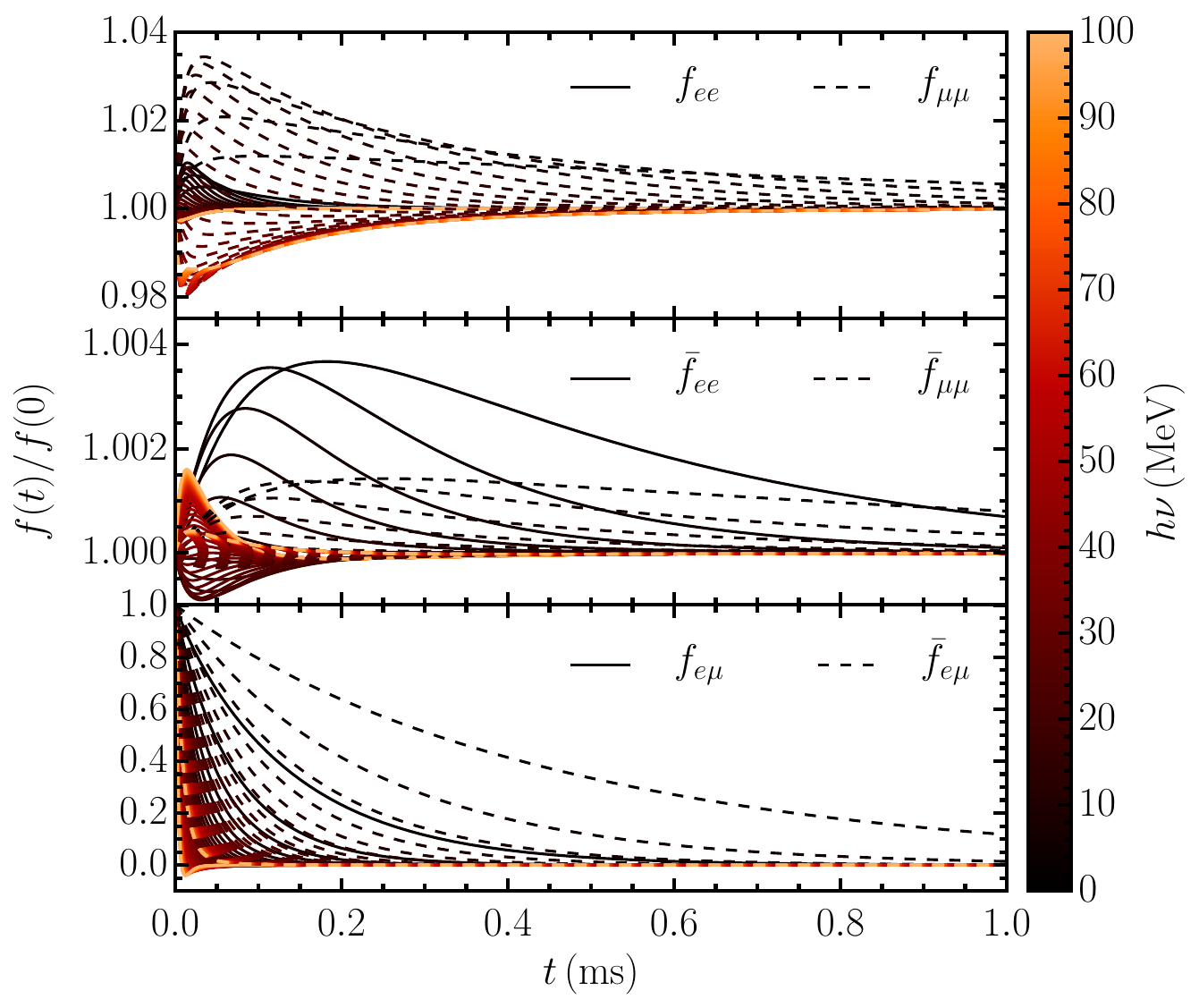}
    \caption{\textit{Inelastic electron scattering} - Evolution of maximally mixed Fermi-Dirac neutrino distribution [Eq.~\ref{eq:FDmaxmix}] due to inelastic electron scattering ($\nu+e^-\leftrightarrow \nu+e^-$) interactions. The top panel contains the flavor-diagonal neutrino distribution components, the middle panel constains the flavor-diagonal antineutrino distribution components, and the bottom panel contains the flavor-off-diagonal neutrino and antineutrino components. Interaction rates are based on a background described by $\rho=10^{12}\,\mathrm{g\,cm}^{-3}$, $T=10\,\mathrm{MeV}$, and $Y_e=0.3$. The charged-current parts of the electron scattering interaction drive the off-diagonal elements to zero, and the diagonal elements redistribute in energy to non-Fermi-Dirac values as long as the off-diagonal elements are nonzero.}
    \label{fig:noosc_iscat}
\end{figure}
Given this intuition, we can now understand the action of the inelastic scattering kernel on our fiducial maximally mixed Fermi-Dirac distribution [Eq.~\ref{eq:FDmaxmix}]. The top and middle panels of Fig.~\ref{fig:noosc_iscat} depict evolution of the flavor-diagonal components for both neutrinos and antineutrinos. The first thing to notice is that even though the flavor-diagonal components are in Fermi-Dirac distributions, the flavor coherence temporarily drives them out of equilibrium. Inspecting Eq.~\ref{eq:C_escat}, it becomes apparent that the only terms that modify the flavor-diagonal components are the $a=b\neq c$ part of the blocking terms $\varsigma^\pm_{ab}$. This can be seen after noting that the flavor-diagonal components of the scattering kernel are $R_{aa}=\langle R_{aa}\rangle=R_{(\nu_a)}$ and that the first terms in Eq.~\ref{eq:C_escat} combined with the $a=b=c$ part of the $\varsigma^\pm_{ab}$ terms constitute the standard flavor-diagonal collision term for incoherent neutrinos. The $a=b=c$ part of $\varsigma^\pm_{ab}$ is the blocking term for incoherent neutrinos, and $\varsigma_{ab}^\pm$ itself is just the extension of the blocking term to non-flavor-diagonal neutrino distributions. Eq.~\ref{eq:vector_identities} shows that the blocking terms, which are quadratic in $f$, have an effect smaller than the standard flavor-incoherent terms when $f$ and $f'$ are antialigned and a larger effect when aligned. Thus, what we see in Fig.~\ref{fig:noosc_iscat} is that the Fermi-Dirac distribution, though an equilibrium for incoherent neutrinos, is not an equilibrium in this mixed state. The blocking terms redistribute neutrinos in energy toward a new mixed equilibrium that has more low-energy neutrinos and fewer high-energy neutrinos. However, after a short period of time the terms linear in $f$ (i.e., the regular scattering terms without blocking) drive the distribution to be flavor-diagonal, since $R^+_{ab}$ is present in the inscattering terms but $\langle R\rangle^-_{ab}$ is present in the outscattering terms. This is demonstrated by the sharp decline of the off-diagonal components of both neutrino and antineutrino distributions in the bottom panel of Fig.~\ref{fig:noosc_iscat}. As the flavor-off-diagonal components decay, the blocking terms increasingly resemble the classic flavor-incoherent terms and the distribution settles back down to a flavor-diagonal Fermi-Dirac distribution. As expected, neutrinos interact more strongly than antineutrinos and high-energy neutrinos interact more strongly than low-energy neutrinos, and stronger interaction rates lead to quicker returns to the flavor-diagonal Fermi-Dirac distribution.

Many modern neutrino transport codes treat electron scattering as elastic for computational efficiency. If the collisions are treated as elastic,the scattering kernels simplify as $R^+_{ab}=R^-_{ab}$. The collision integral greatly reduces to
\begin{equation}
\label{eq:C_elasticscat}
\begin{aligned}
C_{ab} =\frac{1}{4\pi}\int d\Omega'[(&\kappa_{0ab}+\kappa_{1ab}\cos\theta) f'_{ab} \\
- (&\langle\kappa\rangle_{0ab} + \langle\kappa\rangle_{1ab}\cos\theta) f_{ab}],
\end{aligned}
\end{equation}
where the elastic but anisotropic scattering opacity is
\begin{equation}
\label{eq:kappa_decomposition}
    \frac{d\kappa_{(\nu_a)}}{d\Omega'} = \frac{1}{4\pi}(\kappa_{0(\nu_a)} + \kappa_{1(\nu_a)} \cos\theta)
\end{equation}
and the matrix form of the opacities are constructed in the same way as the scattering kernels in Eq.~\ref{eq:escat_Rdeconstruct}.

\begin{figure}
    \centering
    \includegraphics[width=\linewidth]{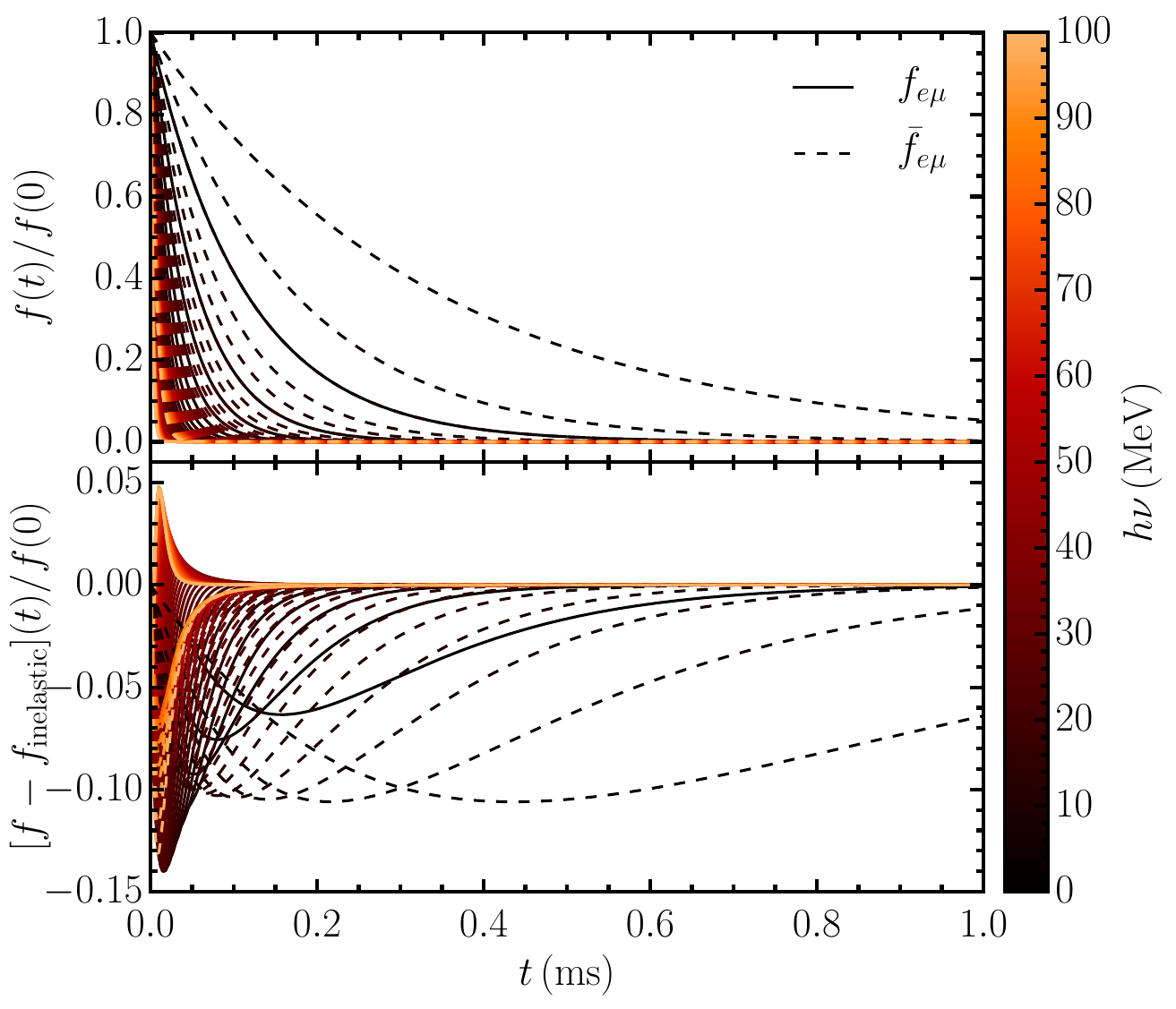}
    \caption{\textit{Elastic electron scattering} - Evolution of maximally mixed Fermi-Dirac neutrino distribution [Eq.~\ref{eq:FDmaxmix}] due to elastic electron scattering ($\nu+e^-\leftrightarrow \nu+e^-$) interactions. The top panel contains the flavor-off-diagonal neutrino and antineutrino components. The bottom panel shows the difference between the elastic ($f$) and inelastic ($f_\mathrm{inelastic}$ from Fig.~\ref{fig:noosc_iscat}) results. On-diagonal components are not shown since they simply remain at their initial values. Interaction rates are based on a background described by $\rho=10^{12}\,\mathrm{g\,cm}^{-3}$, $T=10\,\mathrm{MeV}$, and $Y_e=0.3$. Though these approximate results are qualitatively similar to the bottom panel of Fig.~\ref{fig:noosc_iscat}, there are significant quantitative differences.}
    \label{fig:noosc_escat}
\end{figure}
The top panel of Fig.~\ref{fig:noosc_escat} shows the evolution of the flavor off-diagonal parts of a maximally mixed Fermi-Dirac distribution under this assumption of elastic electron scattering. The on-diagonal components $f_{aa}$ are not shown, since the assumption of elastic scattering causes $\varsigma^\pm_{ab}$ (blocking) terms to cancel and the on-diagonal components do not change. We see that the elastic treatment produces results rather consistent with the full inelastic treatment. The bottom panel shows the difference between the elastic and inelastic treatments, and in this particular choice of fluid parameters elastic scattering causes the distribution to evolve more slowly, leading to differences of at most $\sim15\%$. Of course, greater differences could occur for other fluid and neutrino distributions.

\subsection{Nucleon scattering}
\label{sec:nucscat}
\begin{figure}
\begin{center}
\includegraphics[width=0.65\linewidth]{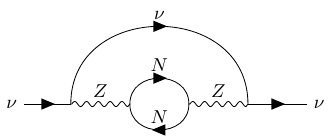}
\end{center}
\caption{Two-point diagram for neutrino-nucleon scattering.This is very similar to Fig.~\ref{fig:feynman_escat}, but the lack of charged-current processes makes the flavor structure less complex.}
\label{fig:feynman_nucscat}
\end{figure}

Next, we consider scattering by nucleons and nuclei (Fig.~\ref{fig:feynman_nucscat}), which at the two-loop level only undergo scattering reactions with neutrinos through neutral-current processes. The outscattering and inscattering self-energies are identical in form to Eq.~\ref{eq:escatPiR}, though we note that due to the neutral current nature of nucelon scattering that the flavor structure of the scattering kernels is simpler ($R_{ab}^\pm=\langle R\rangle_{ab}^\pm$ and $\widetilde{R}_{ab}^\pm=0$). Thus, the scattering kernels do not themselves impart any flavor information on the collision term and could be represented by a scalar instead of a flavor matrix (i.e., $R^\pm_{ab} = R^\pm$). However, in order to minimize the number of representations used in this paper we leave in the flavor indices. Note that weak magnetism corrections cause the interactions to be weaker for antineutrinos than for neutrinos \cite{Horowitz2002}, but leave the flavor structure is unaffected.

The resulting self-energies are then identical in form to Eq.~\ref{eq:escatPiR}. The contribution to the collision term is then also given by Eq.~\ref{eq:C_escat}. Matching the diagonal elements with $f_{ex}=f_{ex}'=0$ to the known scattering interaction rates for incoherent neutrinos, we see that $R^{\pm}_{aa}$ are the standard scattering kernels. Though inelastic nucleon scattering is not yet implemented in {\tt NuLib}, we note that in the electron scattering case the finite value of $\widetilde{R}^\pm_{ab}$ is what drives the distribution to a flavor equilibrium and leads to interesting structure in Figs.~\ref{fig:iscat_Phitilde0} and the center panels of Fig.~\ref{fig:iscat_Phitilde1}. When $\widetilde{R}^\pm_{ab}=0$ as it is for nucleon scattering the ratio $\Phi^\pm_{ab}/\langle\Phi\rangle^\pm_{ab}$ is everywhere identically 1. Thus, inelastic nucleon scattering will redistribute neutrinos and flavor coherence in energy and direction, but will not cause the flavor coherence to decay.

If we consider the collisions to be elastic, which is a much better approximation for nucleon or nucleus scattering than electron scattering because the masses are much larger, then $R^+=R^-$. Once again, the corresponding collision term is identical to that for electron scattering given by Eq.~\ref{eq:C_elasticscat}, though we note that for neutral-current scattering $\kappa_{ab}=\langle\kappa\rangle_{ab}$ and $\widetilde{\kappa}_{ab}=0$. For an isotropic radiation field and elastic nucleon scattering, $f=f'$ and $\mathcal{C}=0$. Thus, we do not plot the uninteresting case of evolving an isotropic mixed distribution function in the presence of elastic nucleon scattering because all quantities remain constant.

\subsection{Pair annihilation}
\label{sec:pair}
Electron-positron pair annihilation ($e^- + e^+ \leftrightarrow \nu + \bar{\nu}$) is a cross diagram of electron scattering and the collision terms have a similar structure. That is \cite{Blaschke2016}, 
\begin{equation}
\begin{aligned}
    &\begin{aligned}
    \Pi^+ = \int \frac{d^3 \nu'}{c^4} \{&A^+ Y_L (1-\bar{f}')Y_L + B^+Y_R(1-\bar{f}')Y_R \\ &+C^+[Y_L(1-\bar{f}')Y_R+Y_R(1-\bar{f}')Y_L]\}\,,\\
    \end{aligned} \\
    &\begin{aligned}
    \Pi^- = \int \frac{d^3 \nu'}{c^4} \{&A^- Y_L \bar{f}' Y_L + B^-Y_R\bar{f}' Y_R \\ &+C^-[Y_L\bar{f}' Y_R+Y_R\bar{f}' Y_L]\}\\
    \end{aligned}\\
\end{aligned}
\end{equation}
where $\bar{f}$ is the antineutrino distribution function. Once again, we neglect the third term in each line. One can go through the term-matching procedure in Sec.~\ref{sec:escat} and show that $A^\pm$ and $B^\pm$ are the same as in the electron scattering case [Eq.~\ref{eq:escatAB}], but with values of $R^\pm_{(\nu_a)}$ from annihilation processes. This leads us to
\begin{equation}
\label{eq:Pi_pair}
\begin{aligned}
    \Pi^+_{ab} &= \int \frac{d^3 \nu'}{c^4} R^+_{ab} (\delta_{ab}-\bar{f'_{ab}})\\
    \Pi^-_{ab} &= \int \frac{d^3 \nu'}{c^4} R^-_{ab} \bar{f'_{ab}}\\
\end{aligned}
\end{equation}
Plugging this into the collision term [Eq.~\ref{eq:collisionintegral}], we arrive at
\begin{equation}
\label{eq:C_pair}
\begin{aligned}
    C_{ab}^{+} &= \int \frac{d^3\bar{\nu}'}{c^4} \left[
    R^+_{ab}\delta_{ab} - \langle R\rangle^+_{ab}f_{ab} -R^+_{ab}\bar{f}_{ab}' + \varsigma^+_{ab}\right]\\
    C_{ab}^{-} &= \int \frac{d^3\bar{\nu}'}{c^4} \varsigma^-_{ab}\,,
\end{aligned}
\end{equation}
where $R^\pm_{ab}$ and $\langle R\rangle^\pm_{ab}$ are defined with respect to values of $R^\pm_{(\nu_a)}$ from annihilation processes in the same way as in Eq.~\ref{eq:escat_Rdeconstruct}. The $\varsigma^\pm_{ab}$ is also similar to the electron scattering case [Eq.~\ref{eq:varsigma}], but replacing $f'\rightarrow \bar{f}'$ since we are integrating over the antineutrino distribution instead of an outgoing neutrino distribution. The first term on the first line of Eq.~\ref{eq:C_pair} is the emission term in the absence of Fermi blocking, which creates flavor-diagonal pairs of neutrinos. Similarly, the second line describes the neutrino pair annihilation rate, where Fermi blocking of final state electrons and positrons is already accounted for in the calculation of $R^-_{ab}$. We can then understand the final three terms on the first line as accounting for neutrino Fermi blocking. The second term in $C^+_{ab}$ behaves like outscattering and will always decrease the magnitude of the flavor-off-diagonal components. The third term has flavor structure like the inscattering term in Eq.~\ref{eq:C_escat}, but acts with the opposite sign. 

\begin{figure}
    \centering
    \includegraphics[width=\linewidth]{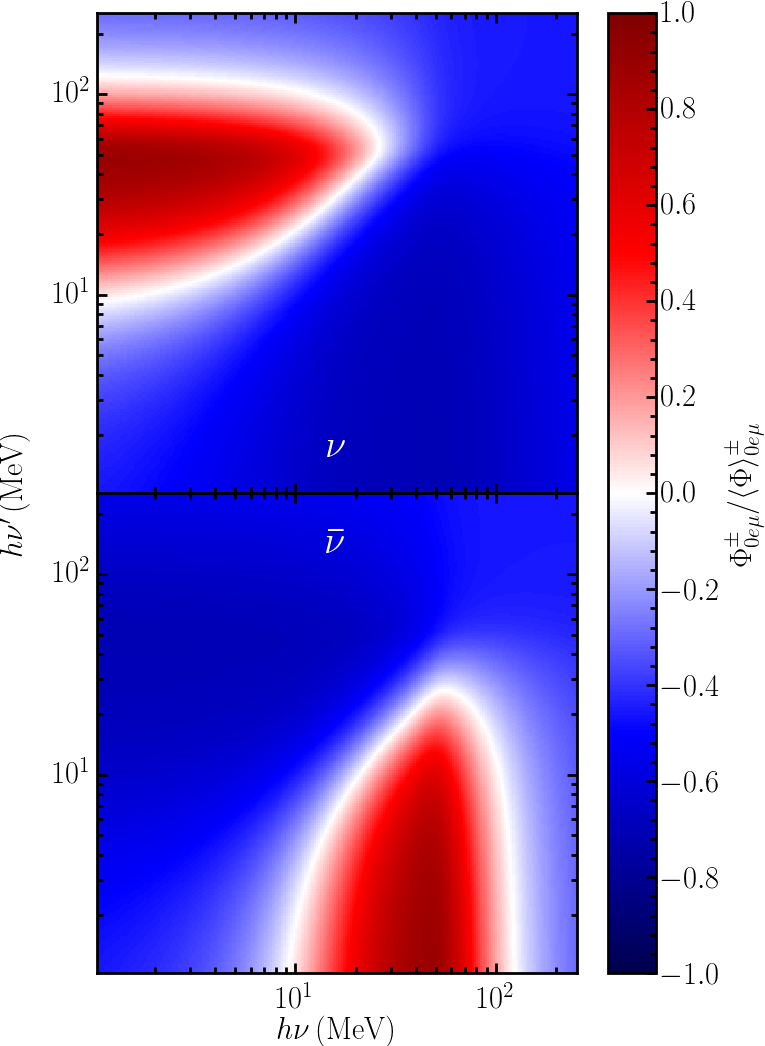}
    \caption{Flavor structure of the isotropic part $\Phi_0^\pm$ of the pair annihilation kernel $R^\pm$ as used in Eq.~\ref{eq:C_pair}. The reaction rates are calculated at an electron chemical potential of $\mu_e=17.8\,\mathrm{MeV}$ and a temperature of $kT=2.74\,\mathrm{MeV}$ using $\theta_W=0.2223$. The top plot is for neutrinos and the bottom for antineutrinos. Red and dark blue regions show where electron degeneracy has a significant impact on the flavor structure of the scattering kernel.}
    \label{fig:pair_kernels0}
\end{figure}
Fig.~\ref{fig:pair_kernels0} shows the isotropic part [see Eq.~\ref{eq:legendre}] of the pair production kernel $R^\pm_{e\mu}$, similar to Fig.~\ref{fig:iscat_Phitilde0}. This term does not affect the unblocked pair production or annihilation rates, but does describe how the presence of neutrinos and antineutrinos contribute to the blocking term differently and how they annihilate. When the ratio $\Phi^\pm_{e\mu}/\langle\Phi\rangle^\pm_{e\mu}$ is negative (blue in the top panel of the figure), the flavor off-diagonal component of the blocking term will increase the neutrino distribution in the direction of the antineutrino distribution's off-diagonal component. Similarly, when the ratio is positive (red in the top panel of the figure), the neutrino's off-diagonal components will be increased in the opposite direction of the antineutrino's off-diagonal components. The same argument can be applied to antineutrinos being blocked by neutrinos (bottom panel of the figure). Notice that the antineutrino collision term is simply the transpose in ingoing/outgoing energy of the neutrino collision term. This is due to the fact that the interaction diagrams are closely related to the interaction diagrams for antineutrinos, so $R^\pm_{ab}(\nu,\bar{\nu},\cos\theta,x^\mu)=\bar{R}^\pm_{ab}(\bar{\nu},\nu,\cos\theta,x^\mu)$. 

\begin{figure*}
    \centering
    \includegraphics[width=\linewidth]{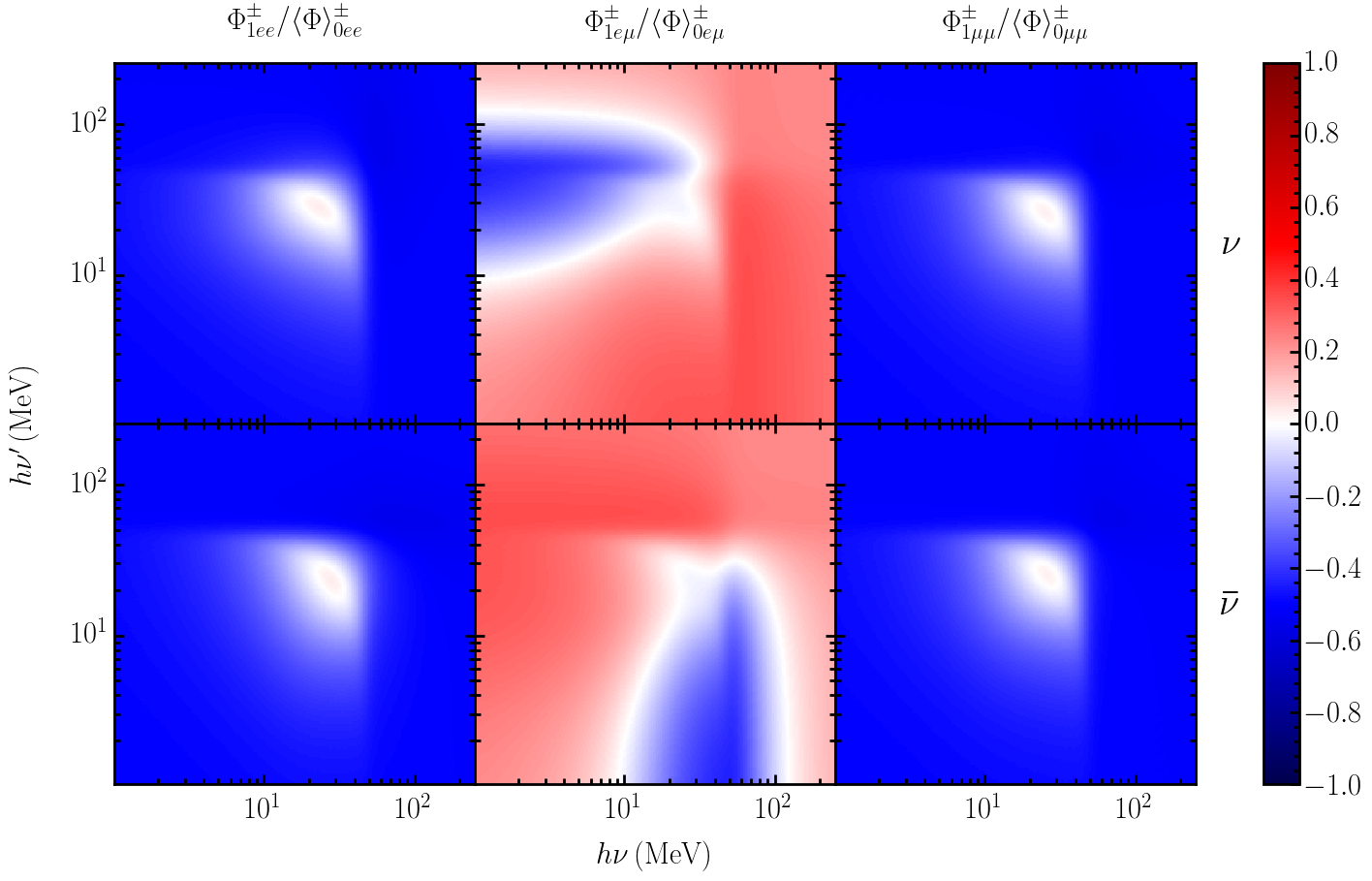}
    \caption{Flavor structure of the linearly anisotropic part $\Phi_1^\pm$ of the pair annihilation kernel $R^\pm$ as used in Eqs.~\ref{eq:C_pair}. The reaction rates are calculated at an electron chemical potential of $\mu_e=17.8\,\mathrm{MeV}$ and a temperature of $kT=2.74\,\mathrm{MeV}$ using $\theta_W=0.2223$. The top plot is for neutrinos and the bottom for antineutrinos. The anisotropic annihilation term for off-diagonal elements (middle panels) reflects both the anisotropy of the flavor-diagonal parts (left and right panels) and the flavor structure of Fig.~\ref{fig:pair_kernels0}.}
    \label{fig:pair_kernels1}
\end{figure*}
Similarly, Fig.~\ref{fig:pair_kernels1} shows the linearly anisotropic part of the electron-positron pair annihilation kernels. The left and right panels show the classic incoherent kernels for electron neutrinos (left) and heavy lepton neutrinos (right) and for antineutrinos (bottom left and right). The predominantly blue color means that neutrinos annihilate more strongly when colliding at large angles, except for a small range of energies near the electron degeneracy energy where dependence on the collision angle is much weaker. This feature is only present when the electron degeneracy is larger than the fluid temperature. The middle plots for neutrinos (top) and antineutrinos (bottom) show the off-diagonal part of the kernel, which is a convolution of angular information in the plots to the left and right and the flavor information in Fig.~\ref{fig:pair_kernels0}. In the blue regions, the blocking term from neutrino-antineutrino pair creation will increase the neutrino's off-diagonal components in the same (flavor) direction as the antineutrino's if they are born moving in the same (spatial) direction, and will do the opposite if born moving in opposite (spatial) directions. In red regions, the opposite is true.

Note that the features in Figs.~\ref{fig:pair_kernels0} and \ref{fig:pair_kernels1} only appear when the electrons are very degenerate. We chose very degenerate parameters to show these effects, but when $kT\gtrsim \mu_e$ there is almost no dependence of the flavor structure on either neutrino or antineutrino energy (the plots are a solid color). There are two relevant energy scales. The first is the inner edge of the features at around $10\,\mathrm{MeV}$ in the figures. This is related to the temperature, since in the range of $0<h(\nu+\nu')\lesssim 10\,\mathrm{MeV}$ a significant fraction of the annihilating electrons will have energies below our example fluid temperature of $2.74\,\mathrm{MeV}$. The second is the outer edge of the features at around $100\,\mathrm{MeV}$ in the figures. This energy scale is related to the electron degeneracy, since in the range of $10\lesssim h(\nu+\nu')\lesssim 100\,\mathrm{MeV}$ a significant fraction of the reacting electrons will be degenerate.

\begin{figure}
    \centering
    \includegraphics[width=\linewidth]{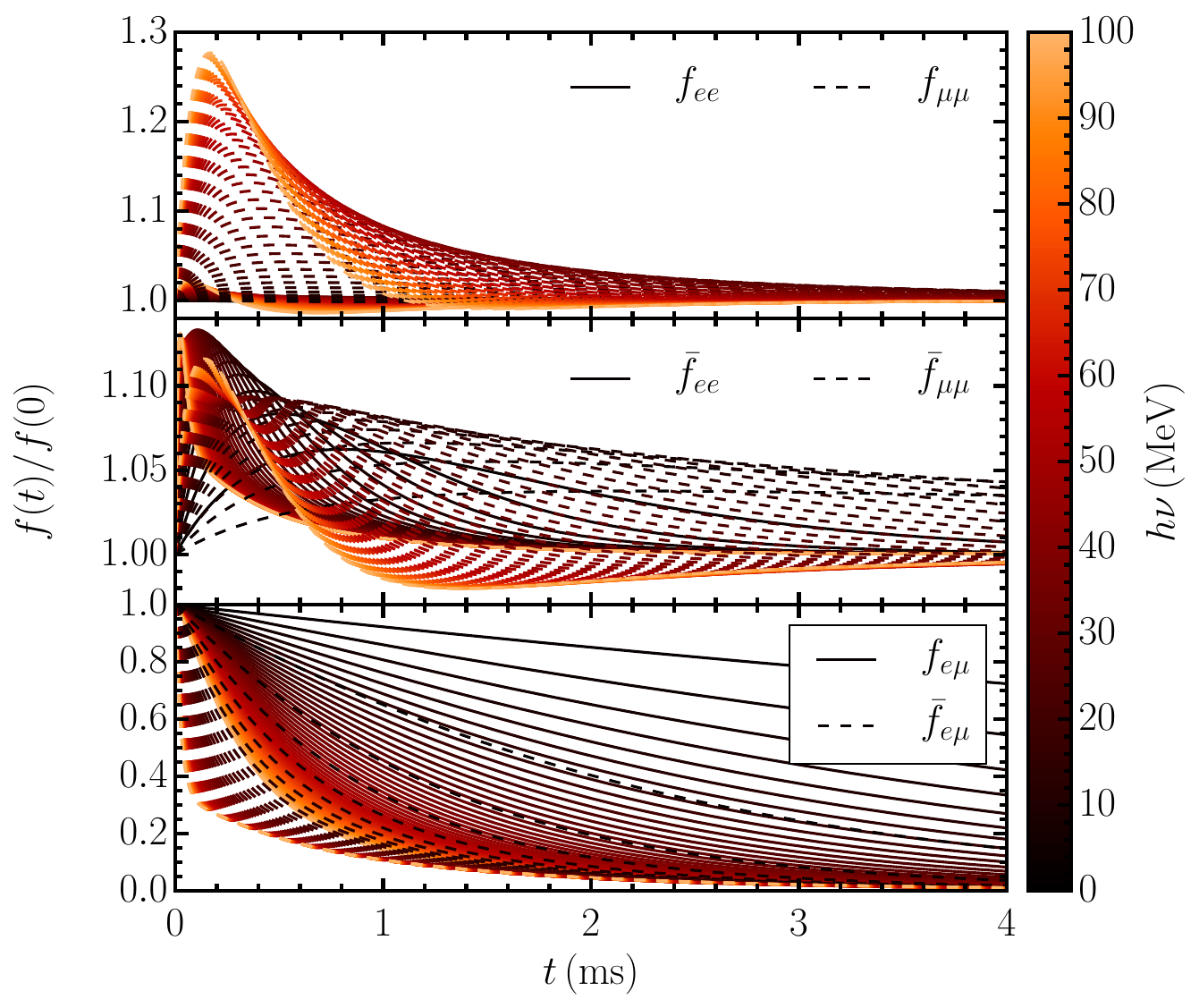}
    \caption{\textit{Pair Annihilation} - Evolution of maximally mixed Fermi-Dirac neutrino distribution [Eq.~\ref{eq:FDmaxmix}] due to electron-positron pair ($e^++e^-\leftrightarrow \nu+\bar{\nu}$) interactions. The top panel contains the flavor-diagonal neutrino distribution components, the middle panel contains the flavor-diagonal antineutrino distribution components, and the bottom panel contains the flavor-off-diagonal neutrino and antineutrino components. Interaction rates are based on a background described by $\rho=10^{12}\,\mathrm{g\,cm}^{-3}$, $T=10\,\mathrm{MeV}$, and $Y_e=0.3$. The nonlinear terms cause the flavor-diagonal elements to redistribute away from Fermi-Dirac values as long as the off-diagonal elements are nonzero. The long evolution timescale makes this process a subdominant driver of flavor decoherence.}
    \label{fig:noosc_pair}
\end{figure}
Fig.~\ref{fig:noosc_pair} shows the evolution of our isotropic maximally mixed Fermi-Dirac distribution of neutrinos [Eq.~\ref{eq:FDmaxmix}] due to electron-positron pair production and annihilation only. Since the distribution is isotropic, only the isotropic part of the kernel $\Phi_0$ enters into the calculation. Similar to the electron scattering case, we see that the pair production kernels modify both the flavor-diagonal (top two panels of Fig.~\ref{fig:noosc_pair}) and the flavor-off-diagonal (bottom panel of Fig.~\ref{fig:noosc_pair}) components of the distribution. Once again, only the nonlinear $\varsigma^\pm_{ab}$ terms lead to the initial change in the on-diagonal components, since the other terms cancel out by construction of the initial Fermi-Dirac distribution. Unlike electron scattering, electron-positron pair production and annihilation can change the total number of neutrinos present, which occurs as all flavor-diagonal components grow in the first $\sim0.1\,\mathrm{ms}$. The $ee$ component of both the neutrino and antineutrino distributions (solid lines in the top two panels) grow on a shorter timescale than the $\mu\mu$ components (dashed lines in the top two panels), since the collision rates for those components are larger due to the contribution of charged-current processes. While this rapid change occurs, the off-diagonal components (bottom panel) of both neutrino and antineutrino distributions decline, and decline more rapidly for high-energy anti/neutrinos.

For all components, the antineutrino distribution evolves more quickly than the neutrino distribution. The initial neutrino distribution has a higher average energy and a higher overall number density than the initial antineutrino distribution due to the large degeneracy of the electrons. This causes both $R^\pm_{ab}$ and $\bar{R}^\pm_{ab}$ to be larger when $\nu>\bar{\nu}$ than when $\nu<\bar{\nu}$. If we look at the third term in Eq.~\ref{eq:C_pair}, since the average energy of $\bar{f}$ is lower than that of $f$ and $R^+_{ab}$ is smaller when $\bar{\nu}<\nu$, the contribution of that term to the integral is relatively small. Similarly, the corresponding term for antineutrinos is an integral over $f$, which has a larger average energy than $\bar{f}$, and $\bar{R}^+$, which is larger when $\nu'>\bar{\nu}$. Thus, it is natural to  expect a larger collision term for antineutrinos than neutrinos and correspondingly more rapid evolution of the antineutrino distribution.

It is common for neutrino transport codes to treat pair processes as an effective absorption and emission process by applying Kirchoff's law to the pair emissivity under the assumption of no Fermi blocking. This allows the neutrino distributions to reach the correct equilibrium and is less computationally expensive. To extend this idea to coherent flavor transport, we take the flavor structure of the term linear in $f$.
\begin{equation}
\label{eq:C_fakepair}
    C_{ab} = j_{(\nu_a)} \delta_{ab} - (\langle j\rangle_{ab}+\langle\kappa\rangle_{ab})f_{ab}
\end{equation}
where 
\begin{equation}
\label{eq:eff_abs}
\begin{aligned}
    j_{(\nu_a)} &= \int d^3\nu' \Phi^+_{0(\nu_a)} \\
    \kappa_{(\nu_a)} &= \frac{j_{(\nu_a)}}{1-FD(T,\mu_{\nu_a},\nu)}\,.
\end{aligned}
\end{equation}

\begin{figure}
    \centering
    \includegraphics[width=\linewidth]{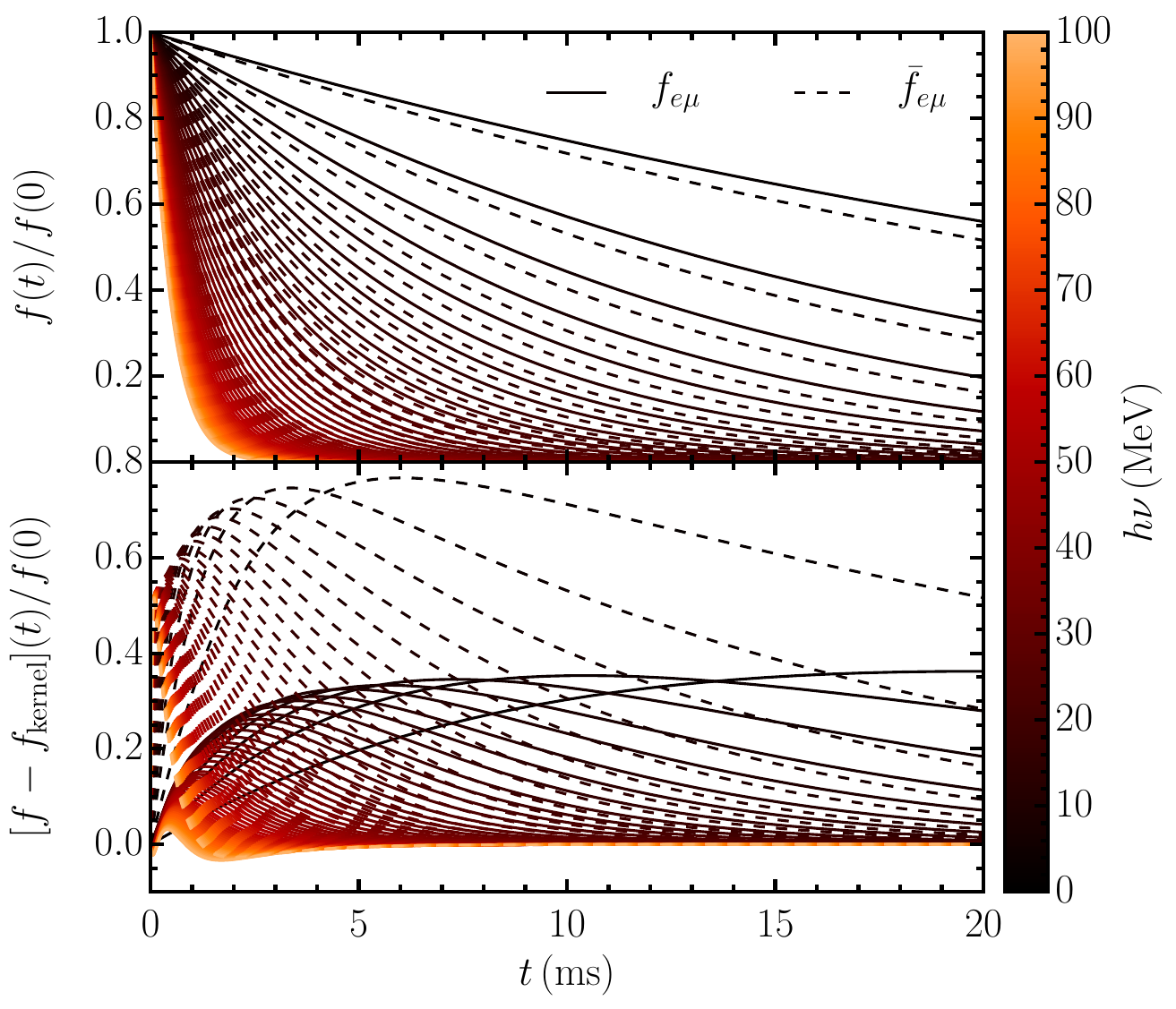}
    \caption{\textit{Effective absorption pair annihilation} - Evolution of maximally mixed Fermi-Dirac neutrino distribution [Eq.~\ref{eq:FDmaxmix}] due to electron-positron pair ($e^++e^-\leftrightarrow \nu+\bar{\nu}$) interactions treated as an effective absorption/emission process. The top panel contains the flavor-off-diagonal neutrino and antineutrino components. The bottom panel shows the difference between the elastic ($f$) and inelastic ($f_\mathrm{inelastic}$ from Fig.~\ref{fig:noosc_iscat}) results. On-diagonal components are not shown since they simply remain at their initial values. Interaction rates are based on a background described by $\rho=10^{12}\,\mathrm{g\,cm}^{-3}$, $T=10\,\mathrm{MeV}$, and $Y_e=0.3$. There are significant differences between how an effective absorption treatment leads to flavor decoherence compared to the full treatment.}
    \label{fig:noosc_fakepair}
\end{figure}
Fig.~\ref{fig:noosc_fakepair} shows the evolution of the same initial condition under this treatment of pair processes. The on-diagonal components are in equilibrium and do not change due to the lack of the nonlinear $\varsigma^\pm_{ab}$ terms in Eq.~\ref{eq:C_fakepair}, and so are not plotted. The top panel shows the evolution of the off-diagonal components of the neutrino (solid lines) and antineutrino (dashed lines) distributions relative to their initial values. We see that they evolve on a timescale that is significantly longer than that seen in the bottom panel of Fig.~\ref{fig:noosc_pair}. When we subtract the full kernel solution from this approximated solution (bottom panel of Fig.~\ref{fig:noosc_fakepair}), we see that the solutions differ by as much as $\sim80\%$. The effective absorption approach is problematic, but we will see in future sections that pair annihilation is not a dominant source of flavor decoherence for these background fluid parameters.

\subsection{Nucleon-nucleon bremsstrahlung}
\label{sec:brems}
\begin{figure}
\begin{center}
\includegraphics[width=0.65\linewidth]{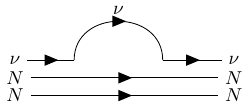}
\end{center}
\caption{Two-point diagram for nucleon-nucleon bremsstrahlung radiation. This diagram represents many diagrams with various permutations of $Z$ bosons connecting the lines. A single internal neutrino line makes the structure similar to electron-positron pair annihilation (Fig.~\ref{fig:feynman_escat}).}
\label{fig:feynman_brems}
\end{figure}

Comparing nucleon-nucleon bremsstrahlung to electron-positron pair annihilation is very similar to comparing nucleon scattering to electron scattering. Nucleon-nucleon bremsstrahlung is a neutral-current process that requires too many diagrams to show them all here. The general structure of the diagrams is depicted in Fig.~\ref{fig:feynman_brems}, but without the gauge bosons connecting the lines in many permutations (see, e.g., \cite{Hannestad1998,Li2015} and references therein). However, in all diagrams there is a single unbroken neutrino line, so the self-energy is linear in $f'$. The resulting expressions for the self-energies are identical in form to Eq.~\ref{eq:Pi_pair}, except that the purely neutral-current nature of the interactions leads to the simplifications $R^\pm_{ab}=\langle R\rangle^\pm_{ab}$ and $\widetilde{R}^\pm_{ab} = 0$. The resulting collision term is then identical to Eq.~\ref{eq:C_pair}.

However, many codes treat nucleon-nucleon bremsstrahlung as an effective absorption and emission process by applying Kirchoff's law to the pair emissivity under the assumption of no Fermi blocking. This leads to reaction rates that are approximately correct, settle to the correct equilibrium, and are much more computationally efficient. To extend this idea to coherent flavor transport, we take the flavor structure of the term linear in $f$, which results in a collision term as in Eq.~\ref{eq:C_fakepair}. With this construction, all species have the same emissivity but different absorption opacities such that the collision term drives the distribution to Fermi-Dirac values.

\begin{figure}
    \centering
    \includegraphics[width=\linewidth]{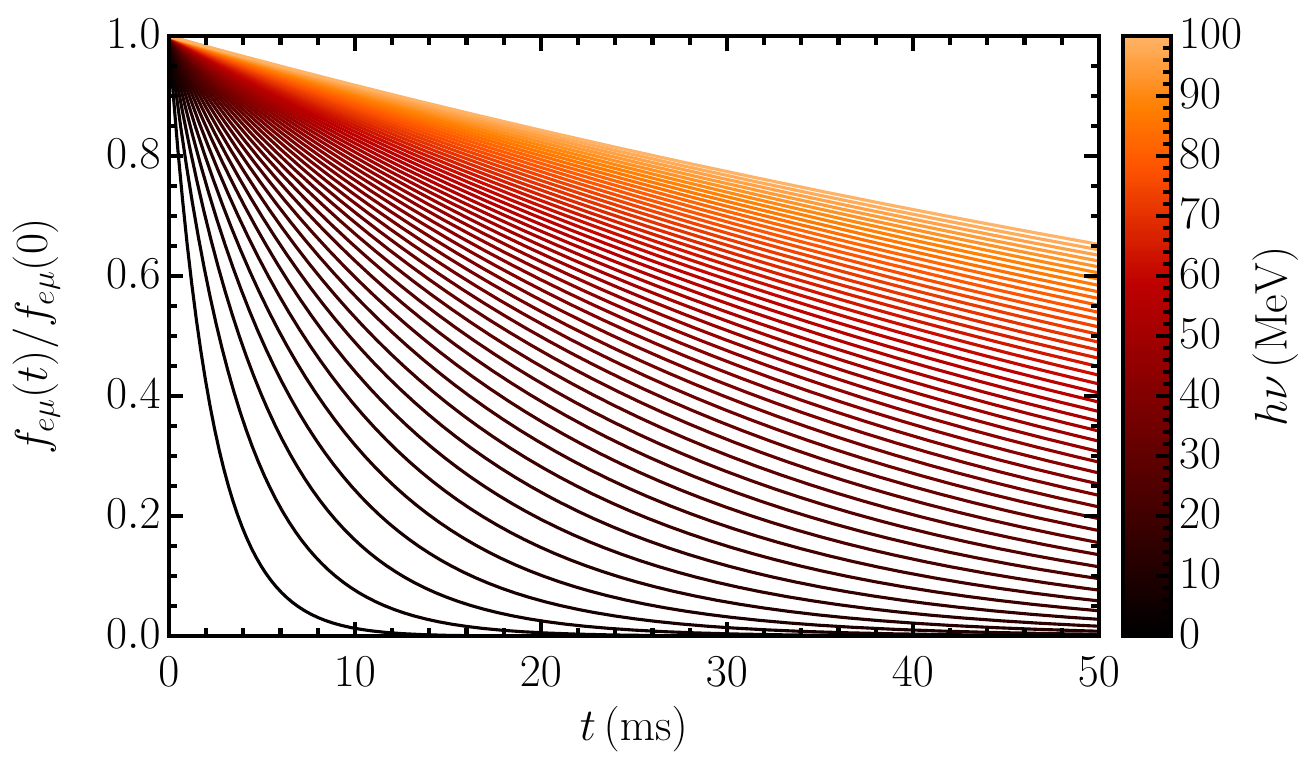}
    \caption{\textit{Effective absorption bremsstrahlung} - Evolution of the flavor-off-diagonal components of a maximally mixed Fermi-Dirac neutrino distribution [Eq.~\ref{eq:FDmaxmix}] due to nucleon-nucleon bremsstrahlung ($N+N\leftrightarrow N+N+\nu+\bar{\nu}$) interactions. On-diagonal components are not shown since they simply remain at their initial values. Interaction rates are based on a background described by $\rho=10^{12}\,\mathrm{g\,cm}^{-3}$, $T=10\,\mathrm{MeV}$, and $Y_e=0.3$. Bremsstrahlung processes decohere low-energy neutrinos more rapidly than high-energy neutrinos and become a dominant driver of decoherence at nuclear densities. The significant errors associated with the effective absorption treatment of pair processes (Fig.~\ref{fig:noosc_fakepair}) calls for a more realistic treatment of bremsstrahlung processes.}
    \label{fig:noosc_fakebrems}
\end{figure}
Fig.~\ref{fig:noosc_fakebrems} shows the evolution of our maximally mixed Fermi-Dirac distribution [Eq.~\ref{eq:FDmaxmix}] using an approximate absorption-emission treatment, since full kernels are not readily available. As with any simple absorption process (e.g., Sec.~\ref{sec:abs}), the off-diagonal component decays exponentially. However, we see that high-energy neutrinos evolve more slowly than low-energy neutrinos, since the bremsstrahlung emissivity decreases with neutrino energy (e.g., \cite{Burrows2006}). The neutrino distribution at a given energy also evolves more slowly than the antineutrino distribution because the values of $\mathrm{FD}(\bar{\nu},-\mu_{\nu_e},T)$ that go into Eq.~\ref{eq:eff_abs} are smaller than $\mathrm{FD}(\nu,\mu_{\nu_e},T)$ at the same neutrino energy. In any case, the effect of bremsstrahlung collision processes on neutrino flavor coherence is much weaker than other processes for this choice of fluid parameters. However, the effects can be much more conspicuous at the nuclear densities in a protoneutron star.

\subsection{Four-neutrino processes}
\label{sec:nunuscat}
\begin{figure}
\begin{center}
\includegraphics[width=0.65\linewidth]{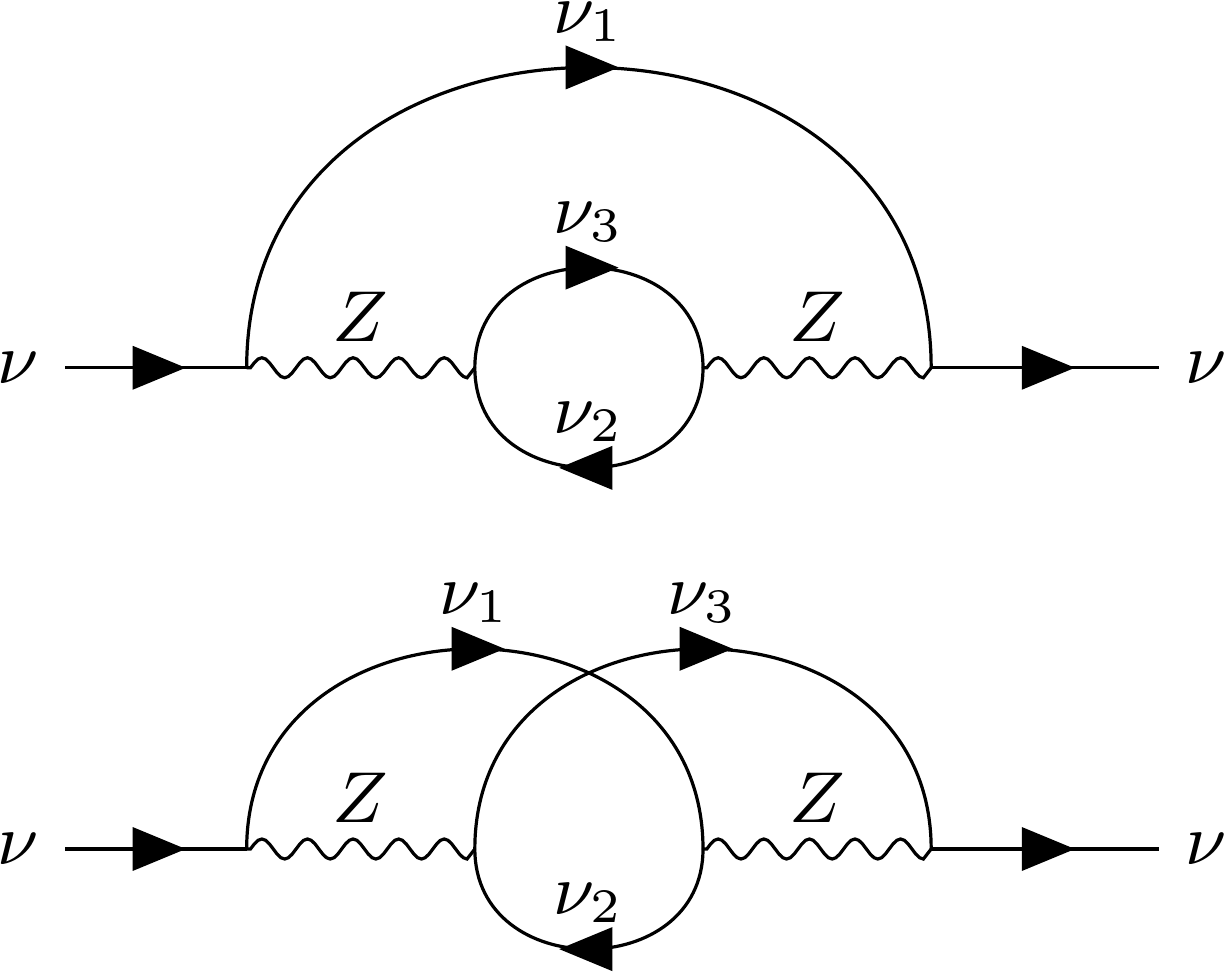}
\end{center}
\caption{Two-point diagrams for neutrino-neutrino scattering and pair annihilation. The multiple internal neutrino lines make this a difficult process to treat in the QKEs. The lack of charged-current interactions precludes phase decoherence.}
\label{fig:feynman_nunuscat}
\end{figure}

Finally, we describe the contribution of four-neutrino processes to the collision term, the diagrams for which are shown in Fig.~\ref{fig:feynman_nunuscat}. The full integrals \cite{Blaschke2016} are rather complicated and we do not reproduce them here. Since this reaction requires integrating over all of the neutrino distributions, we do not extend existing reaction rates and instead directly integrate Eq.~96 in \cite{Blaschke2016}.

\begin{figure}
    \centering
    \includegraphics[width=\linewidth]{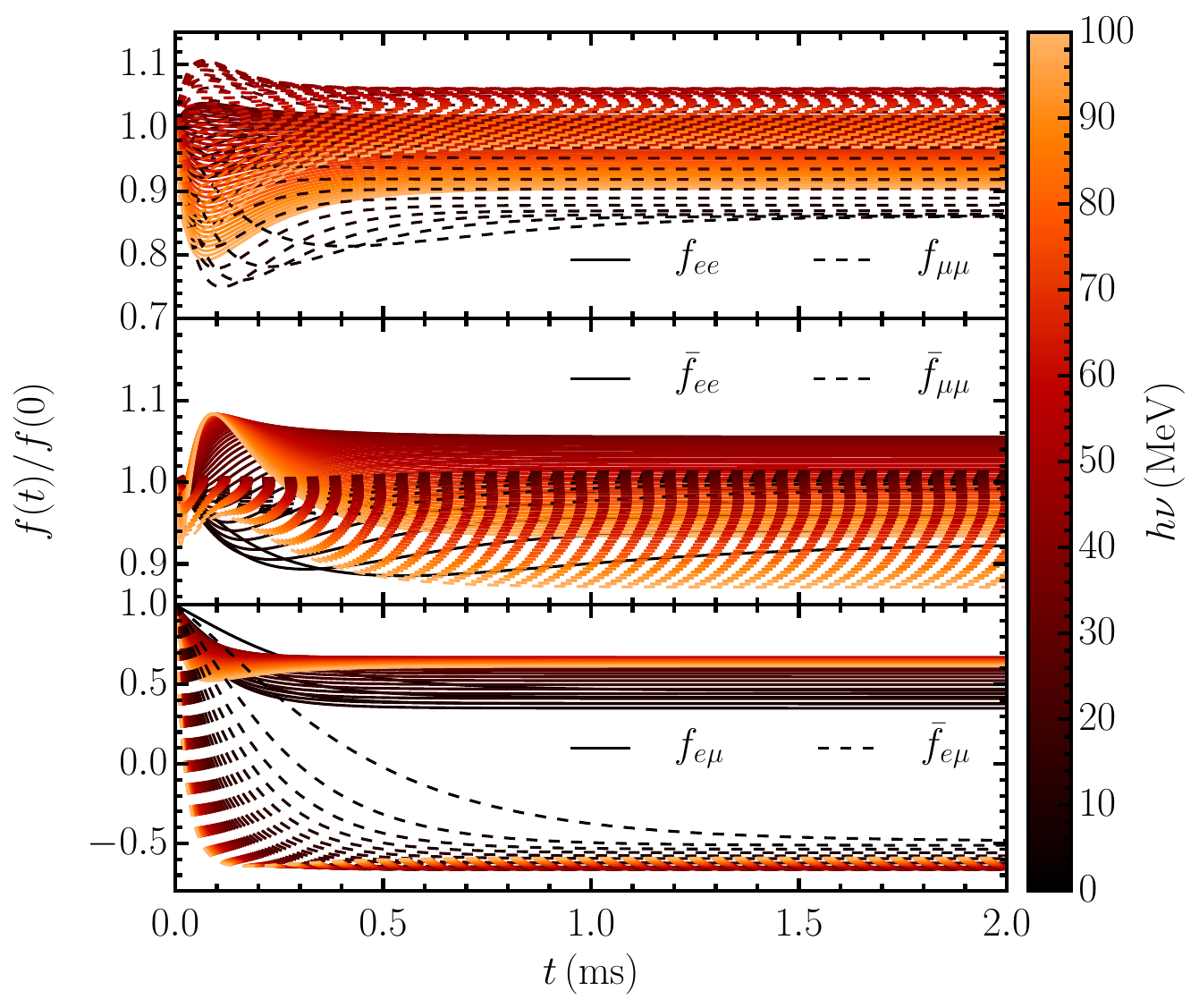}
    \caption{\textit{Four-neutrino processes} - Evolution of maximally mixed Fermi-Dirac neutrino distribution [Eq.~\ref{eq:FDmaxmix}] due to neutrino-neutrino scattering ($\nu+\nu\leftrightarrow \nu+\nu$) and pair ($\nu+\bar{\nu}\leftrightarrow \nu+\bar{\nu}$) interactions. The top panel contains the flavor-diagonal neutrino distribution components, the middle panel contains the flavor-diagonal antineutrino distribution components, and the bottom panel contains the flavor-off-diagonal neutrino and antineutrino components. Interaction rates are based on a background described by $\rho=10^{12}\,\mathrm{g\,cm}^{-3}$, $T=10\,\mathrm{MeV}$, and $Y_e=0.3$. The neutrinos never decay to flavor-diagonal Fermi-Dirac distributions, but instead redistribute such that the distribution flavor vector at all neutrino energies are aligned, and the neutrino vectors are antialigned with the antineutrino vectors.}
    \label{fig:noosc_nu4both}
\end{figure}
Fig.~\ref{fig:noosc_nu4both} shows the evolution of our fiducial maximally mixed Fermi-Dirac distribution [Eq.~\ref{eq:FDmaxmix}] due to neutrino-neutrino scattering and pair processes. It is apparent that the initial distribution is not an equilibrium distribution, and the distribution redistributes itself. Initially, the distribution flavor vector length is 1.6-2.7 (depending on the neutrino energy) times the length of the flavor vector of a flavor-diagonal Fermi-Dirac distribution due to our imposed off-diagonal components [Eq.~\ref{eq:FDmaxmix}]. However, over time the flavor vector length decays down to a nearly constant factor of 1.1-1.4 times the Fermi-Dirac values (again, depending on the neutrino energy). This is done by redistributing flavor coherence between neutrinos and helicities. Indeed, the initial flavor phase angle (the angle between the distribution flavor vector and the positive flavor axis) is initially energy and helicity dependent, but at the end of the calculation the distribution flavor vector is constant in energy at $50.6^\circ$ and the antineutrino distribution flavor vector at $230.6^\circ$. Intuitively, neutrino-neutrino scattering will redistribute flavor phase within a helicity to make the flavor phase angle constant in energy and neutrino-neutrino pair processes will redistribute flavor phase across helicity to make the neutrino and antineutrino flavor phase angles antiparallel. The direction of this equilibrium distribution flavor vector can be in any direction and depends on the initial conditions. 

If we assume that distribution functions 2 and 3 are flavor-diagonal Fermi-Dirac distributions (similar to what is done by \cite{Buras2003a} for standard neutrino transport), the self-energies reduce to
\begin{equation}
\begin{aligned}
    \Pi^+_{ab} &= \int \frac{d^3 \nu_1'}{c^4} \langle R\rangle^+_{ab} f'_{1ab} \\
    \Pi^-_{ab} &= \int \frac{d^3 \nu_1'}{c^4} \langle R\rangle^-_{ab} (\delta_{ab} - f'_{1ab})
\end{aligned}
\end{equation}
where
\begin{equation}
\begin{aligned}
    R^+_{(\nu_a)} &= \sum_{c} \left(1+\delta_{ac}\right) \int \frac{d^3\nu_2'}{c^3} \frac{d^3\nu_3'}{c^3}\\
    &\times r_{(p_1+p_3 \rightarrow p+p_2)} (1-f'_{2cc})f'_{3cc}\\
    R^-_{(\nu_a)} &= \sum_{c}\left(1+\delta_{ac}\right) \int \frac{d^3\nu_2'}{c^3} \frac{d^3\nu_3'}{c^3} \\
    &\times r_{(p+p_2 \rightarrow p_1+p_3)} f'_{2cc}(1-f'_{3cc})
\end{aligned}
\end{equation}
are the ordinary Boltzmann neutrino-neutrino scattering rates. This would lead to similar evolution timescales, but assuming that one of the reacting distributions is flavor diagonal breaks the flavor symmetry that allows the full scattering and annihilation kernels to relax the distributions to an arbitrary flavor phase angle. Instead, the neutrino and antineutrino distributions would always be driven to flavor-diagonal Fermi-Dirac distributions.

\section{Results}
\label{sec:results}

Armed with an understanding of the effects of individual processes, we can make sense of more complete QKE simulations. We will first outline the major points before taking a deeper dive into the results. We have published the most important data from this section on {\tt Zenodo}\footnote{\url{https://doi.org/10.5281/zenodo.3237245}}, and the rest is available on request.

In Sec.~\ref{sec:QKE_noosc} we will combine all of the interactions in Sec.~\ref{sec:collisionterms} while still neglecting oscillations. We demonstrate for our fiducial fluid parameters of $\rho=10^{10}\,\mathrm{g\,cm}^{-3}$, $T=10\,\mathrm{MeV}$, and $Y_e=0.3$ that neutrino flavor coherence decays on timescales near a microsecond depending on energy, as seen in the bottom panel of Fig.~\ref{fig:noosc_full}. Bremsstrahlung processes drive the decoherence of the lowest energy neutrinos, absorption and electron scattering drive decoherence of the highest energies, and inelastic scattering and pair processes drive changes in the flavor-diagonal components.

The keystone to our discussion of the QKEs is the oscillation term, which we discuss in Sec.~\ref{sec:QKE_osc}. We show in Fig.~\ref{fig:osc} that for our fiducial fluid parameters there is a nutation due to the total neutrino density (top two panels) and a precession due to the electron density, both of which occur on timescales of around a picosecond. Extending this calculation by several microseconds, we see the combined action of oscillations and collisions in Fig.~\ref{fig:isospinL}. The neutrino distributions (thick shaded regions covering the oscillation amplitude) decohere to flavor-diagonal Fermi-Dirac distributions on timescales similar to but measurably different from the calculations without oscillations (green curves).

Given this result, we feel justified to do a wider parameter sweep in Sec.~\ref{sec:supernova} using the less-expensive nonoscillating QKE calculations to understand flavor decoherence in core-collapse supernovae. We take background matter parameters from a one-dimensional (1D) CCSN simulation (Fig.~\ref{fig:1dbackground}) and determine the decoherence time using an isotropic QKE calculation at each radial point (Fig.~\ref{fig:decay_time}). We show the results using different interaction sets in different panels to demonstrate that in the PNS decoherence is dominated by electron scattering, nucleon-nucleon bremsstrahlung, and neutrino-neutrino scattering, while near the shock decoherence is dominated by absorption, and in the decoupling region all of these processes except elastic nucleon scattering and neutrino-neutrino processes have a significant impact. Finally we are able to demonstrate that an empirically determined effective decoherence opacity $\kappa_\mathrm{effective}$ [Eq.~\ref{eq:kappa_eff}] is a much better predictor of flavor decoherence rates than the mean free path. In Fig.~\ref{fig:meanfreepath}, one can see that the decoherence timescale from $\kappa_\mathrm{effective}$ matches the simulated one generally to within a factor of 10 in the PNS and to within 20\% outside of the PNS.

\subsection{Quantum kinetics without oscillations}
\label{sec:QKE_noosc}
\begin{figure}
    \centering
    \includegraphics[width=\linewidth]{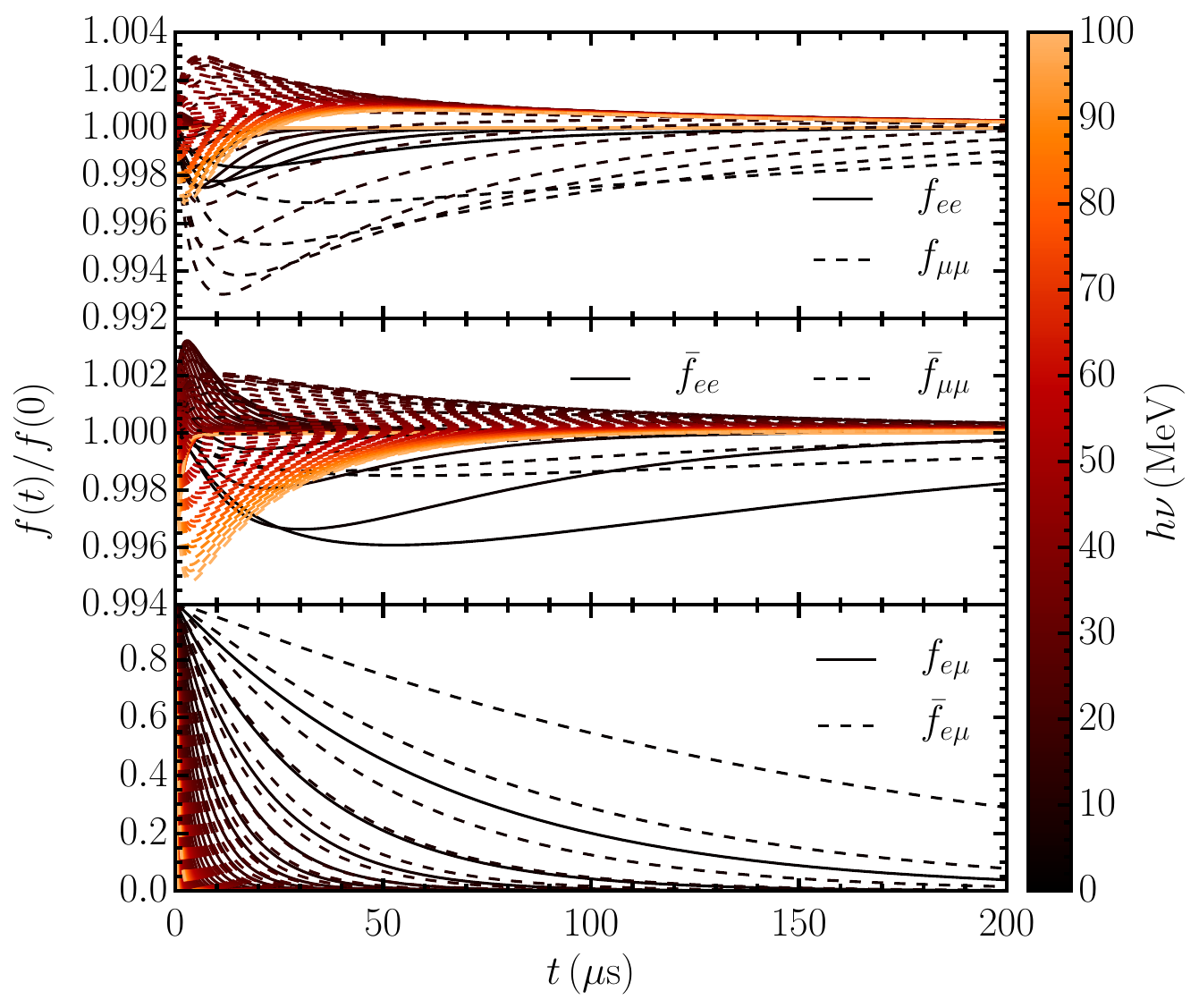}
    \caption{\textit{All processes} - Evolution of maximally mixed Fermi-Dirac neutrino distribution [Eq.~\ref{eq:FDmaxmix}] due to absorption/emission, inelastic electron scattering, electron-positron pair annihilation, nucleon-nucleon bremsstrahlung, neutrino-neutrino scattering, and neutrino-antineutrino pair interactions. The top panel contains the flavor-diagonal neutrino distribution components, the middle panel constrains the flavor-diagonal antineutrino distribution components, and the bottom panel contains the flavor-off diagonal neutrino and antineutrino components. Interaction rates are based on a background described by $\rho=10^{12}\,\mathrm{g\,cm}^{-3}$, $T=10\,\mathrm{MeV}$, and $Y_e=0.3$. This quantitatively provides an estimate of decoherence timescales on the order of a millisecond for these parameters depending on energy.}
    \label{fig:noosc_full}
\end{figure}
First, we will discuss how neutrinos evolve with a full suite of collision processes but without oscillations to provide a basis for understanding what happens when evolving the full QKEs. We begin with the same initial maximally mixed neutrino distribution [Eq.~\ref{eq:FDmaxmix}] and background matter parameters ($\rho=10^{12}\,\mathrm{g\,cm}^{-3}$, $T=10\,\mathrm{MeV}$, $Y_e=0.3$) used in Sec.~\ref{sec:collisionterms}. The resulting evolution of the neutrino distributions is shown in Fig.~\ref{fig:noosc_full}, and it is immediately clear that it behaves unlike that due to any single collision process. 

The initial reactions of the flavor-diagonal elements (top two panels of Fig.~\ref{fig:noosc_full}) are very reminiscent of the evolution due to neutrino-neutrino reactions (Fig.~\ref{fig:noosc_nu4both}), though with  different amplitudes and on a shorter timescale. In all but the lowest energy electron neutrinos (solid lines, top panel of Fig.~\ref{fig:noosc_full}), the impact of electron scattering (Fig.~\ref{fig:noosc_iscat}) is apparent. Similarly, at intermediate energies, the electron and muon antineutrino curves (center panel of Fig.~\ref{fig:noosc_full} evolve similar to those caused by pair annihilation in Fig.~\ref{fig:noosc_pair}. The off-diagonal elements (bottom panel) show that high-energy neutrinos lose flavor coherence more quickly than low-energy neutrinos and that neutrinos lose flavor coherence more quickly than antineutrinos. This reflects the actions of absorption of neutrinos on nucleons (Fig.~\ref{fig:noosc_abs}) and electron scattering (Fig.~\ref{fig:noosc_iscat}). Even though neutrino-neutrino interactions themselves support long-lived flavor coherence, when combined with coherence-destroying reactions they simply accelerate the demise of coherence. The rapid initial adjustment from neutrino-neutrino interactions quickly shuffles neutrinos between energy bins, and energy is also shuffled between neutrinos and antineutrinos, allowing flavor-destructive processes to operate more efficiently.

\subsection{Quantum kinetics with oscillations}
\label{sec:QKE_osc}
\begin{figure}
    \centering
    \includegraphics[width=\linewidth]{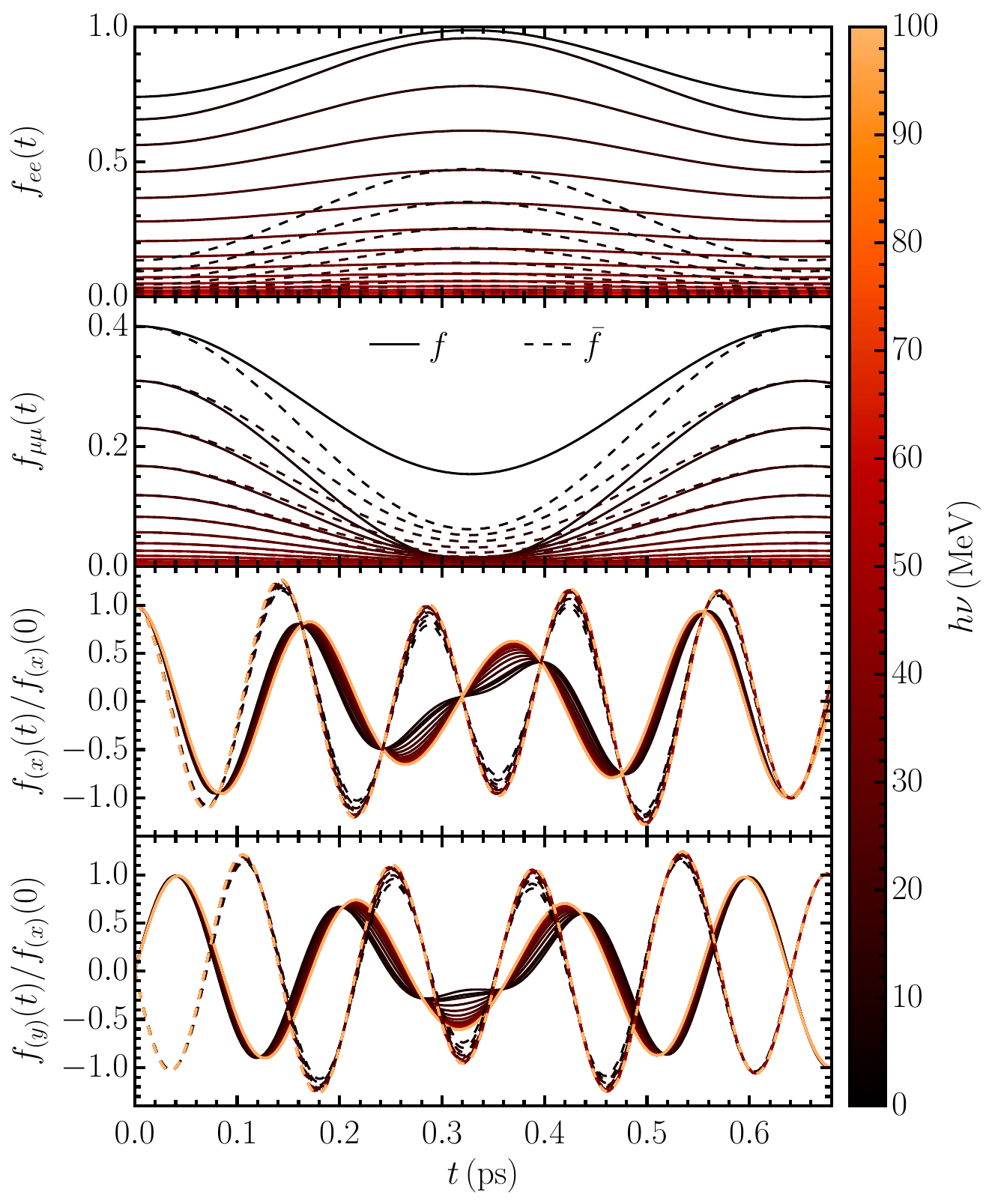}
    \caption{\textit{Oscillation terms} - Evolution of all components of a two-flavor neutrino distribution function due to the action of the oscillation Hamiltonian in Eq.~\ref{eq:QKE} in a background fluid described by $\rho=10^{12}\,\mathrm{g\,cm}^{-3}$, $T=10\,\mathrm{MeV}$, and $Y_e=0.3$. There is a short precession timescale associated with the number density of electrons visible in the flavor-off-diagonal components (bottom two panels) and a longer nutation timescale associated with the total neutrino density most obvious in the flavor-diagonal components (top two panels). Both are much shorter than the collision timescale.}
    \label{fig:osc}
\end{figure}
Now we can finally add the oscillation term back in. Before discussing the combined effects of oscillations and collisions, we note that the oscillations occur on an extremely short timescale of $\sim10^{-16}\,\mathrm{s}$. Fig.~\ref{fig:osc} shows this in detail. The top two panels show the evolution of the flavor-diagonal components of the neutrino (solid) and antineutrino (dashed) distributions. When one flavor becomes more abundant, the other flavor becomes less abundant, conserving the total number of neutrinos. Lower energy neutrinos (dark colors) are more degenerate and have larger occupation probabilities. The lowest-energy electron neutrinos (black solid curves at the top of the top panel) actually oscillate into a completely degenerate state with occupation numbers of 1. This emphasizes the importance of taking care to choose only physically motivated initial conditions since a larger amount of initial flavor mixing in the initial conditions than used here would allow the oscillation term to push the occupation numbers in these degenerate states higher than 1. With our choice of initial conditions, the blocking terms in our calculations naturally prevent the neutrino distribution from developing into a state with $f>1$ in any basis. For the realization of the QKEs \cite{Cirigliano2015} and our adaption of them, we have not, however, proven that the collision terms are incapable of developing such an unphysical distribution under all circumstances. This is trivially true for absorption and elastic scattering, for which evolution of the diagonal components is unchanged by flavor coherence, but the nonlinear processes are less straightforward.

The bottom two panels of Fig.~\ref{fig:osc} show the real and imaginary parts of the distribution, respectively. First, we notice that the real and imaginary parts are out of phase with each other, indicating a circular oscillation of the distribution flavor vector around the flavor axis. Also, the direction of the antineutrino flavor vector oscillation is opposite to the the neutrino vector, indicated by the opposite sign of the imaginary components. This is expected based on the relationship between the neutrino and antineutrino matter Hamiltonians (Sec.~\ref{sec:QKE}). Finally, we see that when the difference between the values of the on-diagonal components (the $z$ component of the distribution flavor vector) grows toward $t=0.3\,\mathrm{ps}$, the magnitude of the off-diagonal components ($x$ and $y$ components of the distribution flavor vector) shrink. This is an expression of the fact that the oscillation term is unable to change the length of the distribution flavor vector. Effects from interactions are technically present in these data, but they do not have have significant impact over such a short time.

The rapid oscillations in the bottom two panels show the effect of the matter potential. At the relevant density of $10^{12}\,\mathrm{g\,cm}^{-3}$ and electron fraction of 0.3, the matter potential [Eq.~\ref{eq:hamiltonian}] is $0.023\,\mathrm{eV}$, a factor of $1.8\times10^7$ larger than the vacuum potential for the $1\,\mathrm{MeV}$ bin. Because of the large matter potential, the oscillation axis is nearly aligned with the flavor axis. One can also see that the period of the oscillations is approximately $h/\sqrt{2}G_F n_e=1.8\times10^{-13}\,\mathrm{s}$. Deviations from this are due to the additional contribution of the neutrino self-interaction potential. For the diagonal components in the top panel, the period of oscillation, also due to the neutrino potential, is $h/\sqrt{2} G_F n_\nu=6.9\times10^{-13}\,\mathrm{s}$, where $n_\nu$ is the total neutrino number density.

\begin{figure*}
    \centering
    \includegraphics[width=\linewidth]{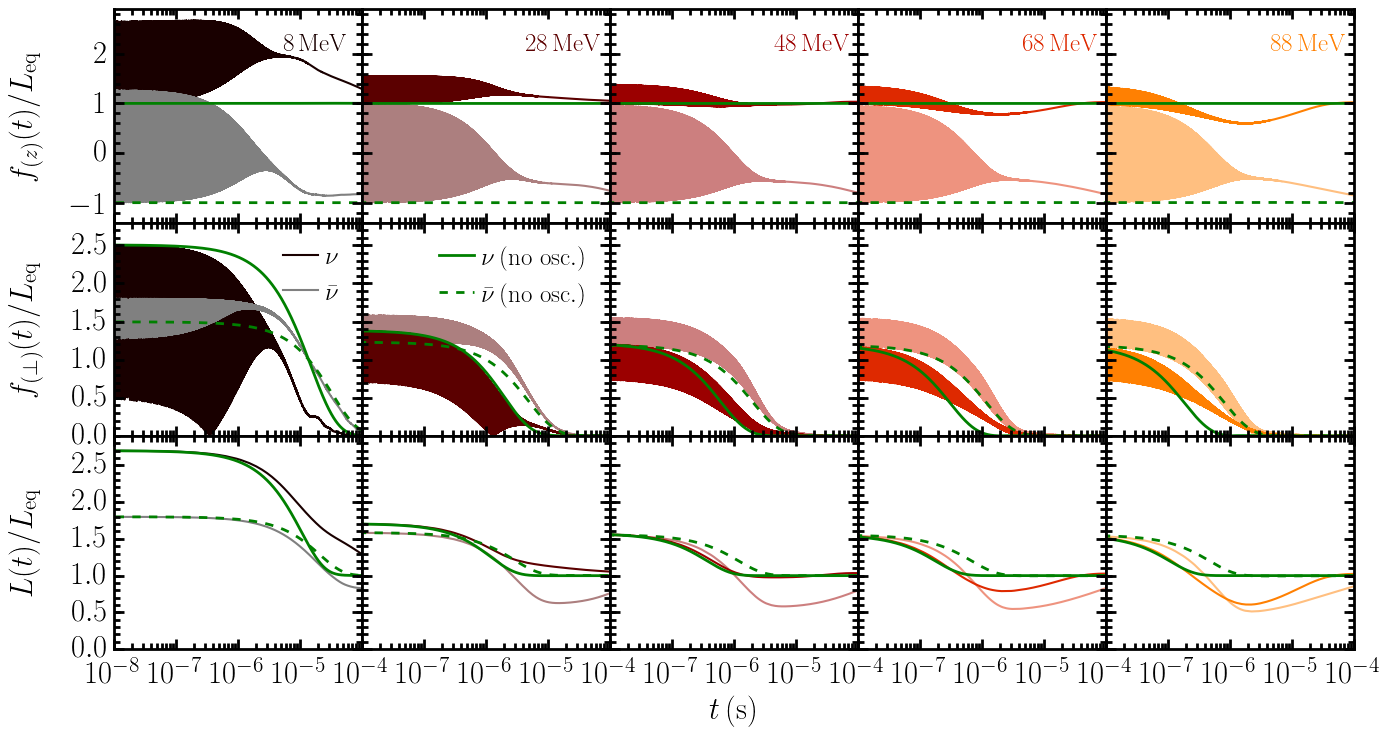}
    \caption{\textit{Top panels}: Evolution of the $z$ component of the neutrino distribution flavor vectors relative to the equilibrium vector length at several neutrino energies. Dark (light) shades show (anti)neutrino values. The equivalent values from the nonoscillating calculations are plotted in green for neutrinos (solid lines) and antineutrinos (dashed lines). The spread is due to violent oscillations on much shorter timescales, and one can see collisions decreasing the amplitude of the oscillations (width of the spread) and driving the distributions toward the flavor axis. \textit{Center panels}: Component of the distribution flavor vector perpendicular to the $z$ direction [Eq.~\ref{eq:fperp}] relative to the equilibrium vector length using the same color/line conventions as above. The spread is due to the same oscillations that cause the spread in the top panels. The results roughly agree with those of the nonoscillating calculations, plotted as solid green for neutrinos and dashed green for antineutrinos. \textit{Bottom panels}: Length of the neutrino distribution flavor vector relative to the equilibrium length using the same color and line conventions as above. There is not spread because the distribution flavor vector length is not directly sensitive to oscillations.}
    \label{fig:isospinL}
\end{figure*}
Only on timescales much longer than the oscillation time do collisions have a chance to impact the neutrino distribution, so we continue the simulation for an additional $100\,\mu\mathrm{s}$. The top left panel of Fig.~\ref{fig:isospinL} makes it clear that the amplitude of the nutations at neutrino energies of $8\,\mathrm{MeV}$ are damped on a timescale of around a microsecond and that the distributions are driven to flavor states on a timescale of several microseconds. In particular, this panel shows the $z$ component of the neutrino (dark) and antineutrino (light) distribution flavor vectors relative to their equilibrium Fermi-Dirac values. In flavor-diagonal equilibrium, the distribution flavor vector is entirely in the $z$ direction with a magnitude of $L_\mathrm{eq}$. As a reminder, the initial conditions were constructed as distribution flavor vectors described by $f_{(z)}(0)=L_\mathrm{eq}$ and an imposed large $x$ component, resulting in a value of $L(0)>L_\mathrm{eq}$. On the left side of the plot, the oscillations from the top two panels of Fig.~\ref{fig:osc} are visible as a vertical spread in the colored regions. The neutrinos oscillate from their initial latitude (bottom edge of the dark region) toward the electron flavor axis (top edge of the dark region), indicated by a $z$ component that grows to values of $f_{(z)}/L_\mathrm{eq}>1$. Antineutrinos start with a negative latitude (more muon antineutrinos that electron antineutrinos, bottom edge of the light region) and oscillate across the equator a bit beyond the opposite latitude (top edge of the light region). The oscillation term drives the flavor-diagonal components much farther from their initial values than the collision terms themselves were able to do in the absence of oscillations (plotted in solid green for neutrinos and dashed green for antineutrinos). As time progresses, two effects are evident. First, the amplitude of the oscillations in the $z$ direction decreases as the shaded areas compress to lines. Second, the location of the average value of $f_{(z)}$ decays to the equilibrium value. Stated another way, the distribution flavor vectors nutate within a band between two latitudes, and with time the width of the band (amplitude of the nutations) shrinks and the average latitude approaches the (muon) electron flavor axis for (anti)neutrinos.

The top left plot only reflects the behavior of the flavor-diagonal components of the distribution. However, in the leftmost center plot in Fig.~\ref{fig:isospinL} we see that oscillations in the flavor off-diagonal components are damped similarly to the flavor-diagonal components and that they damp on roughly the timescale predicted by the nonoscillating calculations. This can be seen as follows. To avoid overloading the plot with the precession motion, we only plot the part of the distribution flavor vector perpendicular to the flavor axis 
\begin{equation}
\label{eq:fperp}
    f_{(\perp)}\coloneqq\sqrt{f_{(x)}^2+f_{(y)}^2}\,,
\end{equation}
still at $8\,\mathrm{MeV}$. Though this quantity also  oscillates at the left-hand side of the plot, the neutrinos (dark) exhibit a much broader band than antineutrinos (light). This corroborates the interpretation of the top panel, namely that the neutrino distribution flavor vector is oscillating from its initial latitude (large $f_{(\perp)}$) to very near the electron flavor axis (small $f_{(\perp)}$) and back. Antineutrinos, on the other hand, oscillate from their initial latitudes across the flavor equator without getting very close to either axis (opposite sign $f_{(z)}$ and small variation in $f_{(\perp)}$). As with the $z$ component, the collisions simultaneously decrease the average value of $f_{(\perp)}$ and the range of the oscillation. At $t\approx0.4\,\mu\mathrm{s}$, the lower edge of the $f_{(\perp)}$ band is able to reach a value of zero, meaning the neutrino distribution flavor vector is able to nutate directly in line with the electron flavor axis. However, the minimum $f_{(\perp)}$ increases once again before slowly decaying back to zero. Compared to the decay of the corresponding quantities in the absence of oscillations (plotted in green), the overall evolution is somewhat more rapid for neutrinos and slower for antineutrinos.

As time progresses, the collisions work to remove off-diagonal components of the neutrino distribution, or equivalently to reduce $f_{(\perp)}$. In the case of antineutrinos, which oscillate back and forth over the flavor equator, this rapidly shortens the total length of the distribution flavor vector. The bottom left panel of Fig.~\ref{fig:isospinL} shows the length of the distribution flavor vector relative to the equilibrium Fermi-Dirac length at the same neutrino energy of $8\,\mathrm{MeV}$. The light curves show that the collisions decrease the antineutrino distribution flavor vector length to less than the equilibrium value, since the flavor-diagonal parts of the collision term are unable to replenish the on-diagonal distribution components as quickly as the off-diagonal parts of the collision term decrease the off-diagonal distribution components. Once $f_{(\perp)}$ is depleted, the diagonal components and hence the vector length slowly decay back to equilibrium values. Since the oscillations cause the neutrino distribution flavor vector to spend more time than the antineutrino vector spends near the flavor axis, the collisions are less effective at shortening the length of the distribution flavor vector and so it decreases more slowly than the case when oscillations are switched off (solid green lines). In contrast, we see that since oscillations cause the antineutrino vector to spend more time near the equator, the vector length decreases more rapidly than when oscillations are not present (dashed green lines).

This story remains true at higher energies as depicted in farther right plots in each row, but at higher energies the rate that at which diagonal components of the collision term equilibrate $f_{(z)}$ to $L_\mathrm{eq}$ become comparable to the rate at which the off-diagonal components equilibrate $f_{(\perp)}$ to zero. This can be seen in the top panels as the centers of the bands for both neutrinos and antineutrinos approach equilibrium more quickly, though the bands in the bottom panel approach 0 at a similar rate. In the case of low-energy neutrinos, the distribution flavor vector decays to the flavor axis, and when $f_{(\perp)}$ is depleted $f_{(z)}>L_\mathrm{eq}$ and the vector decays down the flavor axis until it reaches $f_{(z)}=L_\mathrm{eq}$. On the other hand, at high energies $f_{(z)}$ is driven to $L_\mathrm{eq}$ before $f_{(\perp)}$ reaches 0. As a result, the range of latitudes covered by the oscillations extends below the starting latitude, and the off-diagonal components are able to push the vector toward the flavor axis with both $f_{(z)}$ (top panels) and $L$ (bottom panels) less than $L_\mathrm{eq}$. Switching now to the antineutrinos, at low energies the distribution flavor vector decays to the flavor axis with $f_{(z)}$ and $L$ less than $L_\mathrm{eq}$, as we described earlier. Just as the neutrinos, at higher energies the antineutrinos are driven to $f_{(z)}=L_\mathrm{eq}$ before $f_{(\perp)}$ decays to zero, resulting in values of $L$ that do not dip as far below $L_\mathrm{eq}$ as at low energies. For both neutrinos and antineutrinos, the continued presence of off-diagonal components when $f_{(z)}$ first reaches $L_\mathrm{eq}$ allows continued evolution of $f_(z)$ and $L$ away from $L_\mathrm{eq}$ until $f_{(\perp)}$ decays to 0, and the interesting fluctuations in the top and bottom panels result. These behaviors are apparent in the evolution of the nonoscillating distributions (green curves) as well, but without oscillations driving the distribution flavor vectors through a wide range of latitudes, they tend not to be as pronounced.
  
Finally, the thickness of the bands in the top and center plots are much larger at low energies than at high energies. This is a result of our choice of initial conditions, since the differences between the distributions for different neutrino flavors at low energies permit a larger initial $f_{(\perp)}$ relative to $f_{(z)}$ [Eq.~\ref{eq:FDmaxmix}]. We remind the reader that our initial conditions also place all neutrinos and antineutrinos at the same phase angle (i.e., $f_{(x)}>0$ and $f_{(y)}=0$) and with Fermi-Dirac values for diagonal components. Understanding how the evolution of the neutrino distributions proceeds under different initial conditions is certainly worth further study.

\subsection{Flavor decoherence in core-collapse supernovae}
\label{sec:supernova}
\begin{figure}
    \centering
    \includegraphics[width=\linewidth]{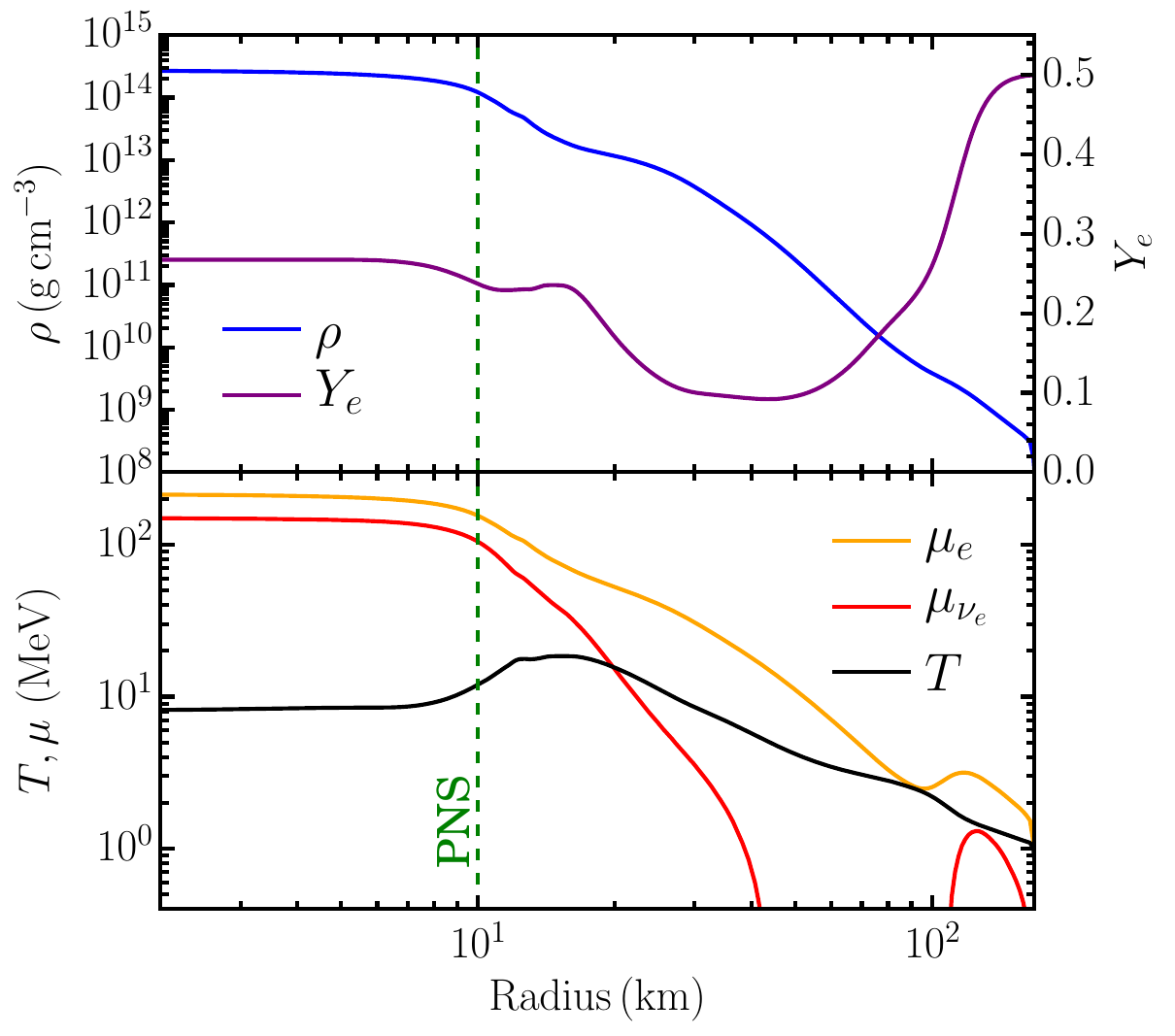}
    \caption{Background fluid snapshot from a one-dimensional neutrino radiation hydrodynamics core-collapse supernova simulation at $100\,\mathrm{ms}$ after core bounce \cite{Nagakura2017a}. We perform separate isotropic, homogeneous quantum kinetics calculations without the oscillation term at each radial point using the matter density $\rho$, electron fraction $Y_e$, and temperature $T$ at that radius as input. The bottom panel also shows for reference the chemical potentials of electrons $\mu_e$ and electron neutrinos $\mu_{\nu_e}$ as determined by the HShen equation of state \cite{Shen2011a}. The location of the shock is at the right edge of the plot at $168\,\mathrm{km}$, and the approximate location of the outer edge of the protoneutron star is shown at $10\,\mathrm{km}$ with a vertical dashed green line.}
    \label{fig:1dbackground}
\end{figure}
\begin{figure*}
    \centering
    \includegraphics[width=\linewidth]{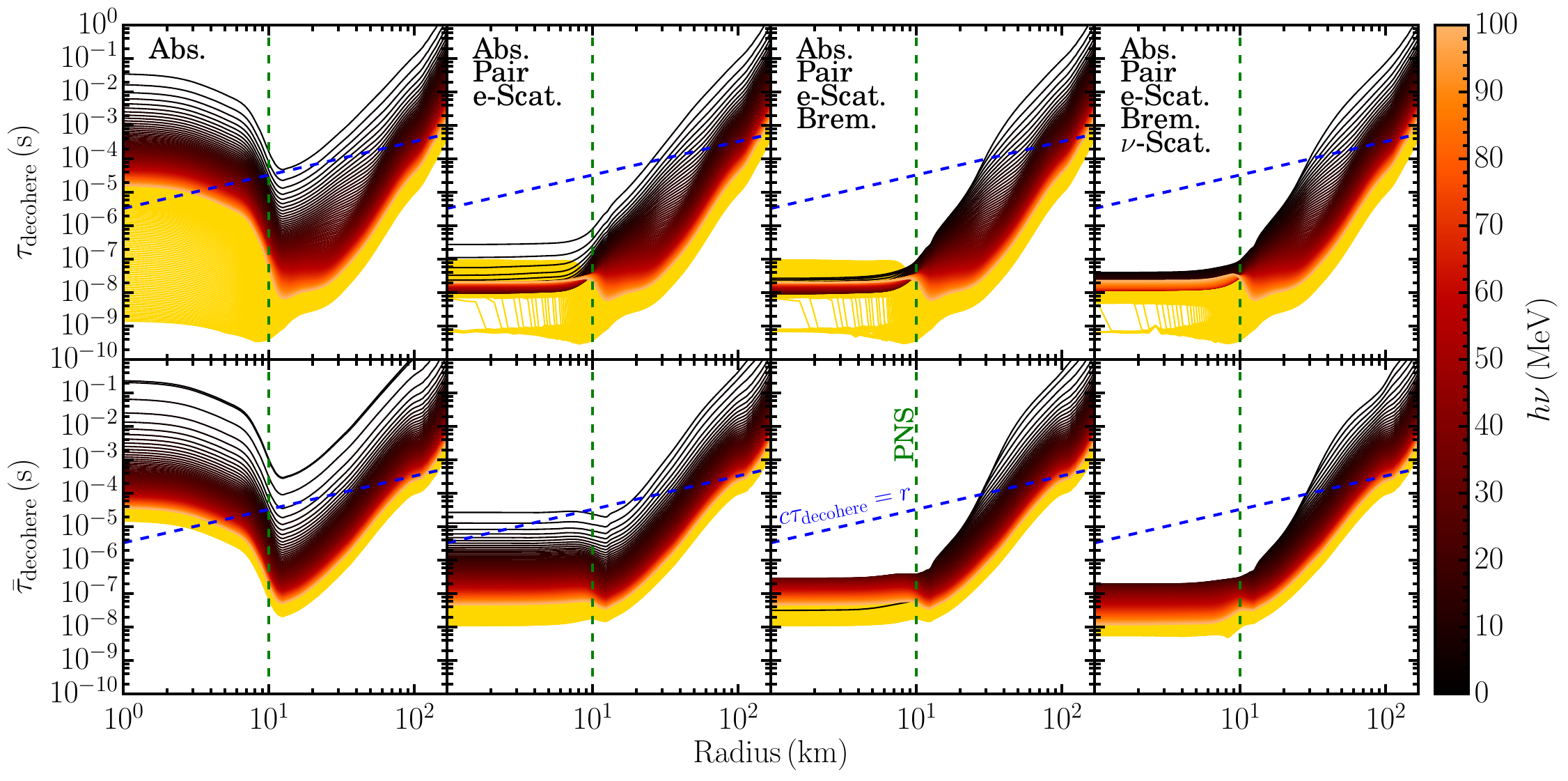}
    \caption{Decoherence times of flavor off-diagonal components of the neutrino distribution function in conditions relevant to CCSNe. We perform an isotropic, homogeneous nonoscillating quantum kinetic calculation using fluid values from each radial point in Fig.~\ref{fig:1dbackground}. Top panels show the decoherence times for neutrinos and bottom panels show those for antineutrinos. Neutrino energies from $1\,\mathrm{MeV}$ to $100\,\mathrm{MeV}$ are colored according to the color bar on the right, while energies from $101$ to $200\,\mathrm{MeV}$ are all colored gold. Each panel shows the results using a different set of interactions as indicated by the text in the panel. The approximate outer edge of the protoneutron star is depicted with a dashed green line at $10\,\mathrm{km}$. For reference, we also show the location where the radial coordinate is equal to the decoherence length scale with a dashed blue line. Electron scattering, bremsstrahlung processes, and four-neutrino processes dominate decoherence in the PNS, while absorption dominates decoherence just under the shock. All processes except neutrino-nucleon elastic scattering and neutrino-neutrino processes contribute in the decoupling region.}
    \label{fig:decay_time}
\end{figure*}

In order to understand flavor decoherence in conditions throughout a CCSN explosion, we perform a series of isotropic homogeneous QKE calculations using a range of input parameters relevant to CCSNe. We see in Fig.~\ref{fig:decay_time} that a detailed accounting of all of the processes described in Sec.~\ref{sec:collisionterms} (except for elastic nucleon scattering) are required to describe flavor decoherence everywhere under the shock. However, before describing the results in detail, we will outline the details of the calculations.

The range of parameters come from a snapshot of a one-dimensional neutrino radiation hydrodynamics CCSN simulation \cite{Nagakura2017a} shown in Fig.~\ref{fig:1dbackground}. This snapshot was taken at $100\,\mathrm{ms}$ after core bounce, by which time the shock has stalled at a radius of around $168\,\mathrm{km}$. The spatial grid of the original simulation has 384 radial points extending out to $5000\,\mathrm{km}$, but we only perform QKE calculations on each of the 241 radial points inside of the shock. Inside of the PNS (dashed green line) the temperatures are only $\sim8\,\mathrm{MeV}$ (black curve), but electrons and electron neutrinos are trapped and very degenerate (yellow and red curves). Just outside of the PNS the temperature reaches a maximum of $18.5\,\mathrm{MeV}$ and then continues to decrease with radius. The equilibrium electron neutrino degeneracy drops more quickly with radius than does the temperature, causing thermal electron neutrinos to become nondegenerate at $r\approx20\,\mathrm{km}$.

We perform an isotropic calculation for the conditions associated with each location in the CCSN shown in Fig.~\ref{fig:1dbackground}, i.e. between the center of the PNS and the position of the shock at $168\,\mathrm{km}$. In these calculations, we use a large energy grid spanning a domain of $200\,\mathrm{MeV}$ with energy grid spacing of $1\,\mathrm{MeV}$ in order to resolve the neutrino distribution anywhere under the shock. The large degeneracies within and near the PNS require an energy domain extending to $200\,\mathrm{MeV}$ to contain the neutrino distributions, and  when the temperature and electron neutrino degeneracy drop at larger radii we need to adequately resolve the more compact distribution. We elect to initialize the distribution function in each calculation with maximally mixed Fermi-Dirac values [Eq.~\ref{eq:FDmaxmix}] based on the background fluid temperature and chemical potentials for the initial neutrino distributions, even though outside of the neutrinosphere the neutrino distribution is more sparse and anisotropic than a thermal distribution. Without spatial transport, even flavor-diagonal distributions will evolve toward Fermi-Dirac distributions if they begin out of equilibrium with the fluid. It would be difficult to disentangle this known effect from the new off-diagonal parts of the collision terms if we did not start with Fermi-Dirac initial conditions on the diagonals. In addition, we do not see a way to consistently map anisotropic distributions into isotropic calculations in a meaningful way, so we construct our initial conditions using Eq.~\ref{eq:FDmaxmix} rather than from output of the CCSN simulation. Effects due to anisotropy should be addressed with anisotropic calculations with spatial transport, which will be the subject of future work.

Using the results of each of these simulations, we can define a flavor decoherence timescale $\tau_\mathrm{decohere}$ as the amount of time it takes for the magnitude of the flavor off-diagonal component of the neutrino distribution to decrease to a factor of $e^{-1}$ of its initial value. We show $\tau_\mathrm{decohere}$ throughout the CCSN profile in Fig.~\ref{fig:decay_time}, color coded by neutrino energy. The left panels show the results when we only consider absorption onto nucleons and nuclei \textit{\`a la} Sec.~\ref{sec:abs}. The interpretation of decay time is especially accurate in this case, since the off-diagonal components truly decay exponentially according to the absorption opacities. As expected, we see once again that neutrinos (top left panel) lose flavor coherence more quickly than antineutrinos (bottom left panel), and high-energy neutrinos decohere faster than low-energy neutrinos. Going inward from the shock (right side of the panel) to the protoneutron star, the decoherence timescales decrease as increasing matter densities and temperatures increase the interaction rates. Progressing into the PNS, the temperature decreases as the electron and neutron degeneracies increase, blocking neutrino and antineutrino absorption processes, respectively. In the PNS the electron neutrinos are also very degenerate and final-state neutrino blocking increases the effective electron neutrino absorption opacity (corrected for stimulated absorption), but the electrons more than compensate for this with an even larger degeneracy that greatly reduces the opacity.

In the second panels from the left of Fig.~\ref{fig:decay_time} we perform the same calculations, but include inelastic electron scattering and electron-positron pair annihilation processes. Outside of the PNS the results are rather similar to those in the first panel, but electron scattering totally dominates absorption and lepton pair processes as a driver of flavor decoherence in the PNS. Just as with the absorption processes, the high electron degeneracy strongly blocks electron-positron pair production, and the lack of positrons due to the same degeneracy makes pair production inefficient in the PNS. Note that the decoherence times for the highest energy neutrinos (gold curves) discontinuously jump within the PNS. This is an artifact of how we measure decoherence times. At the lower branch, the blocking terms in the off-diagonal component of the scattering term redistribute neutrinos in a way that causes the off-diagonal component $|f_{e\mu}(t)|$ to quickly decrease below $f_{e\mu}(0)/e$. On the upper branch, $f_{e\mu(t)}$ decreases past this point, continues through zero, and becomes large and negative; we must then wait for $|f_{e\mu}(t)|$ to decrease below $f_{e\mu}(0)/e$ a second time. With larger radii and smaller degeneracies, this overshoot effect becomes less severe. Eventually we reach a radius where the size of the overshoot is smaller than $1/e$ of the initial value and our measured decoherence time drops to the lower branch. 

In the third panels from the left, we once again repeat the calculations now with the addition of nucleon-nucleon bremsstrahlung interactions, implemented as effective absorption and emission of heavy lepton neutrinos. The bremsstrahlung reaction predominantly decreases the decoherence timescale of low-energy neutrinos and antineutrinos, leaving the high energies for the most part unchanged. In the antineutrino case, the effect is strong enough to make decoherence times increase with energy up to $24\,\mathrm{MeV}$, above which decoherence times again decrease with increasing energy. The importance of the bremsstrahlung reaction increases with density and is especially important within the PNS. More work needs to be done to develop detailed bremsstrahlung kernels to replace the effective absorption treatment, but once developed for flavor-diagonal neutrinos, they can be extended and developed into the QKE collision term in the same way as the electron-positron pair annihilation kernels.

Finally, in the fourth panels from the left we repeat the calculations one more time and include the full suite of reactions described in this paper. Within the PNS, neutrino-neutrino scattering increases the decoherence times of low-energy neutrinos (black curves, top right panel), decreases those of neutrinos above $100\,\mathrm{MeV}$ (yellow curves, top right panel), and decreases those of all antineutrinos (bottom-right panel). Outside of the PNS, there is some overal decrease in decoherence times due to neutrino-neutrino interactions, though primarily at low energies. However, we must once again consider that an isotropic thermal distribution is not realistic outside of the decoupling region, and we have not tested the effects of different initial conditions. More work is required to understand the quantum kinetics of non-equilibrium distributions.

\begin{figure}
    \centering
    \includegraphics[width=\linewidth]{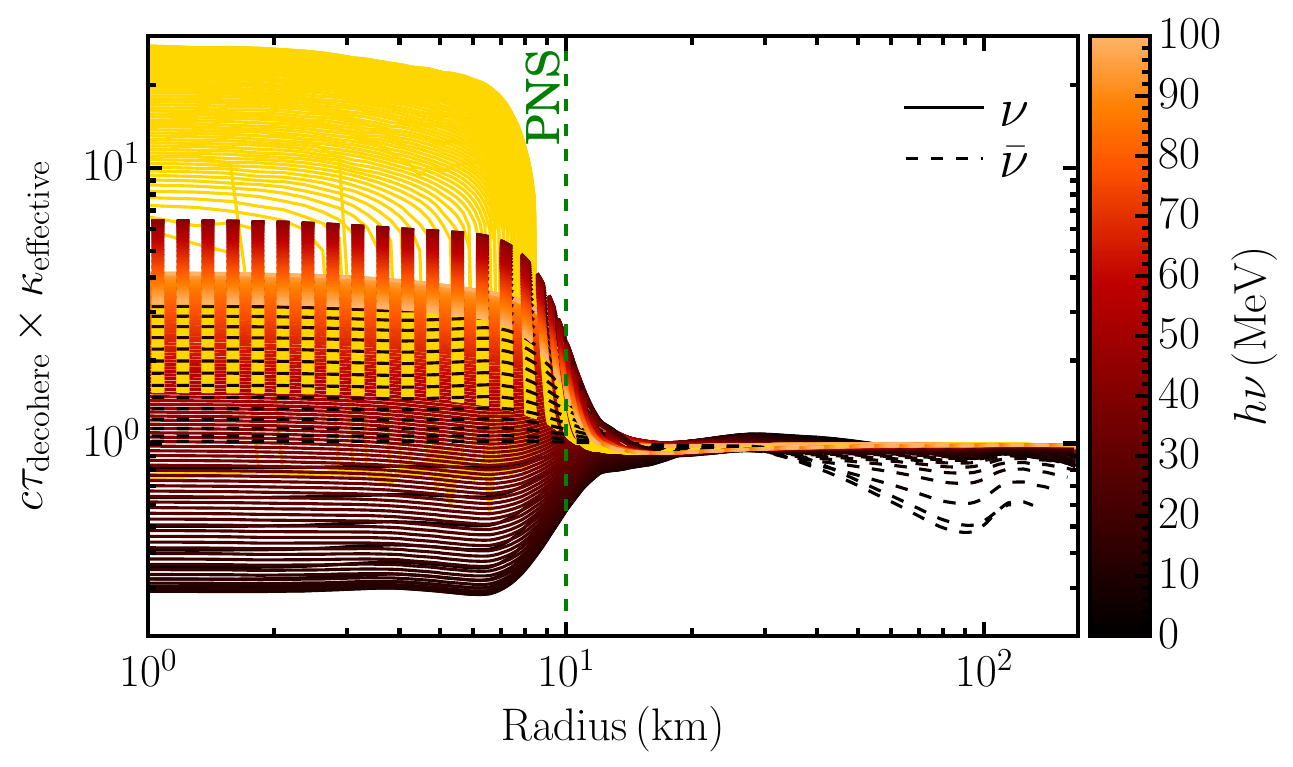}
    \caption{Ratio of the computed flavor decoherence length scale to the effective decoherence opacity [Eq.~\ref{eq:kappa_eff}] in the nonoscillating calculations including absorption, inelastic electron scattering, electron-positron pair annihilation, and effective-absorption nucleon-nucleon bremsstrahlung. This effective opacity is a much better predictor of flavor decoherence rates than the mean free path.}
    \label{fig:meanfreepath}
\end{figure}
The high computational cost of QKE calculations incites us to attempt to predict the flavor decoherence rates using only the flavor-diagonal interaction rates, i.e., without needing to run QKE calculations. Based on our QKE calculations, we find that the neutrino mean free path is not an accurate predictor of decoherence rates. However, we can empirically construct an effective decoherence opacity to reflect flavor decoherence rates as
\begin{equation}
\label{eq:kappa_eff}
    \kappa_\mathrm{effective} \coloneqq \kappa_{\mathrm{abs},e\mu} + \frac{1}{2}\widetilde{\kappa}_{\mathrm{scat},e\mu}\,.
\end{equation}
The value of $\kappa_\mathrm{abs}$ we use here contains bremsstrahlung radiation and electron-positron pair annihilation as effective absorption, and $\kappa_\mathrm{scat}$ is the elastic electron scattering opacity described in Sec.~\ref{sec:escat}. In Fig.~\ref{fig:meanfreepath} we compare the decoherence times from the third panels in Fig.~\ref{fig:decay_time} (i.e., including absorption, inelastic electron scattering, electron-positron pair annihilation, and nucleon-nucelon bremsstrahlung) to those predicted by $\kappa_\mathrm{effective}$. Within the PNS, the effective opacity predicts flavor decoherence rates for the $1-100\,\mathrm{MeV}$ within a factor of 10, while the results from higher energies are thrown off by the overshoot discussed in reference to Fig.~\ref{fig:decay_time}. Outside of $12\,\mathrm{km}$ $\kappa_\mathrm{effective}$ predicts decoherence rates to within $20\%$, except for the four lowest antineutrino energies ($1-4\,\mathrm{MeV}$) at $\sim90\,\mathrm{km}$. One might expect the charged-current component of the total opacity ($\kappa_{\mathrm{abs},e\mu}+\widetilde{\kappa}_{\mathrm{scat},e\mu}$) to describe flavor decoherence, but this combination is only accurate to within a factor of 2 outside of $12\,\mathrm{km}$ and is significantly worse than $\kappa_\mathrm{effective}$ inside the PNS. For completeness, we also examined the absorption opacity by itself and found that it consistently underpredicts the decoherence rates. Another combination, the total off-diagonal opacity ($\kappa_{\mathrm{abs},e\mu} + \kappa_{\mathrm{scat},e\mu}$), is also less successful than $\kappa_\mathrm{effective}$, since it is only good to within a factor of 10 outside of $12\,\mathrm{km}$ and much worse inside the PNS.

\section{Discussion and conclusions}
CCSN simulations have long included neutrino transport with a detailed set of collision rates, but until now the technology to simultaneously and self-consistently also treat neutrino flavor oscillations did not exist. In this work, we take the first steps toward developing the technology to simulate neutrino quantum kinetics in regions relevant to the CCSN explosion mechanism. In Sec.~\ref{sec:collisionterms} we demonstrate a means of converting existing neutrino interaction rates used in CCSN simulations into full collision terms for the quantum kinetic equations. For reference, we also describe the moment-integrated form of these source terms in Appendix~\ref{app:moment} for application to moment-based QKE calculations. To demonstrate the use of these collision terms, we developed a novel method and open-source code {\tt IsotropicSQA} for explicitly evolving both oscillations and collisions using stochastic integration to randomly sample the impact of collisions during the evolution (Sec.~\ref{sec:isotropicsqa}). In our code and derivations, we include absorption of neutrinos by nucleons and nuclei, inelastic scattering by electrons, elastic scattering by nucleons, kernel-based electron-positron pair annihilation, kernel-based neutrino-neutrino scattering and annihilation, and nucleon bremsstrahlung as an effective absorptive process.

In Sec.~\ref{sec:results}, we demonstrated the use of this new method and source terms to perform the first, albeit isotropic and homogeneous, direct evolution of the QKEs in conditions relevant to the CCSN explosion mechanism. In particular, we chose conditions of $\rho=10^{12}\,\mathrm{g\,cm}^{-3}$, $T=10\,\mathrm{MeV}$, and $Y_e=0.3$, and evolved an initially strongly flavor-mixed neutrino distribution for $20\,\mu\mathrm{s}$. This was long enough to demonstrate that the distribution relaxes to a flavor-diagonal Fermi-Dirac distribution on timescales similar to but measurably different from those observed in calculations where the oscillation term is neglected.

Given this insight, we performed a parameter sweep using nonoscillating calculations in conditions relevant to CCSNe. We find that electron scattering, nucleon-nucleon bremsstrahlung, and four-neutrino processes are the dominant drivers of decoherence within the PNS, while just inside the shock front absorption is dominant. In the decoupling region, all of the processes discussed in this paper except for elastic nucleon scattering and neutrino-neutrino processes had a significant impact on flavor decoherence times. This demonstrates the need for a more sophisticated estimate of decoherence rates than the neutrino mean free path. To address this, we defined an effective decoherence opacity in Eq.~\ref{eq:kappa_eff} that predicts decoherence rates within $\sim20\%$ outside of $12\,\mathrm{km}$ and within a factor of 10 everywhere in our parameter set. In the future, we will test this estimate for different stages of a CCSN and under more realistic treatments of neutrino angular distributions.

Numerical neutrino quantum kinetics is a nascent field and there is a great deal of work to do. Our calculations are limited to isotropic and homogeneous neutrino and matter distributions, but we are hopeful that this work will encourage development of QKE codes with spatial transport. This will be required to gain a quantitative understanding of how flavor coherence is transported throughout a CCSN, especially in the decoupling region where neutrino densities and collision rates are high. The importance of nucleon-nucleon bremsstrahlung in and near the PNS begs for an improved kernel-based treatment of these reactions rather than our effective absorption method. Fortunately, the same framework for generating QKE source terms from existing electron-positron pair production rates can also do the same for bremsstrahlung interaction rates. Inelastic nucleon scattering kernels could also be implemented in the framework of inelastic electron scattering. Finally, our treatment of neutrino-neutrino scattering and annihilation could become computationally prohibitive for more large-scale calculations. Since this term requires integrating over the phase space of four neutrino distributions, the cost of calculating this term increases as $N_E^4$, where $N_E$ is the number of energy bins. In addition, our particular discretization of neutrino energy (bins having equal widths centered on integer multiples of the first bin center), though allowing for a straightforward implementation of the four-neutrino processes, does not allow a single energy grid to cover neutrino distributions in the range of temperatures and chemical potentials seen in CCSNe unless an intractable number of energy bins are used. Developing a more efficient way to accurately treat four-neutrino processes will be necessary for QKE calculations covering the wide range of conditions seen in a CCSN.

The full QKE calculation presented in Sec.~\ref{sec:results} required collision term time steps on the order of $2\times10^{-11}\,\mathrm{s}$ and oscillation time steps on the order of $3\times10^{-15}\,\mathrm{s}$, and required about four days on as many cores. The cost of the calculation is in the large number of time steps needed to follow the oscillations and the high accuracy required of each step to prevent numerical artifacts from appearing. Since these isotropic and homogeneous calculations evolve a relatively small number of variables, the parallelizability is limited. However, future calculations including spatial transport will have many more quantities to evolve, and thus will be much more suited to taking advantage of larger computing resources.

While simulating the QKEs in CCSNe presents new challenges, doing so is essential to understanding the neutrino signal from and potentially also to the explosion mechanism of CCSNe. In addition to CCSNe, mergers of two neutron stars or of a neutron star and a black hole are astrophysical environment that are a particularly interesting home to potential quantum kinetic effects, as neutrino flavor transformations have already been suggested to be significant very close to the decoupling region \cite{Zhu2016,Malkus2016,Chatelain2017}. There is much to be explored on the frontier of numerical quantum kinetics.

\section{Acknowledgments}
SR is supported by the N3AS Fellowship under National Science Foundation Grant No. PHY-1630782 and Heising-Simons Foundation Grant No. 2017-228. G.M., A.V. and J.K. are supported at NC State by DOE Grant No. DE-FG02-02ER41216. We thank Hiroki Nagakura for access to core-collapse supernova simulation results. Some computations were performed on the Wheeler cluster at Caltech, which is supported by the Sherman Fairchild Foundation and by Caltech. Feynman diagrams are generated with the {\tt TIKZ-FEYNMAN} package \cite{Ellis2017}.

\bibliography{mendeley_v2}

\appendix

\section{Moment form}
\label{app:moment}
Following \cite{Shibata2011,Cardall2013}, we can take moments of the QKEs in order to generate evolution equations for a small number of angular moments of the distribution function. In order to do this, we decompose the neutrino four-momentum into
\begin{equation}
    p^\alpha = \frac{h\nu}{c}(u^\alpha + l^\alpha)\,,
\end{equation}
where $l^\alpha$ is a unit normal four-vector ($l^\alpha l_\alpha=1$) orthogonal to $u^\alpha$ ($u^\alpha l_\alpha=0$). We can define the first few comoving-frame angular moments (units of $\mathrm{Hz}^3$) of the neutrino distribution function as
\begin{equation}
\begin{aligned}
    J_{ab}(\nu,x^\mu) &\coloneqq \nu^3 \int d\Omega f_{ab}(\nu,\Omega,x^\mu)\\
    H^\alpha_{ab}(\nu,x^\mu) &\coloneqq \nu^3 \int d\Omega f_{ab}(\nu,\Omega,x^\mu)l^\alpha\\
    L^{\alpha\beta}_{ab}(\nu,x^\mu) &\coloneqq \nu^3 \int d\Omega f_{ab}(\nu,\Omega,x^\mu)l^\alpha l^\beta\\
    N^{\alpha\beta\gamma}_{ab}(\nu,x^\mu) &\coloneqq \nu^3 \int d\Omega f_{ab}(\nu,\Omega,x^\mu)l^\alpha l^\beta l^\gamma\\
\end{aligned}\,.
\end{equation}
Note that the oscillation Hamiltonian $H_{ab}$ can be distinguished from the first moment of the radiation field $H^\alpha_{ab}$ by the presence of a spacetime index. The two-moment evolution equations from \cite{Shibata2011} Eq. 3.19 are
\begin{equation}
\label{eq:QKE_moment}
    \nabla_\beta M^{\alpha\beta}_{ab} - \frac{\partial}{\partial \nu} \left(\nu M^{\alpha\beta\gamma}_{ab}\nabla_\gamma u_\beta\right) = S^\alpha_{ab}\,,
\end{equation}
where
\begin{equation}
    \begin{aligned}
    M^{\alpha\beta}_{ab} &= J_{ab} u^\alpha u^\beta+H^\alpha_{ab} u^\beta + H^\beta_{ab} u^\alpha + L^\alpha\beta_{ab} \\
    M^{\alpha\beta\gamma}_{ab} &= J_{ab} u^\alpha u^\beta u^\gamma + H^\alpha_{ab} u^\beta u^\gamma H^\beta_{ab} u^\alpha u^\gamma + H^\gamma_{ab} u^\alpha u^\beta \\
    &+ L^{\alpha\beta}_{ab}u^\gamma + L^{\alpha\gamma}_{ab}u^\beta + L^{\beta\gamma}_{ab}u^\alpha + N^{\alpha\beta\gamma}_{ab}\,.
    \end{aligned}
\end{equation}

In this work, we will focus on the source terms, which can be expressed in moment form as
\begin{equation}
    S^\alpha_{ab}(\nu,x^\mu) \coloneqq \nu^3 \int d\mathbf{\Omega} \left(C_{ab}-\frac{i}{\hbar c}[H,f]_{ab}\right)(u^\alpha + l^\alpha)
    \label{eq:sourceterm}
\end{equation}
(units of $\mathrm{Hz}^3\,\mathrm{cm}^{-1}$).
We leave the full implementation of relativistic QKEs to future work, but note that, when neglecting the $\mathcal{O}(\epsilon^2)$ force and drift term corrections, all of the flavor coherence effects lie in the source term. These evolution equations need to be massaged into a numerically implementable form (e.g., \cite{Shibata2011,Cardall2013,Roberts2016,OConnor2015,OConnor2015a,Foucart2015}).

\textit{Oscillation Term} - Performing the angular integrals on the contribution to the source term from the vacuum and matter potentials yields
\begin{equation}
    S^\alpha_{ab} = -\frac{i}{\hbar c} [H_\mathrm{vacuum} + H_\mathrm{matter},\Psi^\alpha]_{ab}\,.
\end{equation}
Similarly, performing the angular integral on the contribution from the neutrino self-interaction term, we get
\begin{equation}
\begin{aligned}
    S^\alpha_{ab} = -i\frac{\sqrt{2} \hbar^2 G_F}{\nu^3 c} \int \frac{d\nu'}{\nu'}  \{&[(J'-\bar{J}^{\prime *}),\Psi^\alpha]_{ab}\\
    - &[(H'^\beta-\bar{H}^{\prime\beta *}),\Xi^\alpha_{\beta}]_{ab}\}\,,
    \end{aligned}
\end{equation}
where we have defined the following Lorentz-invariant objects (units of $\mathrm{Hz}^3$):
\begin{equation}
\begin{aligned}
\label{eq:helpervars}
    \Psi^\alpha_{ab}(\nu,x^\mu) &\coloneqq J_{ab} u^\alpha + H^\alpha_{ab} \\
    \Xi^\alpha_{\beta ab}(\nu,x^\mu) &\coloneqq g_{\beta\mu}(H^\mu_{ab}u^\alpha + L^{\mu\alpha}_{ab})\,.
\end{aligned}
\end{equation}

\textit{Absorption and Emission} - To get the source term for the two-moment form of the QKEs [Eq.~\ref{eq:QKE_moment}], we perform the integral in Eq.~\ref{eq:sourceterm} for the contribution to the source term in Eq.~\ref{eq:C_abs} and arrive at
\begin{equation}
\label{eq:S_abs}
    S^\alpha_{ab} = 4\pi \nu^3 j_{(\nu_a)}\delta_{ab}u^\alpha - (\langle j\rangle_{ab}+\langle\kappa\rangle_{ab})\Psi^\alpha_{ab}\,.
\end{equation}
The source term for antineutrinos is exactly analogous.

\textit{Scattering} - Angular moments of the inelastic scattering kernel [Eq.~\ref{eq:escat_Rdeconstruct}] can be integrated as
\begin{equation}
\begin{aligned}
    \nu'^3\nu^3\int d\Omega d\Omega' f (u^\alpha + l^\alpha) R^\pm &=  \frac{4\pi\nu'^3}{2}\Phi_0^\pm \Psi^\alpha\\
    \nu'^3\nu^3\int d\Omega d\Omega' f' (u^\alpha + l^\alpha) R^\pm &= \frac{4\pi\nu^3}{2}(\Phi_0^\pm J' u^\alpha + \Phi_1^\pm H'^\alpha)\\
    \nu'^3\nu^3\int d\Omega d\Omega' f f' (u^\alpha+l^\alpha) R^\pm &= \frac{1}{2}\Phi_0^\pm J'\Psi^\alpha \\
    & + \frac{3}{2}\Phi_1^\pm H'^\beta\Xi^{\alpha}_{\beta}.\\
\end{aligned}
\label{eq:angularintegrals}
\end{equation}
The resulting source terms from performing the angular integral in Eq.~\ref{eq:sourceterm} on Eq.~\ref{eq:C_escat} are
\begin{equation}
\label{eq:S_escat}
\begin{aligned}
S^\alpha_{ab} = \int \frac{d\nu'}{c^4\nu'}\{ \frac{4\pi}{2}[&\nu^3(\Phi_{0ab}^+ J'_{ab}u^\alpha + \Phi_{1ab}^+ H'^\alpha_{ab}) \\
- &\nu'^3 \langle\Phi\rangle_{0ab}^-\Psi^\alpha_{ab}]- \zeta^\alpha_{ab}\},\\
\end{aligned}
\end{equation}
where
\begin{equation}
\label{eq:zeta}
\begin{aligned}
    \zeta^\alpha_{ab} &\coloneqq \nu'^3\nu^3\int d\Omega d\Omega' (\varsigma_{ab}^+-\varsigma^-_{ab}) (u^\alpha + l^\alpha)\\
    &=\frac{1}{2}\sum_{c} [\frac{1}{2}\left(\Delta\Phi_{0cb}\Psi^\alpha_{ac}J'_{cb}+\Delta\Phi_{0ac}J'_{ac}\Psi^\alpha_{cb}\right)+ \\ &\phantom{\frac{1}{2}\sum_{c} [}\frac{3}{2}\left(\Delta\Phi_{1cb}\Xi^{\alpha}_{\beta ac}H'^{\beta}_{cb}+\Delta\Phi_{1ac}H'^{\beta}_{ac}\Xi^{\alpha}_{\beta cb}\right)],
\end{aligned}
\end{equation}
(units $\mathrm{Hz}^7\mathrm{cm}^3$) and $\Delta\Phi_{ab}\coloneqq\Phi^+_{ab}-\Phi^-_{ab}$. The contribution to $S^\alpha_{aa}$ from the terms inside the square brackets in Eq.~\ref{eq:S_escat} and the $c=a$ terms in $\zeta^\alpha_{aa}$ make the well-known source term for noncoherent neutrinos as in \cite{Shibata2011}. If the scattering process is treated as elastic [Eq.~\ref{eq:C_elasticscat}], the source terms simplify to
\begin{equation}
\label{eq:S_elasticscat}
\begin{aligned}
    S^\alpha_{ab} &= \left[- H^\alpha_{ab} (\langle\kappa\rangle_{0ab} - {\kappa}_{1ab}/3) - J_{ab}\widetilde{\kappa}_{0ab}u^\alpha \right]
    \end{aligned}
\end{equation}

\textit{Pair Processes} - The pair process (electron-positron and nucleon-nucleon bremsstrahlung) source terms can be integrated from Eq.~\ref{eq:C_pair} as
\begin{equation}
\label{eq:S_pair}
\begin{aligned}
S^\alpha_{ab} = \int \frac{d\bar{\nu}'}{\nu'}\{ \frac{4\pi}{2}[&4\pi \nu^3 \bar{\nu}'^3\Phi_{0ab}^+\delta_{ab} \\
-&\nu^3(\Phi_{0ab}^+ \bar{J}'_{ab}u^\alpha/c + \Phi_{1ab}^+ \bar{H}'^\alpha_{ab}) \\
-& \bar{\nu}'^3 \langle\Phi\rangle_{0ab}^+\Psi^\alpha_{ab}] + \zeta^\alpha_{ab}\}\,.\\
\end{aligned}
\end{equation}
$\zeta^\alpha_{ab}$ is defined in Eq.~\ref{eq:zeta}, though for Eq.~\ref{eq:S_pair} one should use values of $R^\pm_{ab}$ for this process, and primed quantities should be primed and barred. The source term from integrating the effective absorption collision term [Eq.~\ref{eq:C_fakepair}] looks identical to Eq.~\ref{eq:S_abs}, but here the emissivities $j_{(\nu_a)}$ and opacities $\kappa_{(\nu_a)}$ for heavy lepton neutrinos are nonzero.

\section{Code tests}
\label{app:tests}
To ensure that IsotropicSQA produces realistic results, we perform a few basic tests. 
\subsection{Equilibrium test}
In the limit of no flavor mixing, the interactions should reduce to the well-known flavor-diagonal interactions, which drive the neutrino distributions to Fermi-Dirac distributions described by the fluid temperature and chemical potentials of $\mu_{\nu_e}=\mu_p + \mu_e - \mu_n$, $\mu_{\bar{\nu}_e}=-\mu_{\nu_e}$, and $\mu_{\nu_x}=0$. Though we see the collision kernels in the main text driving the distribution function to this equilibrium, we quantify how well the collision terms maintain this equilibrium. To test this, we initialize the diagonal components of the distribution function to their thermal equilibrium values, set off-diagonal elements to zero, and evolve including all interactions in Sec.~\ref{sec:collisionterms} and without oscillations. All tests are performed under the conditions of $\rho=10^{12}\,\mathrm{g\,cm}^{-3}$, $T=10\,\mathrm{MeV}$, $Y_e=0.3$, and a neutrino energy domain of $1-101\,\mathrm{MeV}$ as in Sec.s~\ref{sec:collisionterms} and \ref{sec:results}. We run three tests: (a) 50 energy bins and an integration accuracy of $10^{-12}$, (b) 25 energy bins an an integration accuracy of $10^{-12}$, and (c) 50 bins and an integration accuracy of $10^{-11}$. In all cases, within $25\,\mu\mathrm{s}$ the maximum deviation of any component of the distribution function settles to within one part in $2\times10^{-15}$ of its initial value. The 25 bin case briefly has maximum errors that grow to $5\times10^{-12}$ before settling down to the previously mentioned error. We also tested individual processes with similar results. These tests show that the equilibrium Fermi-Dirac distribution is reproduced to within numerical error by our collision processes.

\subsection{Vacuum oscillations}
\begin{figure}
    \centering
    \includegraphics[width=\linewidth]{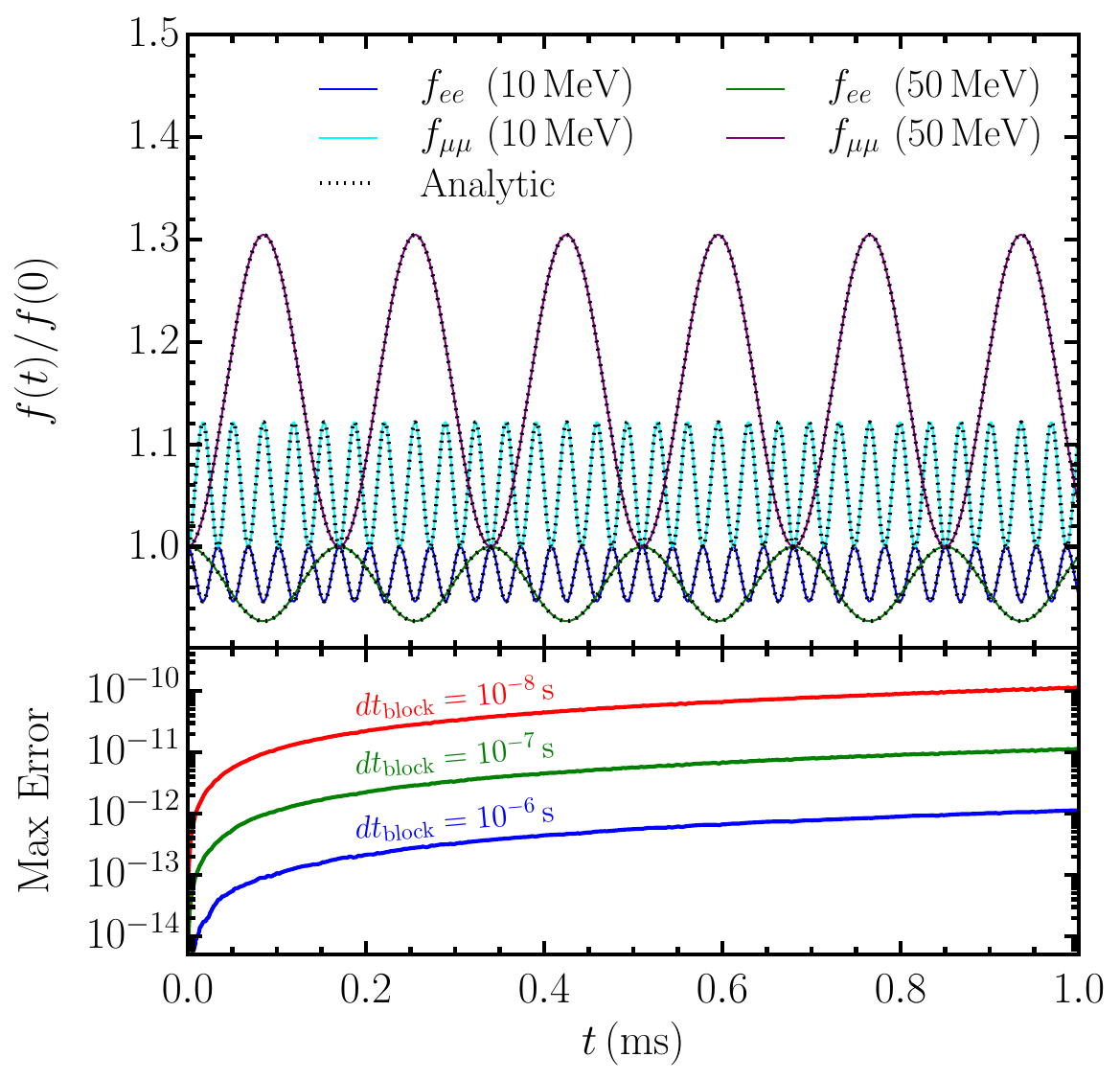}
    \caption{\textit{Vacuum oscillations test}. -- The top panel shows the evolution of the $\nu_e$ and $\nu_\mu$ distribution functions for the $10\,\mathrm{MeV}$ and $49\,\mathrm{MeV}$ energy bins, along with the analytic solution. The bottom panel shows the maximum error in the solution for simulations with three different time step sizes. The error increases with decreasing step size. See text for details.}
    \label{fig:test_vacuum}
\end{figure}
To ensure that the oscillations evolve correctly, we first simulate vacuum oscillations by setting the matter and interaction potentials to zero. The probability of a neutrino transitioning from one flavor to another is given by the well-known formula (e.g., \cite{Bellini2013})
\begin{equation}
    P_T(t) = \sin^2\left(2\theta_{12}\right) \sin^2\left(\frac{c^4 \Delta m_{12}^2 t}{4 E \hbar}\right)
    \label{eq:P_transition}
\end{equation}
Thus, we expect the distribution function values to follow
\begin{equation}
\begin{aligned}
    f_{ee}(t) &= [1-P_T(t)]f_{ee}(0) + P_T(t)f_{\mu\mu}(0) \\
    f_{\mu\mu}(t) &= [1-P_T(t)]f_{\mu\mu}(0) + P_T(t)f_{ee}(0)
\end{aligned}
\label{eq:oscillated_f}
\end{equation}
The good agreement between the computed (colored lines) and analytic (dotted lines) results is shown in the top panel of Fig.~\ref{fig:test_vacuum}. The bottom panel shows the maximum relative error among all energy groups and neutrino species between the computed and analytic solution as a function of the time step size. The error actually increases when the time step decreases. This is due to the hybrid representation of the distribution function matrices. After every oscillation step, the unitary evolution operator is applied to the distribution functions. Though the operator itself is represented very accurately, some accuracy is lost in applying the evolution operator to the distribution function. This emphasizes the importance of applying the operator as sparsely as possible, and it is the reason we allow the evolution operator to evolve for many steps between applications to the distribution function.

\subsection{MSW resonance}
\begin{figure}
    \centering
    \includegraphics[width=\linewidth]{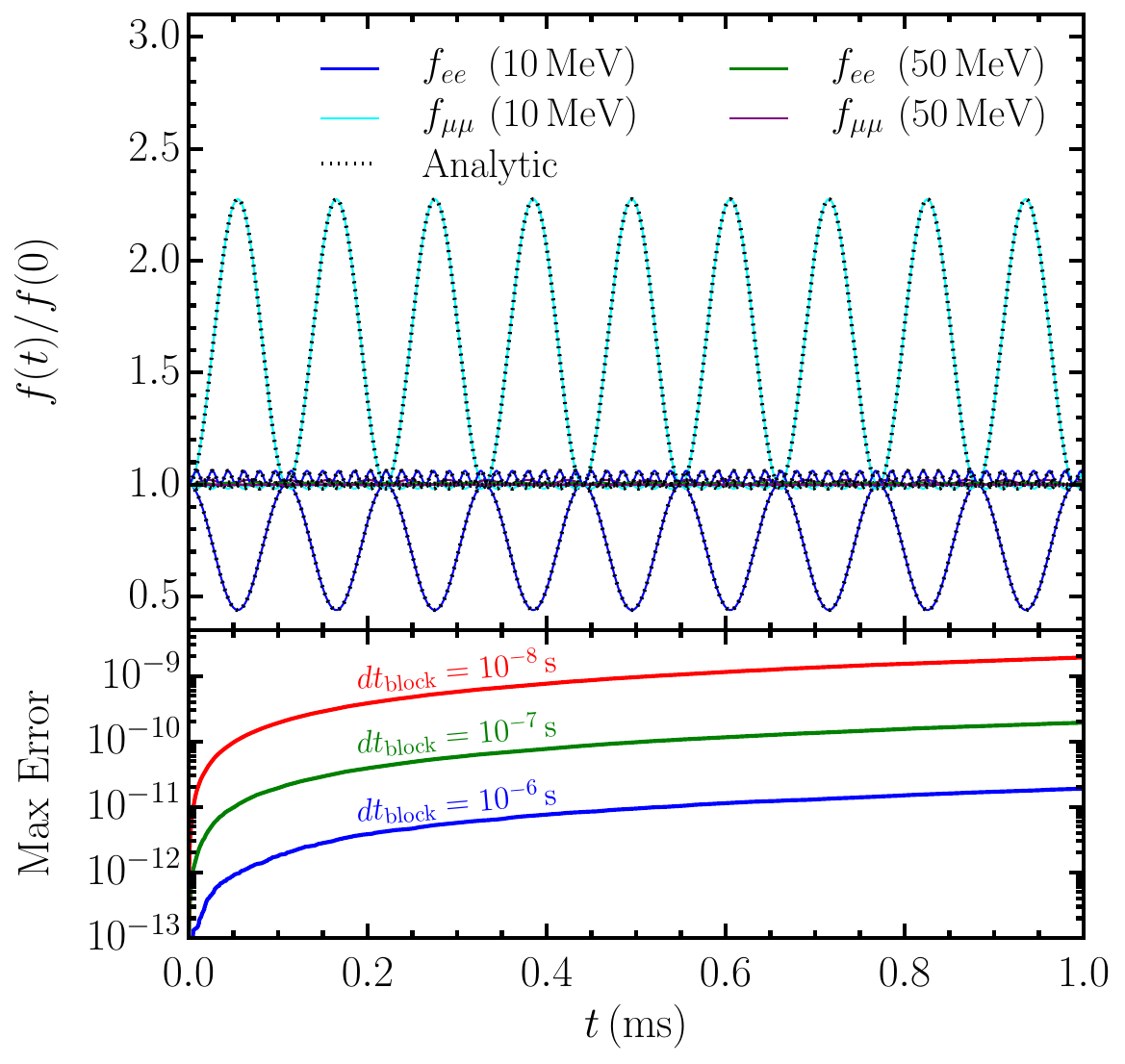}
    \caption{\textit{MSW oscillations test}. -- The top panel shows the evolution of the $\nu_e$ and $\nu_\mu$ distribution functions for the $10\,\mathrm{MeV}$ and $49\,\mathrm{MeV}$ energy bins, along with the analytic solution. The bottom panel shows the maximum error in the solution for simulations with three different time step sizes. The error increases with decreasing step size. See text for details.}
    \label{fig:test_msw}
\end{figure}
Oscillations in a constant matter background are also straightforward to compute. Eq.~\ref{eq:oscillated_f} still applies, but the mixing angle and mass squared difference in Eq.~\ref{eq:P_transition} are replaced by effective values of
\begin{equation}
\begin{aligned}
    \sin^2(2\widetilde{\theta}_{12}) &= \frac{\sin^2(2\theta_{12})}{\sin^2(2\theta_{12}) + C^2} \\
    \Delta \widetilde{m}_{12}^2 &= \Delta m_{12}^2 \sqrt{\sin^2(2\theta_{12}) + C^2} \\
    C &= \cos(2\theta_{12}) - \frac{2V E}{\Delta m_{12}^2 c^4}
\end{aligned}
\end{equation}
where $V=\pm \sqrt{2} G_F \hbar^3 c^3 n_e$ is the matter potential (positive for neutrinos, negative for antineutrinos). Fig.~\ref{fig:test_msw} shows the error associated with evolution in a matter potential that puts the $20\,\mathrm{MeV}$ neutrinos on the Mikheyev–Smirnov–Wolfenstein (MSW) resonance ($C=0$).

\subsection{Bipolar oscillations}
\begin{figure}
    \centering
    \includegraphics[width=\linewidth]{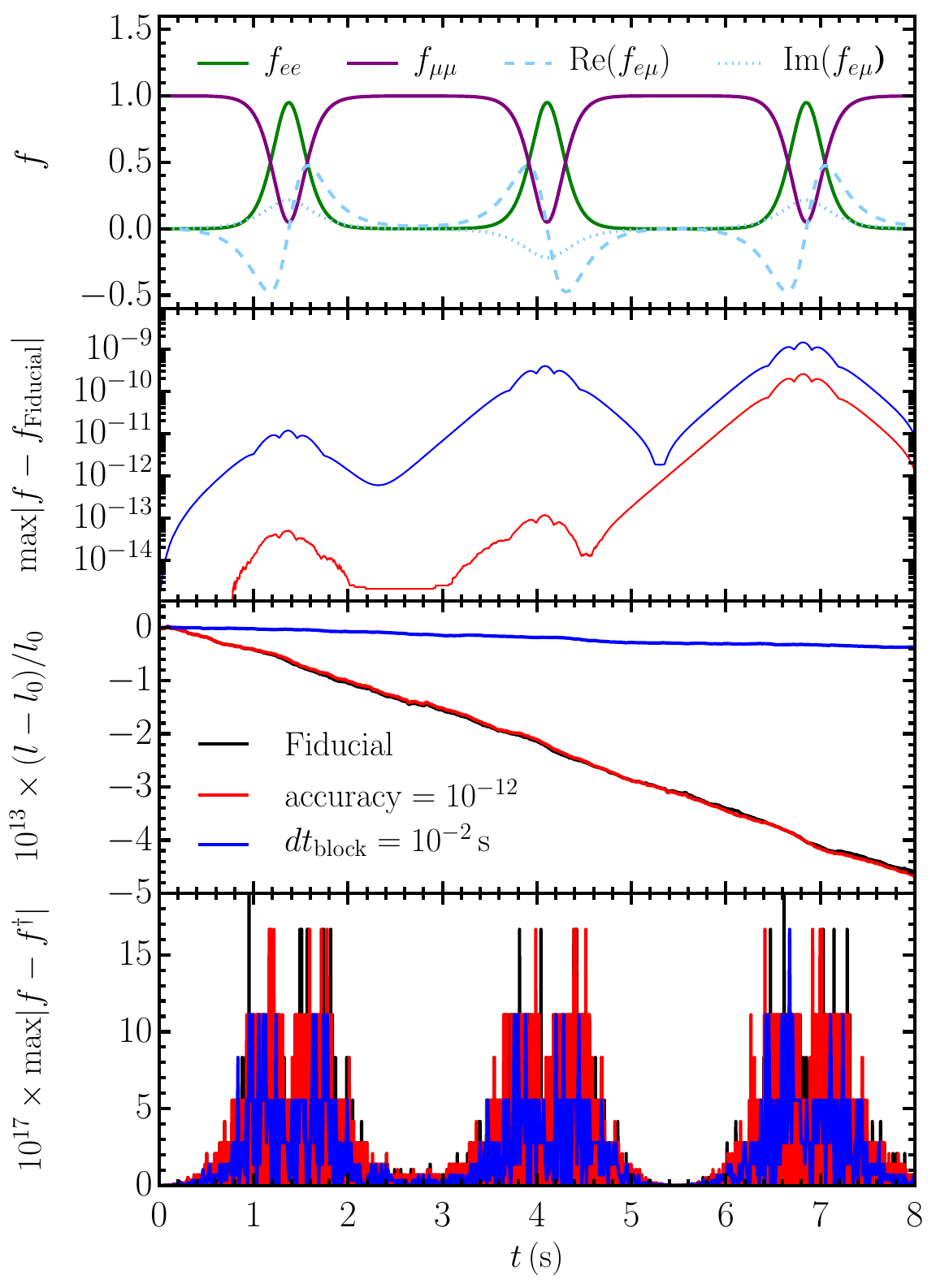}
    \caption{\textit{Bipolar oscillation test}. -- \textit{First panel:} evolution of a neutrino distribution function $f$ initially consisting of pure $50\,\mathrm{MeV}\,\nu_\mu$ and $\bar{\nu}_\mu$ in equal amounts ignoring collision terms. \textit{Second panel:} Maximum error between the fiducial calculation (integration accuracy of $10^{-13}$ and $dt_\mathrm{block}=10^{-3}\,\mathrm{s}$) and that with lower integration accuracy (red line) and longer $dt_\mathrm{block}$ (blue line). \textit{Third panel:}  Relative change in the length of the distribution flavor vector, which should be $0$ since collision terms are set to zero. \textit{Fourth panel:} How non-Hermitian the distribution function becomes at each step before enforcing Hermitivity.}
    \label{fig:bipolar_error}
\end{figure}
Fig.~\ref{fig:bipolar_error} shows a calculation of bipolar oscillations and associated errors. Bipolar oscillations occur for pure electron anti/neutrinos with the inverted mass hierarchy and for pure anti/muon neutrinos with the normal mass hierarchy. In this test, we set up an initial distribution of pure anti/muon neutrinos with $f_{\mu\mu}=\bar{f}_{\mu\mu}=1$ in a single energy bin at an energy of $h\nu$. The background matter density is set to 0, the angle between the mass and flavor bases to $0.01$, and the self-interaction Hamiltonian to be
\begin{equation}
    H_\mathrm{neutrino} = \frac{10 (m_2^2-m_1^2) c^4}{2 h\nu}(f-\bar{f})\,.
\end{equation}
We also set the mass difference to $m_2^2-m_1^2 = (2h\nu)\hbar/c^4$, such that the approximate frequency for bipolar oscillations is then $\kappa\approx0.995\,\mathrm{s}^{-1}$, corresponding to the $\mu=10,\,\omega=1$ case in \cite{Hannestad2006}. The top panel of Fig.~\ref{fig:bipolar_error} shows the evolution of the neutrinos over a period of $8\,\mathrm{s}$. The oscillations do indeed occur on approximately this timescale and match Fig.~1 in \cite{Hannestad2006}. However, the period of the oscillations in the numerical solution decreases over time due to numerical errors caused predominantly by the mapping of the evolution matrix onto the distribution function; applying the mapping less often causes the period to decrease more slowly. The third panel shows the relative change in the length of the distribution flavor vector with time, which should always be zero in this test without collision terms. It is once again clear that a longer $dt_\mathrm{block}$ prevents accumulation of error, and that the integration accuracy parameter does not significantly affect the solution on these timescales. Similarly, the distribution function should always be Hermitian. The non-Hermitivity of $f$ is shown in the bottom panel after each step prior to making $f$ Hermitian again. Once again, the largest errors occur during the flavor transitions. 

\subsection{Resolution}
\begin{figure}
    \centering
    \includegraphics[width=\linewidth]{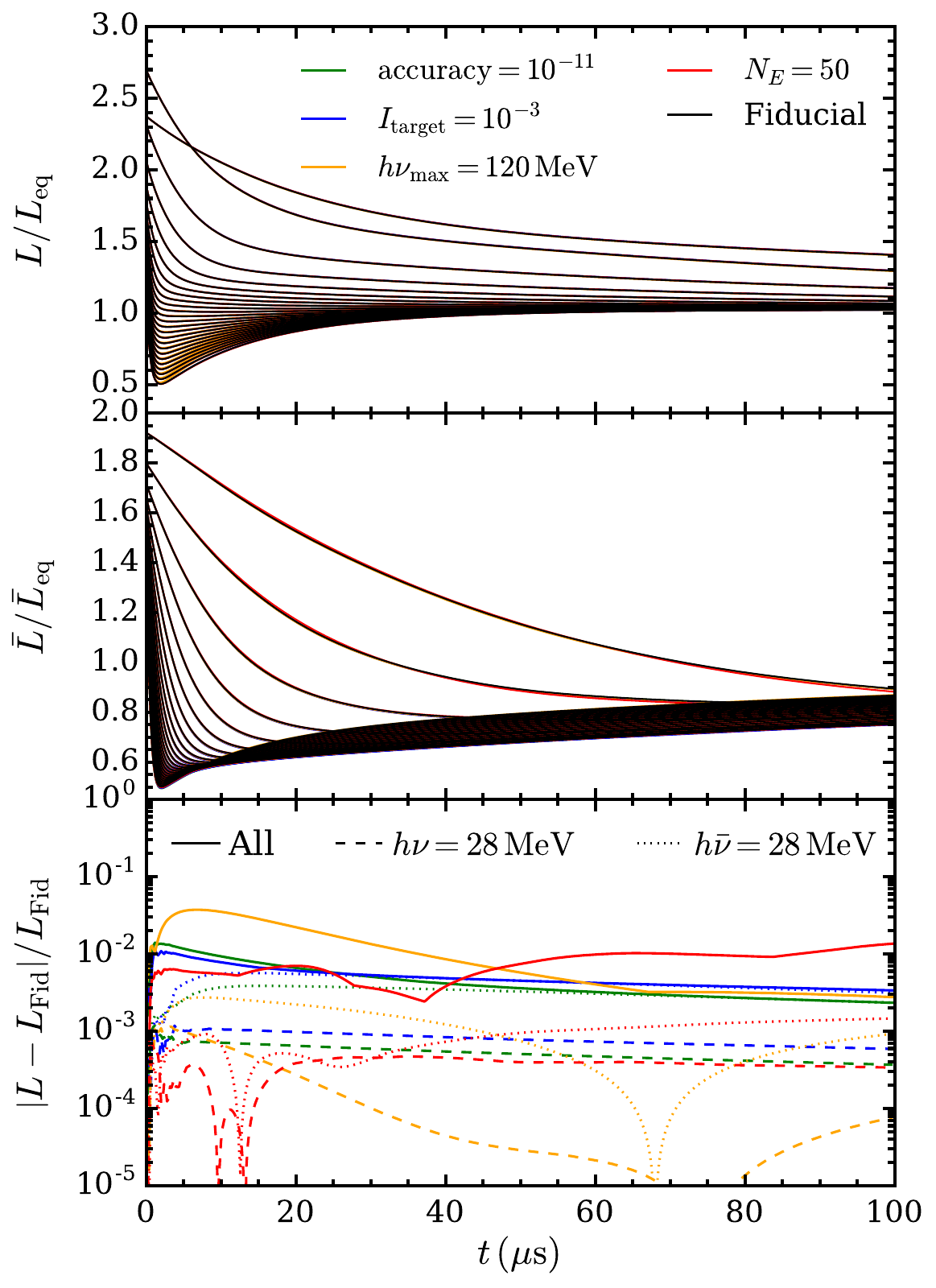}
    \caption{\textit{Fidelity tests}. - Calculations evolving the QKEs including oscillations and collisions. \textit{Top panel:} Neutrino distribution flavor vector length relative to the flavor-diagonal Fermi-Dirac length. \textit{Center panel:} antineutrino distribution flavor vector length. \textit{Bottom panel:} Relative error compared to the fiducial calculation. Black curves show the results of a fiducial calculation using 25 energy bins and an energy grid spacing of $4\,\mathrm{MeV}$, while red curves show the results from Sec.~\ref{sec:results} with 50 energy bins and an energy spacing of $2\,\mathrm{MeV}$. Green curves result from worsening the accuracy of the integrator to $10^{-11}$. Blue curves result from increasing the target impact to $10^{-3}$. Gold curves result from increasing the upper bound of the neutrino energy domain and the number of energy bins by 20\%, keeping the energy grid spacing constant. In the bottom panel, solid lines take the maximum over all energies and helicities, dashed lines show only errors for neutrinos at $28\,\mathrm{MeV}$, and dotted lines show only errors for antineutrinos at $28\,\mathrm{MeV}$.}
    \label{fig:qke_error}
\end{figure}

We demonstrate here that errors associated with our numerical treatment of the QKEs are within acceptable bounds. To do this, we rerun the combined oscillation and collision QKE calculation in Sec.~\ref{sec:results} under varied parameter choices. The black curves in the top two panels of Fig.~\ref{fig:qke_error} show the length of the distribution flavor vectors for neutrinos and antineutrinos, run with 25 energy bins and an energy grid spacing of $4\,\mathrm{MeV}$. Each curve represents one neutrino energy. The distribution flavor vector length is a particularly useful quantity to compare to, since it changes on collision timescales and does not oscillate violently as individual components of the neutrino distribution do. We aim to ensure that this macroscopic behavior converges and do not expect the instantaneous phases of the underlying oscillations to match between calculations.

Underneath the black curves in Fig.~\ref{fig:qke_error} are the corresponding curves from each of the simulations with numerical tweaks. The relative errors between the fiducial and tweaked calculations are shown in the same color in the bottom panel. First we look at the green curves, which result from decreasing the accuracy of the integrator by a factor of 10 from the fiducial case. This is clearly not the dominant source of error. The solid green curve in the bottom panel shows that the relative error maximized over all neutrino energies and both helicities is always less than $1.4\%$. The dashed and dotted lines show the neutrino and antineutrino errors at $28\,\mathrm{MeV}$, respectively, which is near the average energy of the distribution. The errors there are negligibly small.

Next, we increase the target impact by a factor of 10 and show the results as blue curves. These results sampled values of $dt_\mathrm{block}$ that are on average also a factor of 10 larger, causing the collisions to more sparsely sample the rapidly varying distributions. This also leads to at most a 1.3\% error. The sparse sampling seems only to make the evolution of $L$ noisier without significantly changing its long-term evolution.

Of course, we also need to check our energy grid resolution. The red curves (data from Sec.\ref{sec:results}) show that doubling the energy resolution leads to a much more systematic error of at most $\sim1.4\%$, and an even lower error at $28\,\mathrm{MeV}$. Here, we only calculate and plot the errors at neutrino energies at bin centers present in both grids (i.e., integer multiples of $4\,\mathrm{MeV}$). Interestingly, the results seem to be much more sensitive to the size of the energy domain. The gold curves demonstrate the effect of extending the grid out to $120\,\mathrm{MeV}$ with an additional five energy zones in order to keep the grid spacing constant. This leads to significant errors of up to 3.7\%, again where errors are only calculated for neutrino energies contained in both domains. However, we note that the largest errors are at the highest energies close to the energy boundary. If we instead look at the errors at $28\,\mathrm{MeV}$ (near the average neutrino energy of $31.5\,\mathrm{MeV}$), the error is always under 0.27\% for antineutrinos and 0.12\% for neutrinos. The fiducial grid spans a domain of $2-102\,\mathrm{MeV}$, corresponding to $99.7\%$ of the neutrino number and $99.0\%$ of the neutrino energy. The extended grid spans a domain of $2-122\,\mathrm{MeV}$, corresponding to $99.9\%$ of the neutrino number and $99.8\%$of the neutrino energy. Given the small number of neutrinos in these high-energy bins, it is unclear why the energy domain size is so important. 

\end{document}